\documentclass[journal]{IEEEtran}
\usepackage{amsfonts}
\usepackage{cite}
\usepackage{algorithmicx}
\usepackage{algorithm}
\usepackage{algpseudocode}
\usepackage{graphicx}
\usepackage{subfigure}
\usepackage{amsmath}
\usepackage{multirow}
\usepackage{array}
\usepackage{xcolor}
\usepackage{bm}
\usepackage{mathtools}
\usepackage{amssymb}

\newcommand{\myimgsize}{0.6}
\newcommand{\myimgsizer}{1.0}

\begin{document}
\title{Model Inspired Autoencoder for Unsupervised Hyperspectral Image Super-Resolution}

\author{Jianjun~Liu,~\IEEEmembership{Member,~IEEE,}
        Zebin~Wu,~\IEEEmembership{Senior~Member,~IEEE,}
        Liang~Xiao,~\IEEEmembership{Member,~IEEE,}
        and Xiao-Jun~Wu,~\IEEEmembership{Member,~IEEE}
\thanks{This work was supported in part by the National Natural Science Foundation of China under Grant No. 62071204, 61871226 and 61772274,
by the Natural Science Foundation of Jiangsu Province under Grant No. BK20201338 and BK20180018,
by the Jiangsu Provincial Social Developing Project under Grant No. BE2018727,
by the China Postdoctoral Science Foundation under Grant No. 2021M691275,
and by the Jiangsu Postdoctoral Research Funding Program under Grant No. 2021K148B.
}
\thanks{Jianjun Liu and Xiao-Jun Wu are with the Jiangsu Provincial Engineering Laboratory for Pattern Recognition and Computational Intelligence, Jiangnan University, Wuxi, China
(Email: liuofficial@163.com, wu\_xiaojun@jiangnan.edu.cn).}
\thanks{Zebin Wu and Liang Xiao are with the School of Computer Science,
Nanjing University of Science and Technology, Nanjing, China
(Email: zebin.wu@gmail.com, xiaoliang@mail.njust.edu.cn).}
}

\maketitle

\begin{abstract}
This paper focuses on hyperspectral image (HSI) super-resolution that aims to fuse a low-spatial-resolution HSI and a high-spatial-resolution multispectral image to form a high-spatial-resolution HSI (HR-HSI).
Existing deep learning-based approaches are mostly supervised that rely on a large number of labeled training samples, which is unrealistic.
The commonly used model-based approaches are unsupervised and flexible but rely on hand-craft priors.
Inspired by the specific properties of model, we make the first attempt to design a model inspired deep network for HSI super-resolution in an unsupervised manner.
This approach consists of an implicit autoencoder network built on the target HR-HSI that treats each pixel as an individual sample.
The nonnegative matrix factorization (NMF) of the target HR-HSI is integrated into the autoencoder network, where the two NMF parts, spectral and spatial matrices, are treated as decoder parameters and hidden outputs respectively.
In the encoding stage, we present a pixel-wise fusion model to estimate hidden outputs directly, and then reformulate and unfold the model's algorithm to form the encoder network.
With the specific architecture, the proposed network is similar to a manifold prior-based model, and can be trained patch by patch rather than the entire image.
Moreover, we propose an additional unsupervised network to estimate the point spread function and spectral response function.
Experimental results conducted on both synthetic and real datasets demonstrate the effectiveness of the proposed approach.
\end{abstract}


\IEEEpeerreviewmaketitle

\section{Introduction}
\IEEEPARstart{H} hyperspectral image (HSI) is a kind of three dimensional image taken at different spectral bands, with its spectral range covering hundreds of contiguous and narrow bands that span the visible to infrared spectrum. The high spectral resolution of HSIs promotes various applications, such as material identification. Due to the limited incident energy, there is always a tradeoff between spectral resolution, spatial resolution and signal-to-noise ratio of images when designing the imaging sensors \cite{alparone2007comparison,loncan2015hyperspectral,yokoya2017hyperspectral,meng2019review,Dian2021RecentAA,Vivone2021A}. Thus, the spatial resolution of HSIs is usually sacrificed, which impedes the subsequent tasks. Conversely, conventional multispectral images (MSIs) at much lower spectral resolution can be acquired with higher spatial resolution. An economical HSI super-resolution solution is to instead record a low-spatial-resolution HSI (LR-HSI) and a  high-spatial-resolution MSI (HR-MSI), and to fuse them into a target high-spatial-resolution HSI (HR-HSI) \cite{loncan2015hyperspectral,yokoya2017hyperspectral,Dian2021RecentAA}.

HSI super-resolution that fuses a LR-HSI with a HR-MSI has attracted great attention \cite{loncan2015hyperspectral,yokoya2017hyperspectral,Dian2021RecentAA}. This fusion problem arises from the Pansharpening problem that fuses a low-spatial-resolution MSI or HSI with a high-spatial-resolution panchromatic image \cite{alparone2007comparison,meng2019review,Vivone2021A}. Generally, the conventional approaches proposed for the Pansharpening problem can be extended to solve HSI super-resolution, but the fusion process of HSI super-resolution is relatively more complicated than that of Pansharpening due to the rich spectral information. Related fusion approaches can be roughly divided into four categories: component substitution \cite{tu2001new}, multiresolution analysis \cite{nencini2007remote}, model-based approaches and deep learning-based approaches. Among these categories, the research of model-based approaches is the most classic one, and deep learning-based approaches have been the most active one recently.

Model-based approaches consider building optimization models to obtain the target image. Given two observed images, they design fidelity terms and exploit spectral/spatial priors to enforce the desired result.
Some approaches treat the target image as a variable and recover the target image entirely, such as group spectral embedding  \cite{Zhang2017Multispectral}, clustering manifold structure \cite{Zhang2018Exploiting}, nonlocal patch tensor sparse representation \cite{Xu19}, and structured sparse low-rank representation \cite{Xue2021Spatial}.
Most approaches consider separating the target image into parts and regenerating it via the recovered parts. There are many decomposition strategies by making assumptions about the target image.
Examples are, that it lives in a low-dimensional subspace and the subspace-based models are solved by exploiting prior knowledge, such as piecewise smooth \cite{simoes2015convex}, dictionary learning \cite{wei2015hyperspectral}, tensor multi-rank \cite{dian2019hyperspectral}, and truncated matrix decomposition \cite{liu2020truncated};
or that it can be represented linearly by pure spectral signatures and the endmember and abundance matrices are recovered simultaneously \cite{yokoya2011coupled,lanaras2015hyperspectral,lin2018convex,Wu2020Hyperspectral};
or that it can be sparsely represented by an over-complete spectral dictionary and different priors are used to obtain the spectral dictionary and coefficients \cite{dong2016hyperspectral,veganzones2016hyperspectral,han2018self,yi2018hyperspectral,han2020hyperspectral};
or by approaches that separate the target image by tensor decomposition and update each component iteratively \cite{li2018fusing,kanatsoulis2018hyperspectral,Xu2020Hyperspectral,Chen2021Hyperspectral}.
Moreover, there are some approaches that build models to estimate the point spread function (PSF) and spectral response function (SRF) \cite{simoes2015convex,Bungert2018Blind}.
The entire process of model-based approaches is unsupervised. Although these models are flexible and their theory is relatively complete, they rely on hand-craft priors and there are many empirical parameters to tune.

Deep learning-based approaches are data-driven. They build deep neural networks to solve the related fusion problems, and produce the target image by feeding observed images into the network.
Some approaches enhance the ability to fuse images in the network structures, such as 3D convolutional neural networks (CNNs) \cite{palsson2017multispectral}, residual networks \cite{scarpa2018target}, multiscale structures \cite{yuan2018a}, pyramid networks \cite{zhang2019pan}, attention networks \cite{zheng2020hyperspectral,Jiang2020LearningSP}, cross-mode information \cite{Zhang2021SSR}, dense networks \cite{Dong2021Remote,Liu2021A}, and adversarial network \cite{Li2020Hyperspectral,Dong2021Generative}.
Some approaches use detail information from high-spatial-resolution conventional images to improve performance
\cite{yang2017pannet,xie2019hyperspectral,fu2021deep,Deng2021Detail}.
Inspired by the specific properties of model, some form a hybrid of model- and deep learning-based approaches \cite{dian2018deep,shen2019spatial,shen2020a}, and some use the deep unfolding technique to ease the construction of networks \cite{xie2019multispectral,wei2020deep,Dong2021Model}.
These approaches have shown good performance in exploiting the relationship between the observed and target images. However, they are mostly supervised that require plenty of labeled samples to train the networks, which limits their applications in many scenarios.

There are some deep learning-based approaches developed for HSI super-resolution that are performed in an unsupervised manner.
For instance, Dian \emph{et al.} \cite{dian2021regularizing} introduce a CNN denoiser to regularize the fusion model;
Zhang \emph{et al.} \cite{zhang2021deep} integrate the deep image prior into the fusion model, and thereby present a unified unsupervised network for HSI super-resolution;
Qu \emph{et al.} \cite{uSDN} exploit an unsupervised approach composed of two autoencoder networks, which are coupled through a shared decoder;
Wang \emph{et al.} \cite{FusionNet} propose a variational probabilistic autoencoder framework implemented by CNNs for HSI super-resolution;
Yao \emph{et al.} \cite{Yao2020Cross} propose a two-stream convolutional autoencoder framework inspired by coupled spectral unmixing, and introduce a cross-attention module to improve performance;
Uezato \emph{et al.} \cite{GDD} design a network composed of an encoder-decoder network and a deep decoder network;
Zheng \emph{et al.} \cite{HyCoNet} propose a network consisting of three coupled autoencoder networks, inspired by coupled spectral unmixing, where the three autoencoder networks are coupled through two convolutional layers.
Most approaches are built on the autoencoder architecture. Similar to model-based approaches, the construction of networks relies too much on human experience.

Inspired by the specific properties of model, we consider constructing an unsupervised network by referencing some models, and propose a model inspired autoencoder (MIAE) for unsupervised HSI super-resolution. Specifically, we perform nonnegative matrix factorization (NMF) on the target HR-HSI to maintain its intrinsic structure, and thereby propose an implicit autoencoder network for HR-HSI by integrating its NMF model. In the autoencoder network, each hyperspectral pixel is treated as an individual sample, and the two NMF parts of the target HR-HSI, i.e., spectral and spatial matrices, are treated as decoder parameters and hidden outputs respectively. Since the inputs of the autoencoder network are unknown, we take the two observed images as inputs and present a pixel-wise fusion model to estimate each hidden output vector directly. The pixel-wise fusion model is solved by the gradient descent algorithm, and the algorithm is reformulated and unfolded to form the encoder network. The loss function is just built on the mechanism of spectral and spatial degradations, and an additional blind estimation network is proposed to estimate the PSF and SRF. Compared with the existing HSI super-resolution approaches, some of the innovative characteristics of MIAE are highlighted as follows.
\begin{enumerate}
  \item MIAE is an unsupervised deep learning-based approach that involves only one implicit autoencoder. The autoencoder network treats each pixel as an individual sample, and thus the proposed network can be treated as a kind of manifold prior-based model and can be trained patch by patch to accelerate the training process.
  \item MIAE is constructed by referencing models, and thus the construction of the network is relatively concise. The NMF of the target HR-HSI is integrated into the autoencoder, and the encoder network is inspired by the pixel-wise fusion model.
  \item An additional unsupervised network is proposed to estimate the PSF and SRF from the two observed images directly.
\end{enumerate}

The remainder of this paper is organized as follows. Section \ref{sec_prop} proposes the proposed MAIE and its relationship to the model-based approaches, as well as the blind estimation network. In Section \ref{sec_ex}, the effectiveness of MIAE is demonstrated through experiments on three synthetic datasets and one real dataset. Section \ref{sec_con} provides concluding remarks.

\section{Proposed Approach}
\label{sec_prop}
\begin{figure}[!t]
 {\includegraphics[width=0.48 \textwidth]{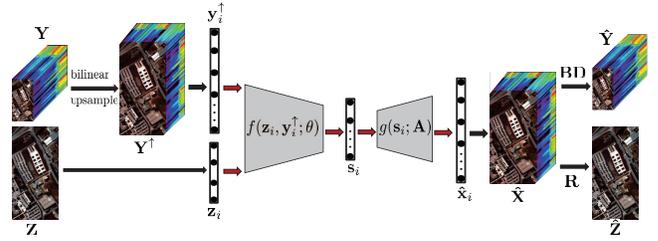}}
\caption{The overall architecture of MIAE.}
\label{fig_main}
\end{figure}

Fig. \ref{fig_main} illustrates the overall architecture of MIAE. The details of the proposed network are described as follows.

\subsection{NMF Inspired Autoencoder}
NMF is a useful dimension reduction method \cite{Lee1999LearningTP}. It can capture the intrinsic structure of the data and represent the data in a sparse manner. The properties of NMF indicate that it can facilitate the inference process of super-resolution if we perform NMF on the target HR-HSI. Let us represent the target HR-HSI as a matrix ${\bf \hat{X}} \in \mathbb{R}^{N_B \times N_H N_W}$, where $N_B$ denotes the spectral band, and $N_H$ and $N_W$ denote the spatial height and width respectively. NMF aims to factor ${\bf \hat{X}}$ into two rank-$J$ ($J < \min\{N_B, N_H N_W\}$) nonnegative matrices, i.e.,
\begin{equation}
  {\bf \hat{X}} \approx {\bf A} {\bf S}, \label{eq_nmf}
\end{equation}
where the spectral matrix ${\bf A} \in \mathbb{R}^{N_B \times J} \succeq 0$ and the spatial matrix ${\bf S} \in \mathbb{R}^{J \times N_H N_W} \succeq 0$ with $\succeq$ being a component-wise inequality.

NMF can be integrated into an autoencoder network \cite{Qu2019uDASAU,Qian2020Spectral,Palsson2021Convolutional}. Equation (\ref{eq_nmf}) can be rewritten as
\begin{equation}
  {\bf \hat{x}_i} \approx {\bf A}{\bf s}_i, \forall i
\end{equation}
where $i=1, 2, \cdots, N_HN_W$, and ${\bf \hat{x}_i} \in \mathbb{R}^{N_B}$ and ${\bf s}_i \in \mathbb{R}^{J}$ are the column vectors of ${\bf \hat{X}}$ and ${\bf S}$, respectively. Let ${\bf \hat{x}}_i$ represent the reconstructed vector and ${\bf s}_i$ represent the hidden output vector, we can construct the following autoencoder network
\begin{equation}
  {\bf x}_i \overset{f(\cdot)} {\longrightarrow} {\bf s}_i \overset{g(\cdot)} {\longrightarrow} {\bf \hat{x}}_i, \forall i \label{eq_ae}
\end{equation}
where the input data ${\bf x}_i \in \mathbb{R}^{N_B}$ is the column vector of the observed HR-HSI ${\bf X} \in \mathbb{R}^{N_B \times N_HN_W}$. The network (\ref{eq_ae}) consists of two networks $f(\cdot)$ and $g(\cdot)$. $f(\cdot)$ is the encoder network with ${\bf s}_i = f({\bf x}_i; \theta)$, where $\theta$ denotes all trainable parameters involved in the network. $g(\cdot)$ is the decoder network with ${\bf \hat{x}}_i = g({\bf s}_i; {\bf A})= \mathcal{C}_0^1({\bf A}{\bf s}_i)$, where ${\bf A}$ is treated as the trainable weight matrix and $\mathcal{C}_0^1(\cdot)$ is a clamp function that forces all elements of the input vector/matrix into the range $[0,1]$.
In hyperspectral unmixing \cite{Qu2019uDASAU,Qian2020Spectral,Palsson2021Convolutional}, sum-to-one constraint is added to enforce the hidden vector, i.e., ${\bf 1}_{J}^T{\bf s}_i=1$ with ${\bf 1}_{J} \in \mathbb{R}^J$ being a vector of all 1s. We do not intend to finish the two tasks of fusion and unmixing at once and only use the nonnegative constraints.
Specifically, ${\bf s}_i$ and ${\bf A}$ are enforced using $\mathcal{C}_0^1({\bf s}_i)$ and $\mathcal{C}_0^1({\bf A})$, when designing the network.
Then, if the input data ${\bf X}$ is given, one can train the network (\ref{eq_ae}) by feeding its $N_HN_W$ hyperspectral pixels.

\subsection{Pixel-Wise Fusion Model Inspired Encoder Network}

\begin{figure*}[!t]
\centering
 {\includegraphics[width=0.7 \textwidth]{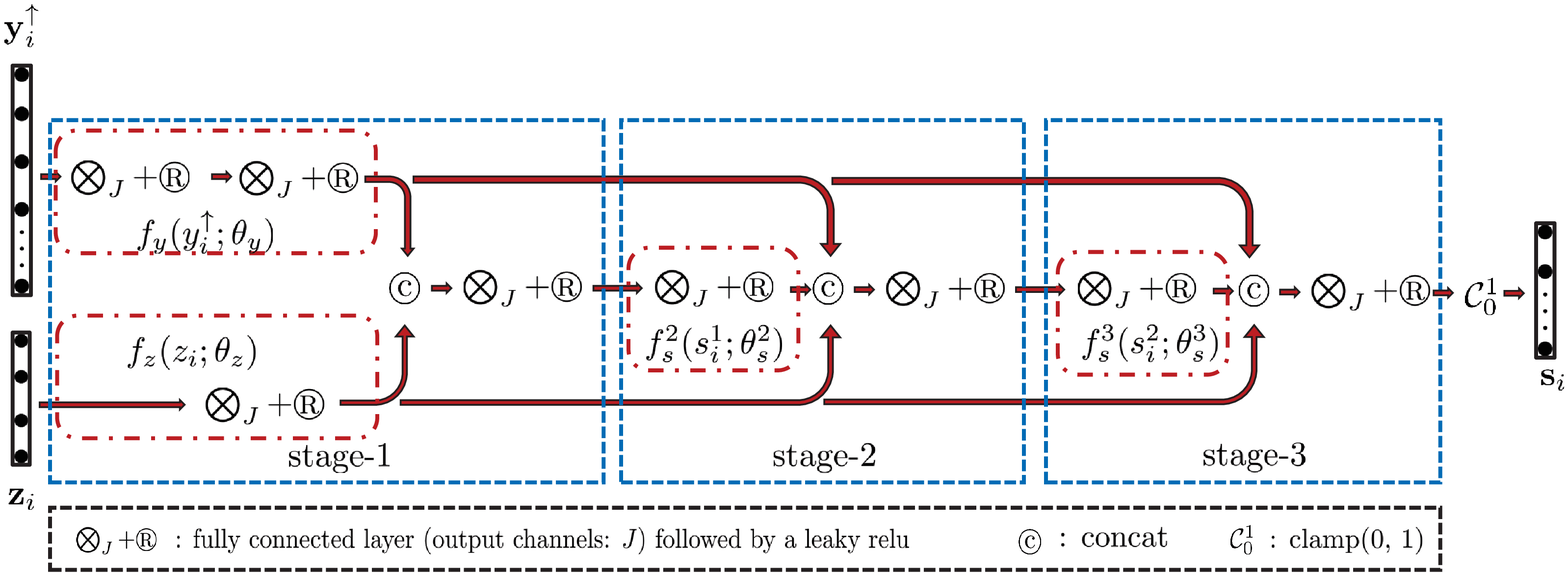}}
\caption{Details of the proposed encoder network when $K=3$.}
\label{fig_encoder}
\end{figure*}

In HSI super-resolution, the input data ${\bf X} \in \mathbb{R}^{N_B \times N_H N_W}$ is not given. One can not train the network (\ref{eq_ae}) directly. Instead, we have two observed (i.e., degenerated) images of ${\bf X}$, a LR-HSI ${\bf Y} \in \mathbb{R}^{N_B \times N_hN_w}$ and a HR-MSI ${\bf Z} \in \mathbb{R}^{N_b \times N_HN_W}$, where $N_b < N_B$ is the multispectral band, and $N_h$ and $N_w$ are the spatial sizes. We assume that $N_H = rN_h$ and $N_W = rN_w$ with $r>1$ being the resolution ratio. The observations ${\bf Y}$ and ${\bf Z}$ can be modeled as spatially degraded and spectrally
degraded versions of ${\bf X}$. Specifically, these two degeneration processes can be written as:
\begin{eqnarray}
  {\bf Y} &\approx& {\bf X}{\bf B}{\bf D} \label{eq_y}\\
  {\bf Z} &\approx& {\bf R}{\bf X} \label{eq_z}
\end{eqnarray}
where the PSF ${\bf B} \in \mathbb{R}^{N_HN_W \times N_HN_W}$ is the spatial blur, ${\bf D} \in \mathbb{R}^{N_HN_W \times N_hN_w}$ is the spatial downsampling, and ${\bf R} \in \mathbb{R}^{N_b \times N_B}$ is the SRF of multispectral sensor.

In (\ref{eq_ae}), the network needs to be trained by feeding the input data pixel by pixel. It is the key to the success of autoencoder. For the degeneration processes, (\ref{eq_z}) can be rewritten as a pixel-wise formulation, i.e., ${\bf z}_i \approx {\bf R} {\bf x}_i$ with ${\bf z}_i \in \mathbb{R}^{N_b}$ being the column vector of ${\bf Z}$, whereas (\ref{eq_y}) can't because of the coupling matrices ${\bf B}$ and ${\bf D}$. We consider resizing ${\bf Y}$ to the same size as ${\bf X}$ using bilinear interpolation, in order to approximate ${\bf X}$ pixel by pixel at the spectral level. ${\bf x}_i$ can be obtained by solving
\begin{equation}
  \min_{{\bf x}_i} \frac{1}{2}\|{\bf z}_i - {\bf R}{\bf x}_i\|_2^2 + \frac{\lambda}{2} \|{\bf y}_i^{\uparrow} -{\bf x}_i\|_2^2, \forall i
\end{equation}
where $\lambda>0$ is the regularization parameter, and ${\bf y}_i^{\uparrow} \in \mathbb{R}^{N_B}$ is the column vector of ${\bf Y}^{\uparrow} \in \mathbb{R}^{N_B \times N_HN_W}$ with ${\bf Y}^{\uparrow}$ being a bilinear interpolated version of ${\bf Y}$.

In order to construct the encoder network, it is unnecessary to solve ${\bf x}_i$ and then design $f(\cdot)$, which will lead to error accumulation. We can treat ${\bf x}_i$ as an implicit variable and design $f(\cdot)$ by using ${\bf z}_i$ and ${\bf y}_i^{\uparrow}$ directly, i.e.,
\begin{equation}
  ({\bf z}_i, {\bf y}_i^{\uparrow}) \overset{f(\cdot)} {\longrightarrow} {\bf s}_i \overset{g(\cdot)} {\longrightarrow} {\bf \hat{x}}_i, \forall i \label{eq_ae2}
\end{equation}
Specifically, we want to obtain the hidden layer output ${\bf s}_i$ by solving the following pixel-wise fusion model
\begin{equation}
  \min_{{\bf s}_i} \frac{1}{2}\|{\bf z}_i - {\bf R}{\bf A}{\bf s}_i\|_2^2 + \frac{\lambda}{2} \|{\bf y}_i^{\uparrow} -{\bf A}{\bf s}_i\|_2^2, \forall i \label{eq_model}
\end{equation}
and design $f(\cdot)$ by unfolding all steps of its algorithm as network layers. Notably, in (\ref{eq_model}) both ${\bf R}$ and ${\bf A}$ are treated as new trainable parameters to facilitate the design of the encoder network. In the model-based HSI super-resolution, our pervious works \cite{liu2020truncated,shen2020a} have shown the effectiveness of (\ref{eq_model}).

Although (\ref{eq_model}) has an analytic solution, it is not suitable as the encoder network and is difficult to implement by a network. (\ref{eq_model}) can be solved by the gradient descent algorithm as
\begin{equation}
  {\bf s}_i^{k} = {\bf s}_i^{k-1} - \eta (\bar{\bf A}^T \bar{\bf A} {\bf s}_i^{k-1} - \bar{\bf A}^T {\bf z}_i + \lambda {\bf A}^T {\bf A} {\bf s}_i^{k-1} - \lambda {\bf A}^T {\bf y}_i^{\uparrow}), \label{eq_grad}
\end{equation}
where $\bar{\bf A} = {\bf R}{\bf A} \in \mathbb{R}^{N_b \times J}$, $\eta > 0$ is the step, and $k=1, 2, \cdots, K$ represents the $k$-th iteration. To better design the network, input data and intermediate variables are distinguished by rewriting (\ref{eq_grad}) as
\begin{equation}
  {\bf s}_i^{k} = ({\bf I} - \eta \bar{\bf A}^T \bar{\bf A} - \eta \lambda {\bf A}^T {\bf A}) {\bf s}_i^{k-1}
  + \eta \bar{\bf A}^T {\bf z}_i + \eta \lambda {\bf A}^T {\bf y}_i^{\uparrow}, \label{eq_grad2}
\end{equation}
where ${\bf I}$ represents the identity matrix.
According to the $K$ iterations of (\ref{eq_grad2}), the proposed encoder network is mainly a structure of $K$ stages.
Fig. \ref{fig_encoder} illustrates the details of $f(\cdot)$ when $K=3$.
In (\ref{eq_grad2}), all variables ${\bf s}_i^{k-1}$, ${\bf z}_i$ and ${\bf y}_i^{\uparrow}$ are left multiplied by a matrix. This process is implemented using a fully connected layer followed by a Leaky ReLU, and (\ref{eq_grad2}) can be rewritten as
\begin{equation}
  {\bf s}_i^{k} = f_s^k({\bf s}_i^{k-1};\theta_s^k) + \eta f_z ({\bf z}_i;\theta_z) + \eta \lambda f_y ({\bf y}_i^{\uparrow};\theta_y), \label{eq_encoder}
\end{equation}
where $\{f_s^k\}_{k=2}^{K}$, $f_z$ and $f_y$ represent the modules designed for the multiplication of matrix and vector, and $\{\theta_s^k\}_{k=2}^{K}$, $\theta_z$ and $\theta_y$ represent the trainable parameters involved in the corresponding networks. The red dotted boxes in Fig. \ref{fig_encoder} show the layers of $\{f_s^k\}_{k=2}^{K}$, $f_z$ and $f_y$, where two fully connected layers are used in $f_y$ for the purpose of feature extraction.
In (\ref{eq_encoder}), the three modules are combined linearly. To break the fixed format of optimization model and provide more flexibility, the linear combination is implemented by concatenating these modules and performing a fully connection and a leaky ReLU. (\ref{eq_encoder}) can be rewritten as
\begin{equation}
  {\bf s}_i^{k} = f^k (f_s^k({\bf s}_i^{k-1};\theta_s^k), f_z ({\bf z}_i;\theta_z), f_y ({\bf y}_i^{\uparrow};\theta_y); \theta^k),
\end{equation}
where $\{f^k\}_{k=1}^{K}$ and $\{\theta^k\}_{k=1}^{K}$ represent the modules and trainable parameters for the linear combination. Finally, by performing $f^k$ from $1$ to $K$, we can obtain the hidden output by ${\bf s}_i = \mathcal{C}_0^1({\bf s}_i^K)$ and have the trainable parameters $\theta = \{\theta_z, \theta_y, \{\theta_s^k\}_{k=2}^{K}, \{\theta^k\}_{k=1}^{K}\}$.

\subsection{Loss Function and Training Strategy}
\label{sec_lf}
To train the autoencoder network (\ref{eq_ae2}), we can't build loss function by using the target ${\bf \hat{x}}_i$ directly, since the input data ${\bf x}_i$ is just an implicit variable. Instead, we have two observation images ${\bf Y}$ and ${\bf Z}$ to work with. Similar to (\ref{eq_y}) and (\ref{eq_z}), the outputs LR-HSI ${\bf \hat{Y}}$ and HR-MSI ${\bf \hat{Z}}$ can be modeled as ${\bf \hat{Y}} \approx {\bf \hat{X}} {\bf BD}$ and ${\bf \hat{Z}} \approx {\bf R}{\bf \hat{X}}$.
To measure the difference between the outputs and observations, the $l_1$-norm is used because it is more robust to outliers than the $l_2$-norm. Then, the overall loss function can be written as
\begin{equation}
  \mathcal{L} = \|{\bf Z} - {\bf \hat{Z}}\|_{1,1} + \|{\bf Y} - {\bf \hat{Y}}\|_{1,1}, \label{eq_loss}
\end{equation}
where $\|\cdot\|_{1,1}$ represents the absolute sum of all the matrix elements.

When training the network (\ref{eq_ae2}) using the loss function (\ref{eq_loss}), one has to combine all pixels ${\bf \hat{x}}_i$ into an image ${\bf \hat{X}}$, since ${\bf \hat{Y}}$ and ${\bf \hat{X}}$ are coupled together by ${\bf BD}$.
In other words, one has to train (\ref{eq_ae2}) by feeding ${\bf Z}$ and ${\bf Y^{\uparrow}}$ entirely.
In spite of this, we can still train (\ref{eq_ae2}) by using small patches to accelerate the training process.
Specifically, we divide ${\bf Z}$ and ${\bf Y^{\uparrow}}$ into overlapped patches so that the patches cover all pixels,
and then discard the pixels affected by spatial blur ${\bf B}$ at the boundaries of these patches,
when computing the loss function (\ref{eq_loss}).

\subsection{Blind Estimation Network}
\label{sec_blind}
\begin{figure}[!t]
 {\includegraphics[width=0.48 \textwidth]{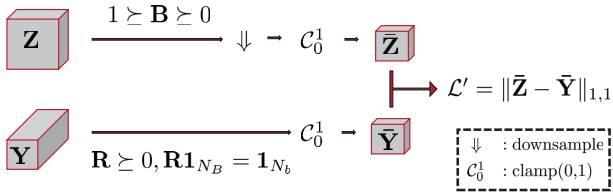}}
\caption{Details of the blind estimation network.}
\label{fig_br}
\end{figure}

The PSF ${\bf B}$ and SRF ${\bf R}$ are required to train the proposed autoencoder network.
By combing (\ref{eq_y}) and (\ref{eq_z}), we have
\begin{equation}
  {\bf Z}{\bf BD} \approx {\bf R}{\bf Y}.
\end{equation}
By imposing some physical constraints, one can obtain ${\bf B}$ and ${\bf R}$ by solving
\begin{eqnarray}
  \min_{{\bf B},{\bf R}} && \|{\bf Z}{\bf BD} - {\bf R}{\bf Y}\|_{1,1} \nonumber \\
  {\rm s.t.} && 1 \succeq {\bf B} \succeq 0, {\bf R} \succeq 0, {\bf R} {\bf 1}_{N_B} = {\bf 1}_{N_b} \label{eq_br}
\end{eqnarray}

Problem (\ref{eq_br}) can be solved by some optimization algorithms. Instead, we solve (\ref{eq_br}) by training a network. Specifically, we treat $({\bf Z}, {\bf Y})$ as inputs and $({\bf B}, {\bf R})$ as trainable parameters. Then, the blind estimation network can be constructed by using the following loss function
\begin{equation}
  \mathcal{L}' = \|{\bf \bar{Z}} - {\bf \bar{Y}}\|_{1,1},
\end{equation}
where ${\bf \bar{Z}} = \mathcal{C}_0^1({\bf Z}{\bf BD})$ and ${\bf \bar{Y}} = \mathcal{C}_0^1({\bf R}{\bf Y})$ represent the output data. The details of the blind estimation network are illustrated in Fig. \ref{fig_br}.

\subsection{Relationship to Model-Based Approaches}
The proposed MIAE can be regarded as a kind of specific fusion model.
By combing (\ref{eq_nmf}), (\ref{eq_ae2}) and (\ref{eq_loss}), MIAE can be rewritten as
\begin{eqnarray}
  \min_{{\bf A},\theta} && \|{\bf Z} - {\bf RAS} \|_{1,1} + \|{\bf Y} - {\bf ASBD} \|_{1,1} \nonumber \\
  {\rm s.t.} && {\bf s}_i = f({\bf z}_i, {\bf y}_i^{\uparrow};\theta), \forall i \label{eq_opt}\\
  &&1 \succeq {\bf A} \succeq 0, 1 \succeq {\bf S} \succeq 0, 1 \succeq {\bf AS} \succeq 0 \nonumber
\end{eqnarray}
In (\ref{eq_opt}), $f(\cdot)$ can be thought of as a nonlinear mapping function,
and each ${\bf s}_i$ is only associated with the input ${\bf z}_i$ and ${\bf y}_i^{\uparrow}$ that correspond to its spatial position.
The constraint ${\bf s}_i = f({\bf z}_i, {\bf y}_i^{\uparrow};\theta)$ acts as a manifold regularization that embeds the combination of ${\bf z}_i$ and ${\bf y}_i^{\uparrow}$ into a low-dimensional space $\mathbb{R}^J$.
${\bf s}_i = f({\bf z}_i, {\bf y}_i^{\uparrow};\theta)$ also acts as a self-supervised deep prior regularization that only uses itself as training data.

\section{Experimental Results and Analysis}
\label{sec_ex}
In this section, experiments on both synthetic and real datasets are conducted to demonstrate the performance of the proposed MIAE. Before the following experiments, all datasets are scaled to the range $[0,1]$.
The quality of the fused images in the synthetic datasets are assessed with root mean squared error (RMSE), peak signal-noise-ratio (PSNR), spectral angle mapper (SAM), relative dimensionless global error in synthesis
(ERGAS), and universal image quality index (UIQI) \cite{loncan2015hyperspectral,yokoya2017hyperspectral,Dian2021RecentAA}.

\subsection{Synthetic Datasets and Implementation Details}
Three real-life HSI datasets, University of Paiva (PaviaU), Kennedy Space Center (KSC), and Washington DC Mall (DC) are manipulated to use as synthetic reference images for the simulation experiments.

\begin{enumerate}
\item The PaviaU dataset is acquired by the Reflective Optics System Imaging Spectrometer (ROSIS), with a spectral range of 0.43 to 0.86 $\mu m$. The ROSIS sensor is characterized by 115 spectral bands and 103 remained after removal of noisy bands. This image, with size of $610 \times 340$ pixels, has spatial resolution of 1.3 $m$ per pixel. We select the up-left $512 \times 256$-pixel part as the reference image.
\item The KSC dataset is acquired by the Airborne Visible/Infrared Imaging Spectrometer (AVIRIS), with a spectral range of 0.4 to 2.5 $\mu m$. The AVIRIS sensor is characterized by 224 spectral bands and the number of spectral bands is reduced to 176 by removing water absorption bands. The size of this image is $512 \times 614$ with a spatial resolution of 18 $m$. We select the up-left $512 \times 512$-pixel part as the reference image.
\item The DC dataset is acquired by the Hyperspectral digital imagery collection experiment (HYDICE) image, with a spectral range of 0.4 to 2.4 $\mu m$. The HYDICE sensor is characterized by 210 spectral bands, and bands in the region where the atmosphere is opaque have been removed, leaving 191 bands. This image, with size of $1208 \times 307$ pixels, has a spatial resolution of about 2.8 $m$. We select a $512 \times 256$-pixel part as the reference image.
\end{enumerate}

For each reference image, we generate the two observation images, LR-HSI and HR-MSI, according to the Wald’s protocol \cite{ranchin2000fusion}. To generate the LR-HSI, we spatially blur the reference image and then downsample it by a factor of 8 ($r=8$) in each direction. A Gaussian blur of $15 \times 15$ pixels, with a mean of 0 and a standard deviation of 3.40, is applied to each band of the reference image. To generate the HR-MSI, ${\bf R}$ is derived from the spectral response of the IKONOS satellite. We generate a 4-band image by averaging the bands of the reference image according to the spectral response profiles of the RGB and NIR bands. To account for ubiquitous noise or error, moderate Gaussian noise is added to the LR-HSI (SNR = 30 dB) and the HR-MSI (SNR = 40 dB).

We implement and train the proposed network and blind estimation network using the PyTorch framework.
As discussed in Section \ref{sec_lf}, we divide the observation images into patches to accelerate the training process. Take the HR-MSI as a reference, $40 \times 40$-pixel overlapping patches with a stride of 24 are extracted for training.
The batch sizes are 25 for the PaviaU and DC datasets, and 50 for the KSC dataset.
An Adam optimizer is used to train the network for 10000 iterations.
The learning rate is initialized as $5 \times 10^{-3}$ and gradually decayed by multiplying $1 - \frac{1}{9000}\max(0, iteration - 1000)$, where `$iteration$' represents the current number of iterations.
As for the blind estimation network, it is trained by feeding the observed images entirely, the total number of iterations is 5000, and the learning rate is set as $5 \times 10^{-5}$.

\subsection{Influence of Parameters}
\begin{figure*}[!t]
\centering
\subfigure[] {\includegraphics[width=0.3 \textwidth]{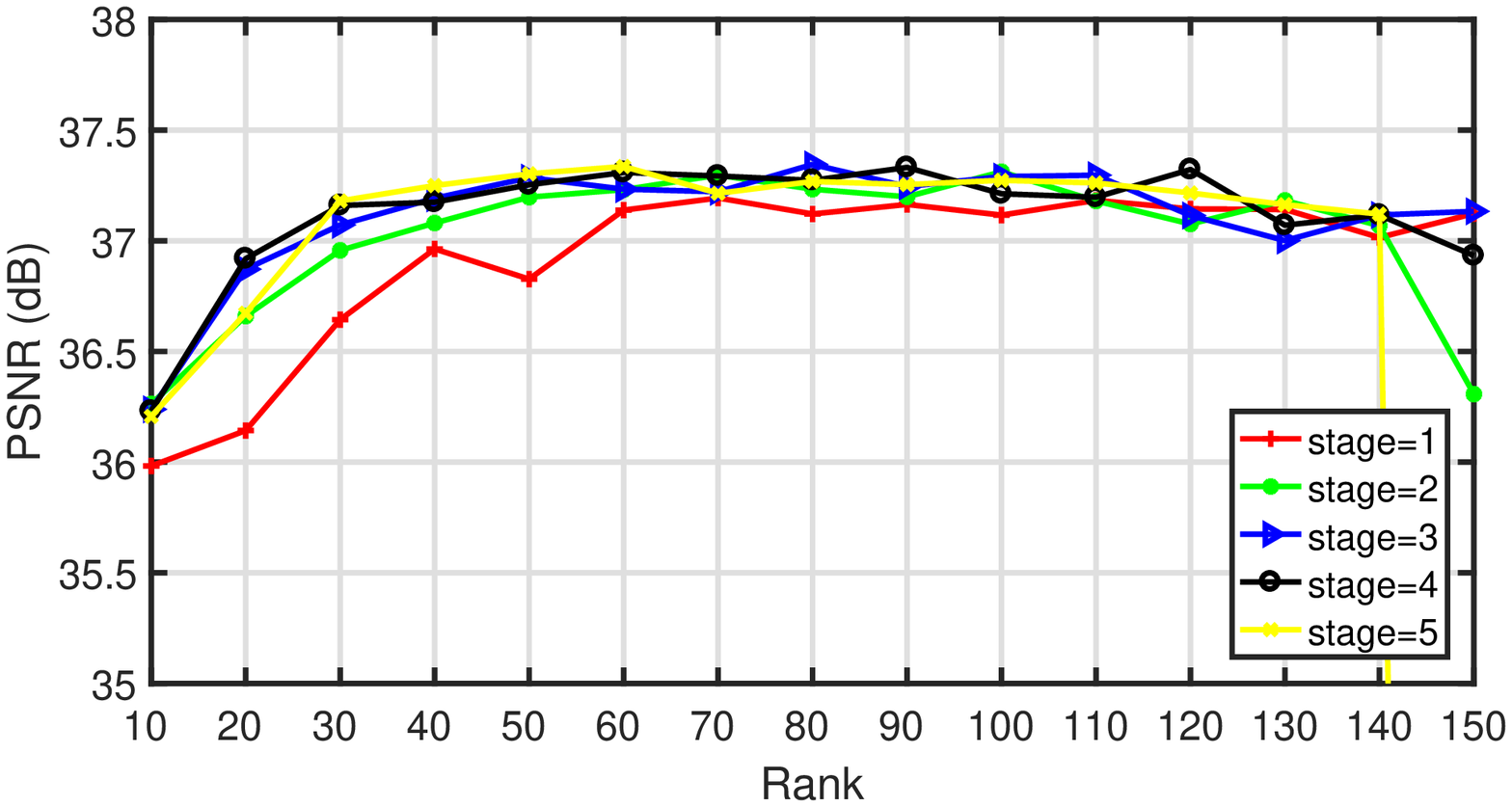}}~~~
\subfigure[] {\includegraphics[width=0.3 \textwidth]{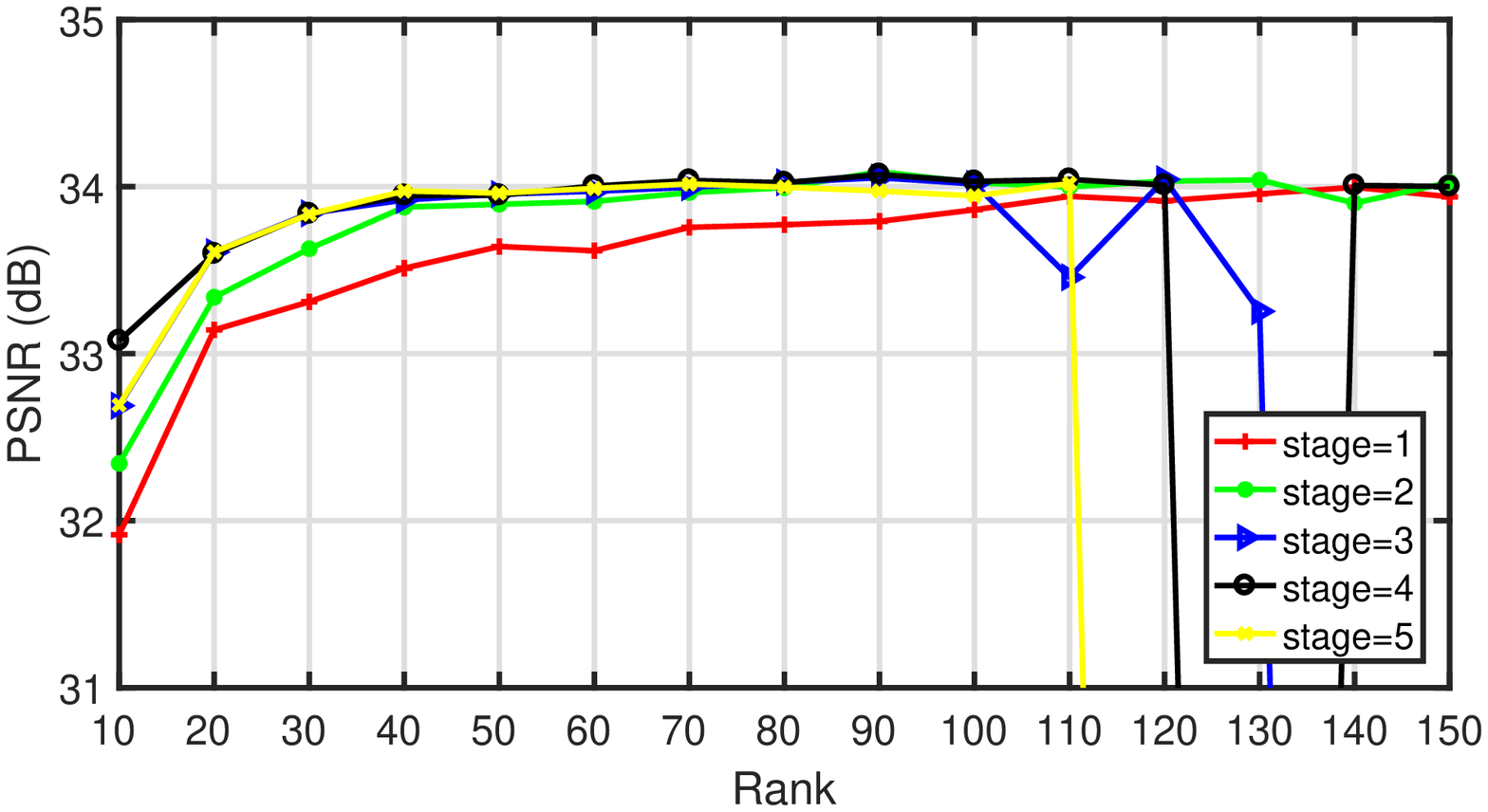}}~~~
\subfigure[] {\includegraphics[width=0.3 \textwidth]{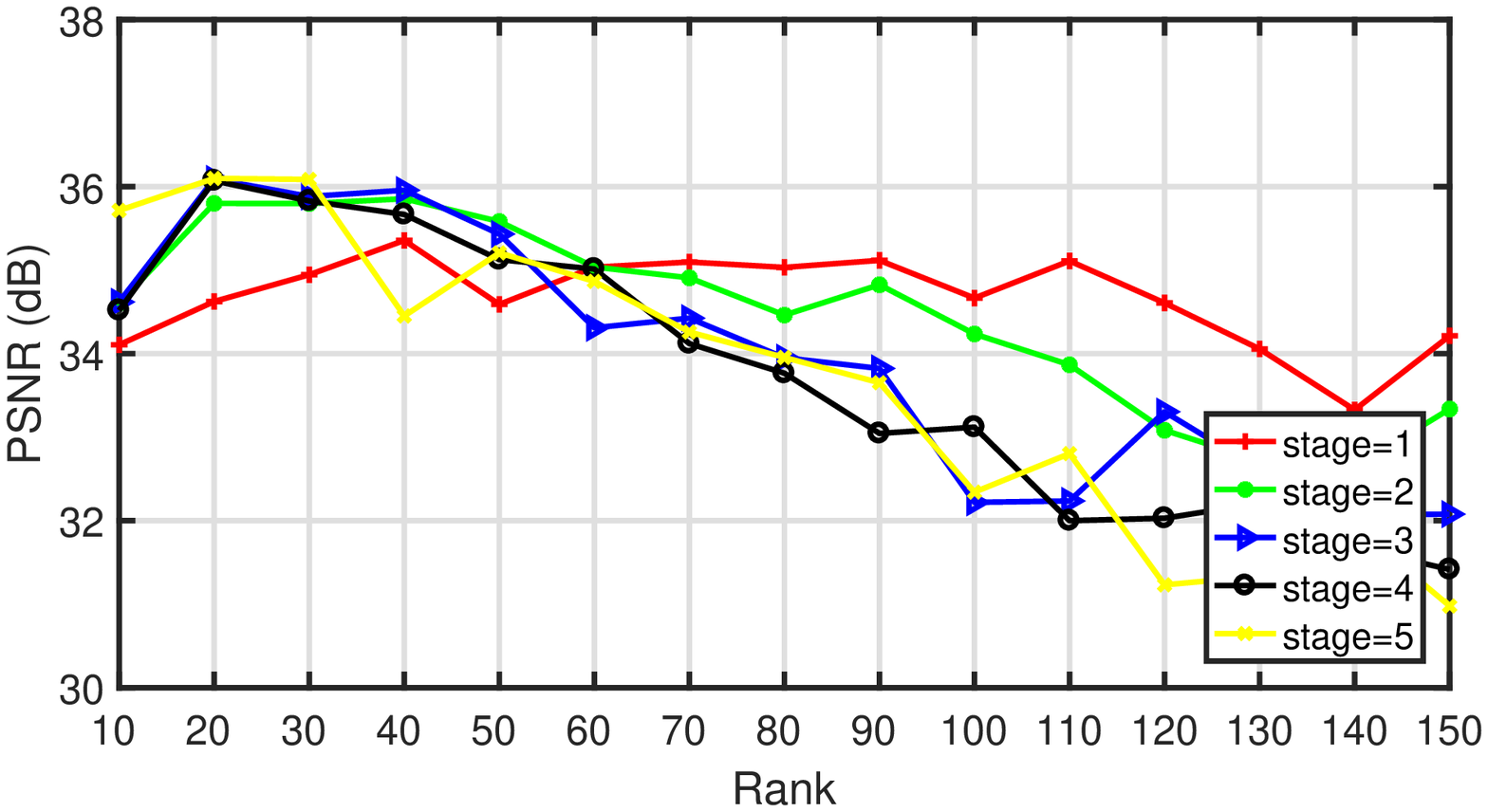}}
\caption{PSNR as a function of rank $J$ when using different stages $K$. (a) PaviaU dataset. (b) KSC dataset. (c) DC dataset.}
\label{fig_edm_stage}
\end{figure*}
Two parameters, rank $J$ and stage $K$, need to be given when constructing the proposed network.
In this set of experiments, we investigate them and show how they impact quality measures of MIAE.
Fig. \ref{fig_edm_stage} illustrates the PSNR results of MIAE as a function of $J$ when $K=1, 2, \cdots, 5$.
It can be seen that, for all datasets, the PSNR performance improves as $J$ increases, but a large $J$ will cause overfitting or performance degradation. Compared with small values of $K$, a moderate $K$ is better and a too large $K$ is prone to overfitting. Thus, $K$ is eventually set as 3 for all datasets, and $J$ is eventually set as 80 for the PaviaU and KSC datasets and 30 for the DC dataset.

\subsection{Experiment Results on Synthetic Datasets}
\subsubsection{Blind and Nonblind}
\begin{table}[!t]
\caption{Quality measures of nonblind and blind MIAE}
\label{tab_blind}
\centering
\tabcolsep = 3.0pt
\begin{tabular}
{c|c||c|c||c|c||c|c}
\hline\hline
\multicolumn{2}{c}{}
&\multicolumn{2}{c}{PaviaU} &\multicolumn{2}{c}{KSC} &\multicolumn{2}{c}{DC}\\
\hline
      &best & nonblind & blind & nonblind & blind & nonblind & blind\\
\hline
      RMSE &0 &0.0169 &0.0172 &0.0426 &0.0427 &0.0127 &0.0145 \\
      PSNR &+$\infty$ &37.57 &37.33 &34.21 &34.05 &37.48 &35.90 \\
       SAM &0 &2.41 &2.43 &6.98 &7.00 &1.56 &1.84 \\
     ERGAS &0 &0.647 &0.656 &3.122 &3.129 &14.184 &14.224 \\
      UIQI &1 &0.988 &0.988 &0.882 &0.882 &0.983 &0.975 \\
\hline\hline
\end{tabular}
\end{table}

Section \ref{sec_blind} presents a blind estimation network for estimating the PSF and SRF.
This experiment is used to evaluate the estimated ${\bf B}$ and ${\bf R}$.
Table \ref{tab_blind} shows the quality measures of the proposed MIAE using the exact and estimated ${\bf B}$ and ${\bf R}$, that is, nonblind and blind cases.
It can be seen that, for the PaviaU and KSC datasets, the performance degradation caused by blind estimation is very small when compared with the nonblind estimation; and for the DC dataset, the performance degradation is also not significant.

\subsubsection{Influence of LR-HSI Interpolation}
\begin{table*}[!t]
\caption{Quality measures of MIAE using different interpolation methods}
\label{tab_upsample}
\centering
\begin{tabular}
{c|c||c|c|c|c||c|c|c|c||c|c|c|c}
\hline\hline
\multicolumn{2}{c}{}
&\multicolumn{4}{c}{PaviaU} &\multicolumn{4}{c}{KSC} &\multicolumn{4}{c}{DC}\\
\hline
      &best & bilinear & nearest & bicubic & spline & bilinear & nearest & bicubic & spline & bilinear & nearest & bicubic & spline\\
\hline
      RMSE &0&0.0172 &0.0172 &0.0171 &0.0172 &0.0427 &0.0431 &0.0428 &0.0426 &0.0145 &0.0151 &0.0148 &0.0142 \\
      PSNR &+$\infty$ &37.33 &37.28 &37.30 &37.27 &34.05 &33.88 &34.02 &34.04 &35.90 &35.61 &34.77 &35.78 \\
       SAM &0 &2.43 &2.43 &2.41 &2.43 &7.00 &7.11 &7.01 &7.01 &1.84 &1.93 &1.83 &1.82 \\
     ERGAS &0 &0.656 &0.656 &0.652 &0.653 &3.129 &3.164 &3.135 &3.130 &14.224 &14.204 &14.349 &14.188 \\
      UIQI &1 &0.988 &0.988 &0.988 &0.988 &0.882 &0.879 &0.882 &0.881 &0.975 &0.974 &0.975 &0.975 \\
\hline\hline
\end{tabular}
\end{table*}

For the proposed MIAE, bilinear interpolation is used to upsample the LR-HSI to the same size of the target HR-HSI. This experiment shows how the interpolation method affects the performance of MIAE.
Four interpolation methods are considered, i.e., bilinear interpolation, nearest interpolation, bicubic interpolation and cubic spline interpolation.
The quality measures to assess the different interpolation methods are given in Table \ref{tab_upsample}.
In most cases, there is no obvious difference between these interpolation methods.
The nearest interpolation performs slightly worse on the KSC dataset, and the bicubic interpolation on the DC dataset.

\subsubsection{Comparison With the State of the Art}
\label{sec_art}
\begin{table*}[!t]
\caption{Quality measures for the PaviaU dataset using different methods (the best values are highlighted)}
\label{tab_pavia}
\centering
\begin{tabular}
{c|c|c|c|c|c|c|c|c|c|c}
\hline\hline
Method & SLYV & CNMF & CSU & NSSR & HySure & NPTSR & CNNFUS & uSDN & HyCoNet & MIAE \\
\hline\hline
      RMSE &0.1072 &0.0196 &0.0231 &0.0236 &0.0194 &0.0186 &0.0237 &0.0258 &0.0188 &\textbf{0.0172} \\
      PSNR &23.79 &35.75 &33.87 &33.93 &36.09 &36.71 &35.24 &32.84 &36.67 &\textbf{37.33} \\
       SAM &12.62 &2.62 &2.89 &3.21 &2.70 &2.64 &3.16 &3.49 &2.66 &\textbf{2.43} \\
     ERGAS &3.646 &0.741 &0.849 &0.871 &0.728 &0.699 &0.825 &0.905 &0.720 &\textbf{0.656} \\
      UIQI &0.853 &0.987 &0.983 &0.983 &0.986 &0.987 &0.984 &0.982 &0.987 &\textbf{0.988} \\
\hline\hline
\end{tabular}
\end{table*}

\begin{figure*}[!t]
\centering
{\includegraphics[width=\myimgsize in]{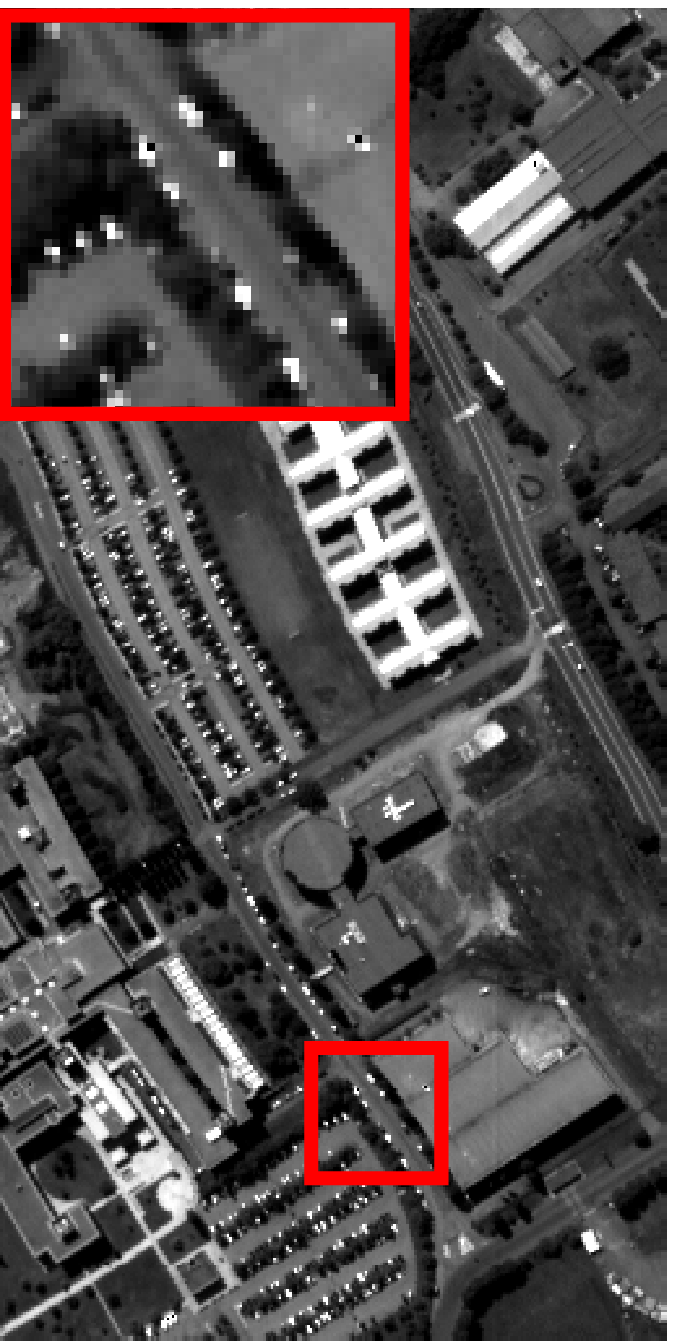}}
{\includegraphics[width=\myimgsize in]{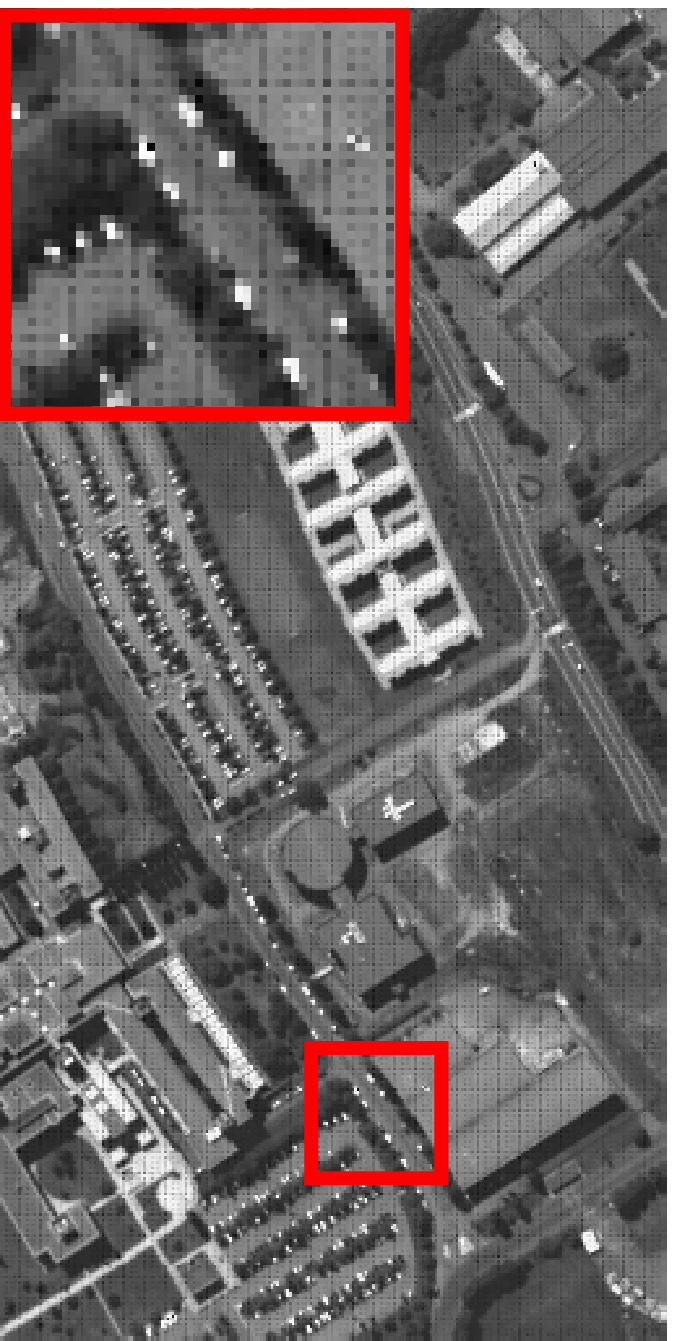}}
{\includegraphics[width=\myimgsize in]{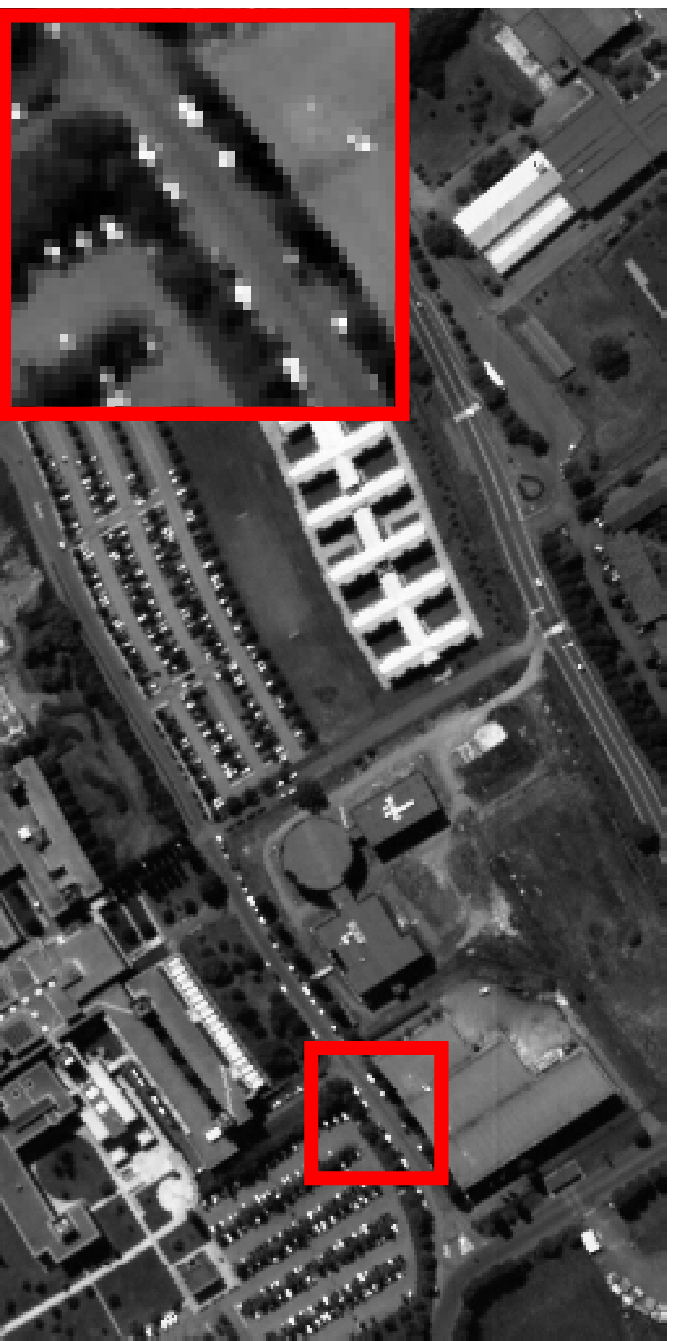}}
{\includegraphics[width=\myimgsize in]{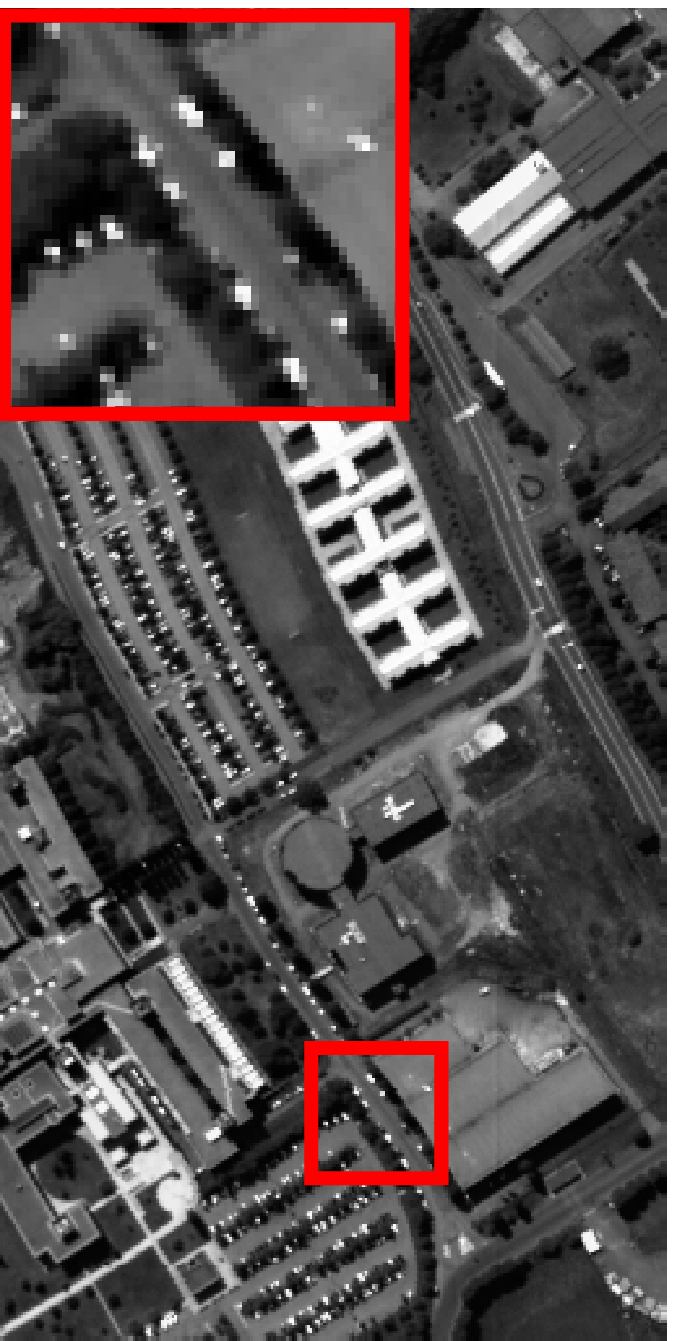}}
{\includegraphics[width=\myimgsize in]{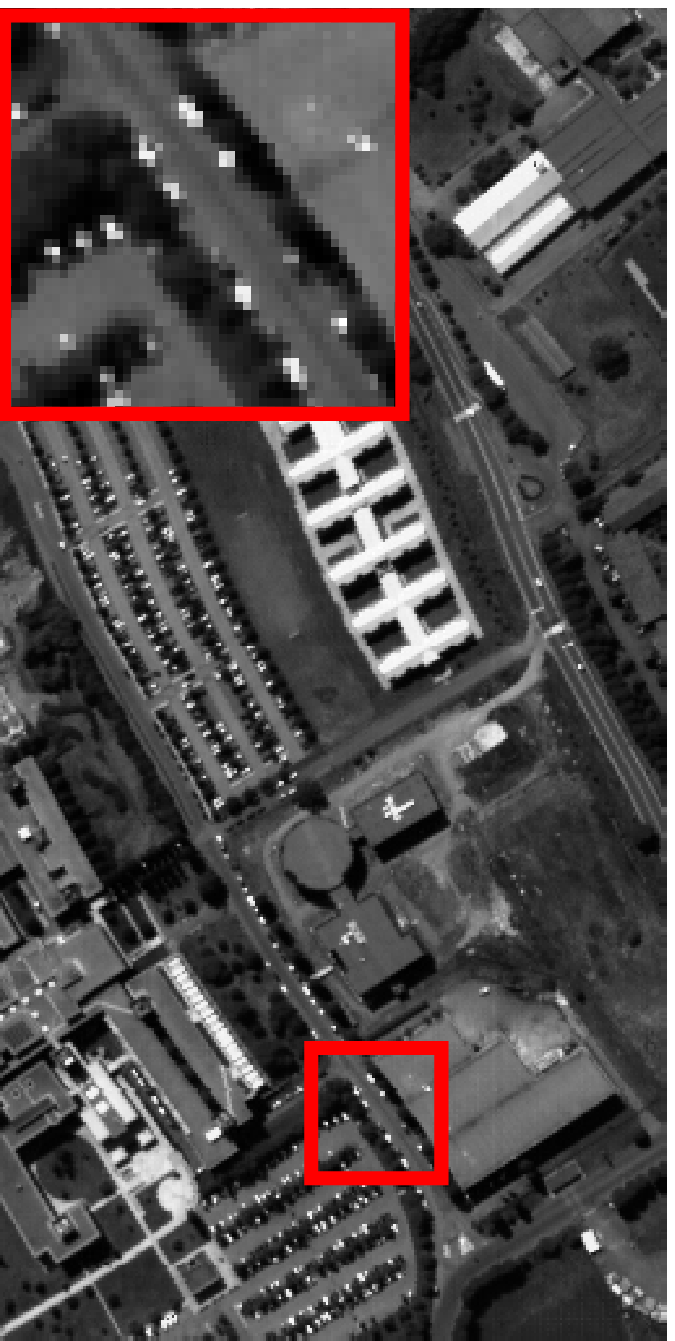}}
{\includegraphics[width=\myimgsize in]{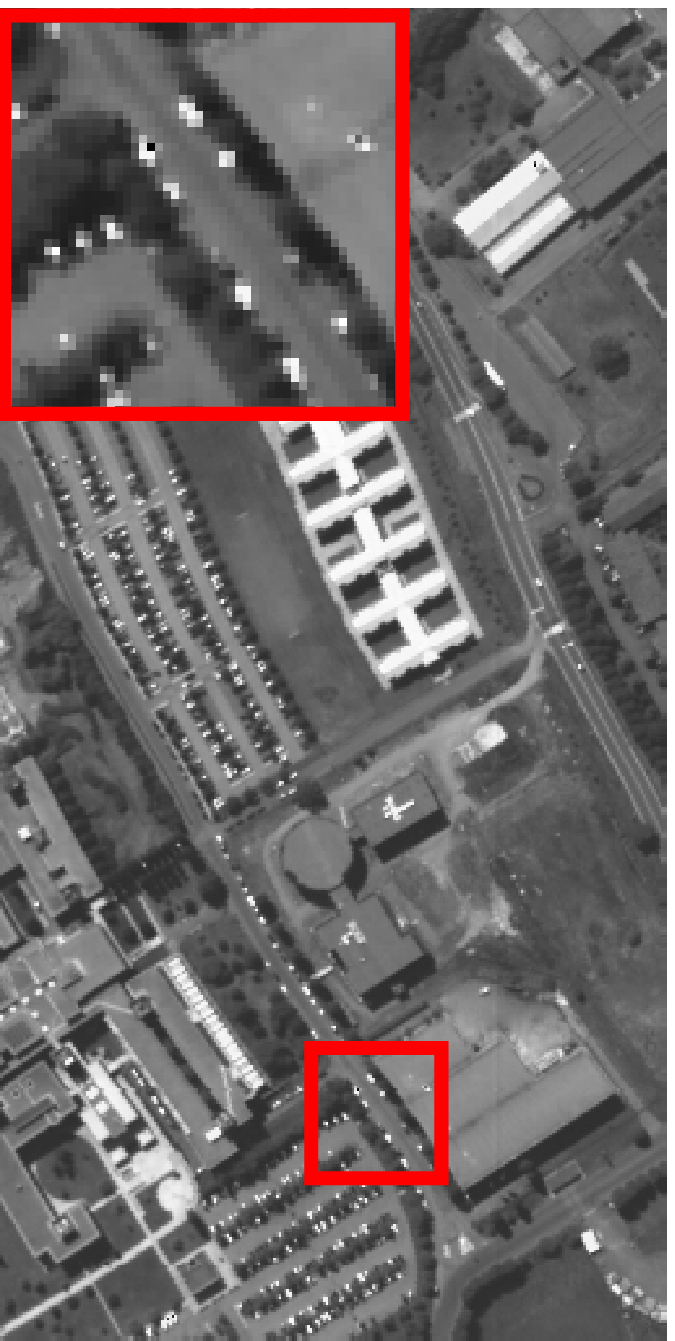}}
{\includegraphics[width=\myimgsize in]{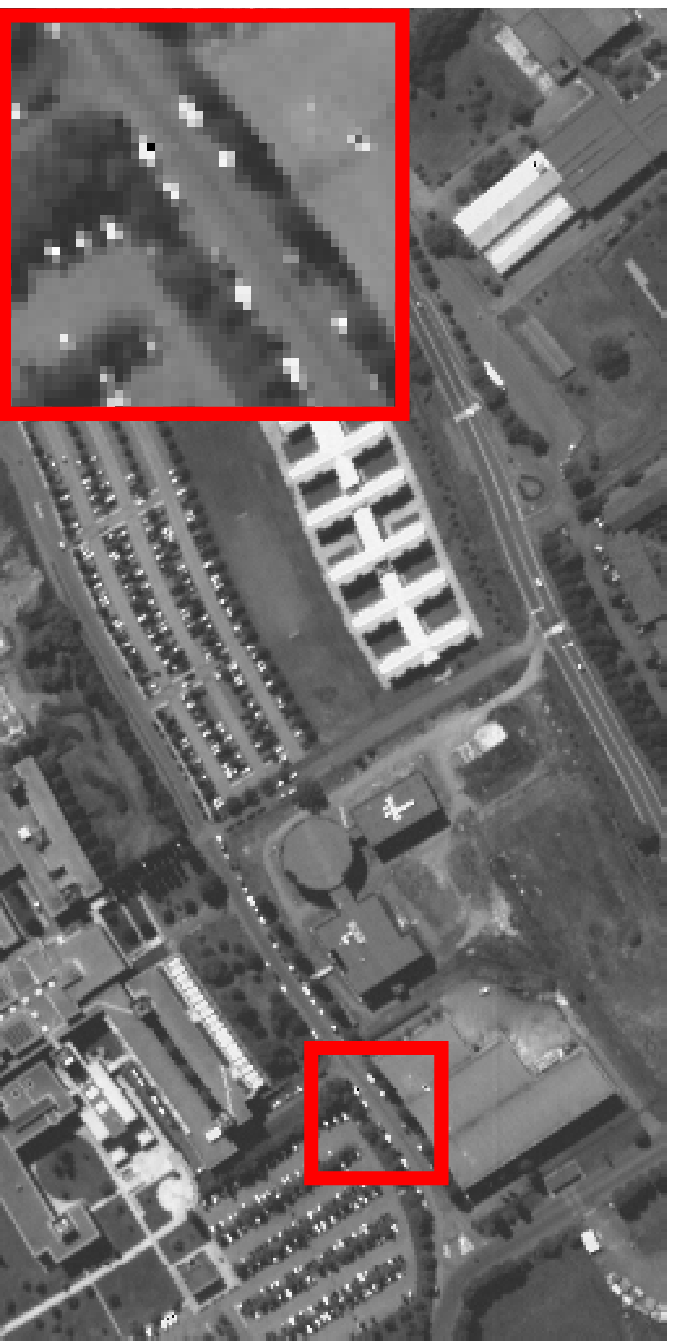}}
{\includegraphics[width=\myimgsize in]{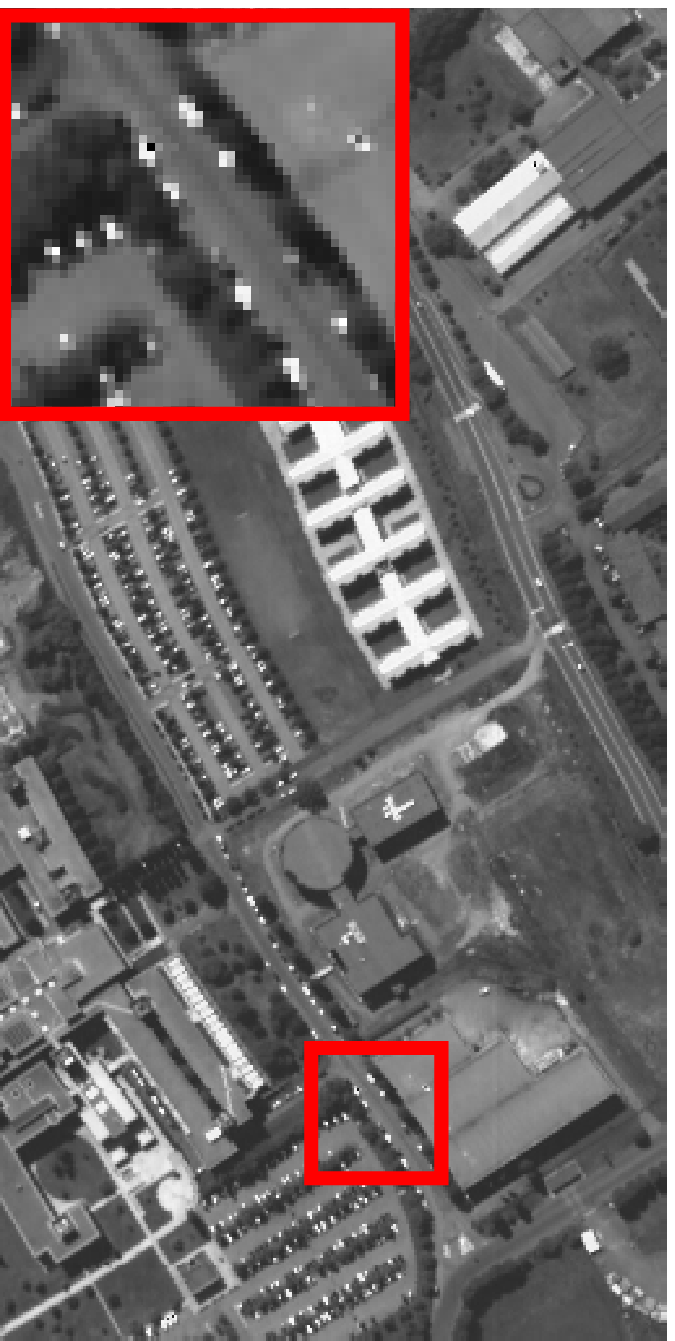}}
{\includegraphics[width=\myimgsize in]{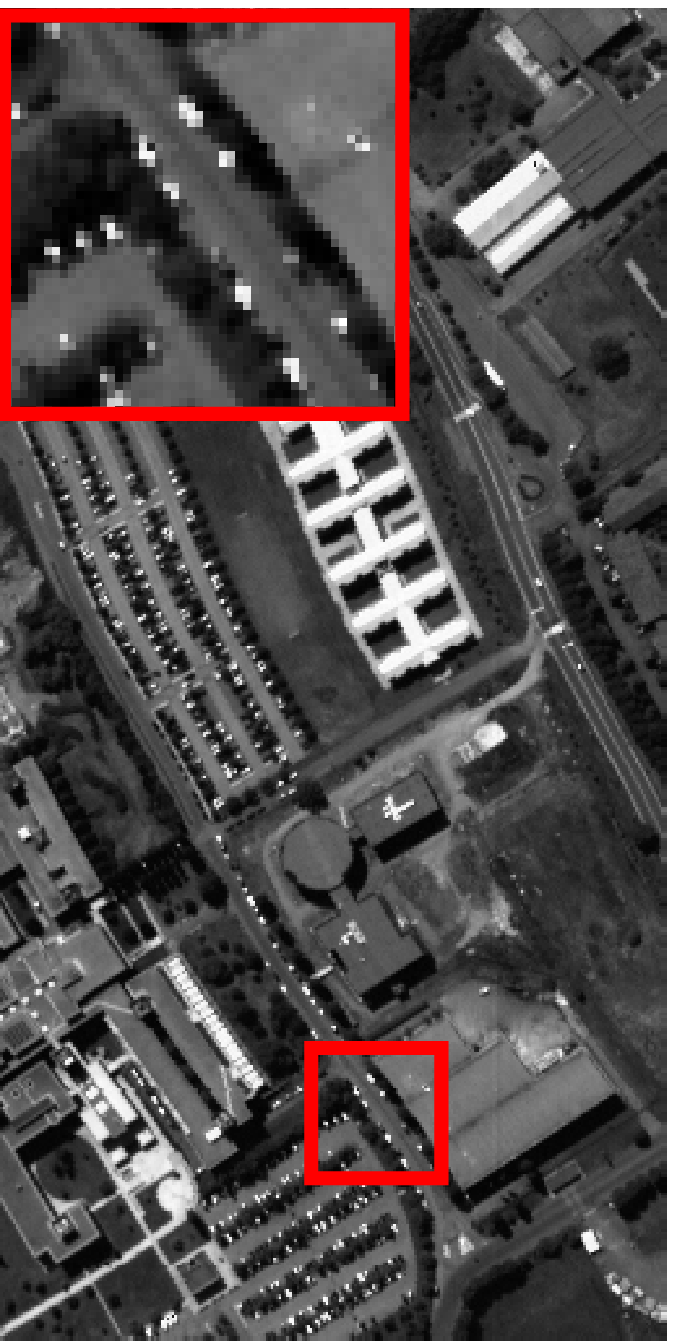}}
{\includegraphics[width=\myimgsize in]{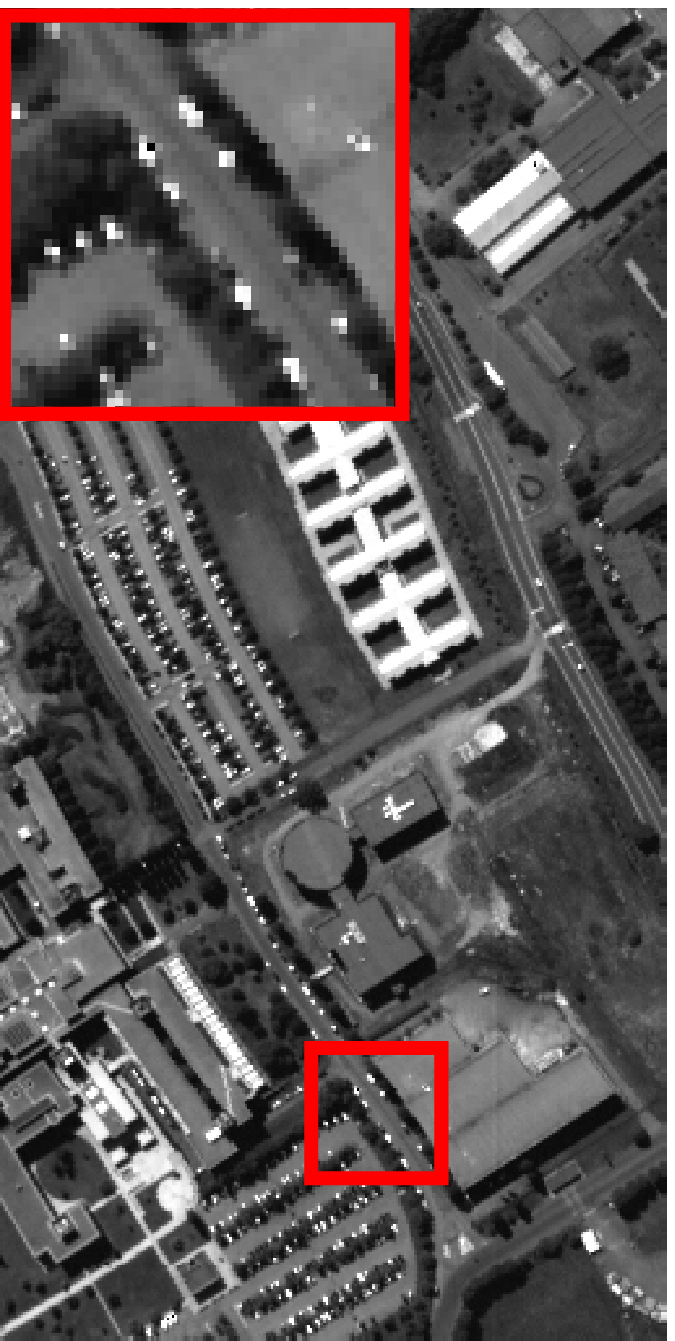}}
{\includegraphics[width=\myimgsize in]{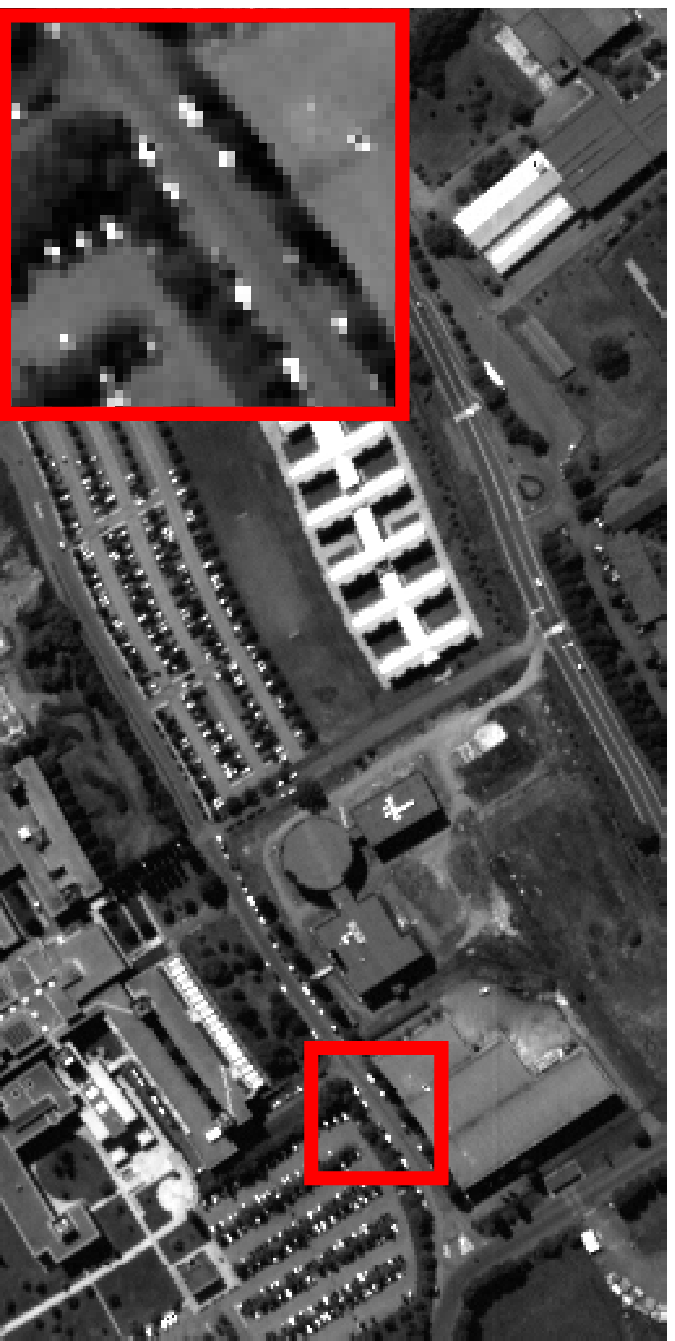}} \\
\subfigure[] {\includegraphics[width=\myimgsize in]{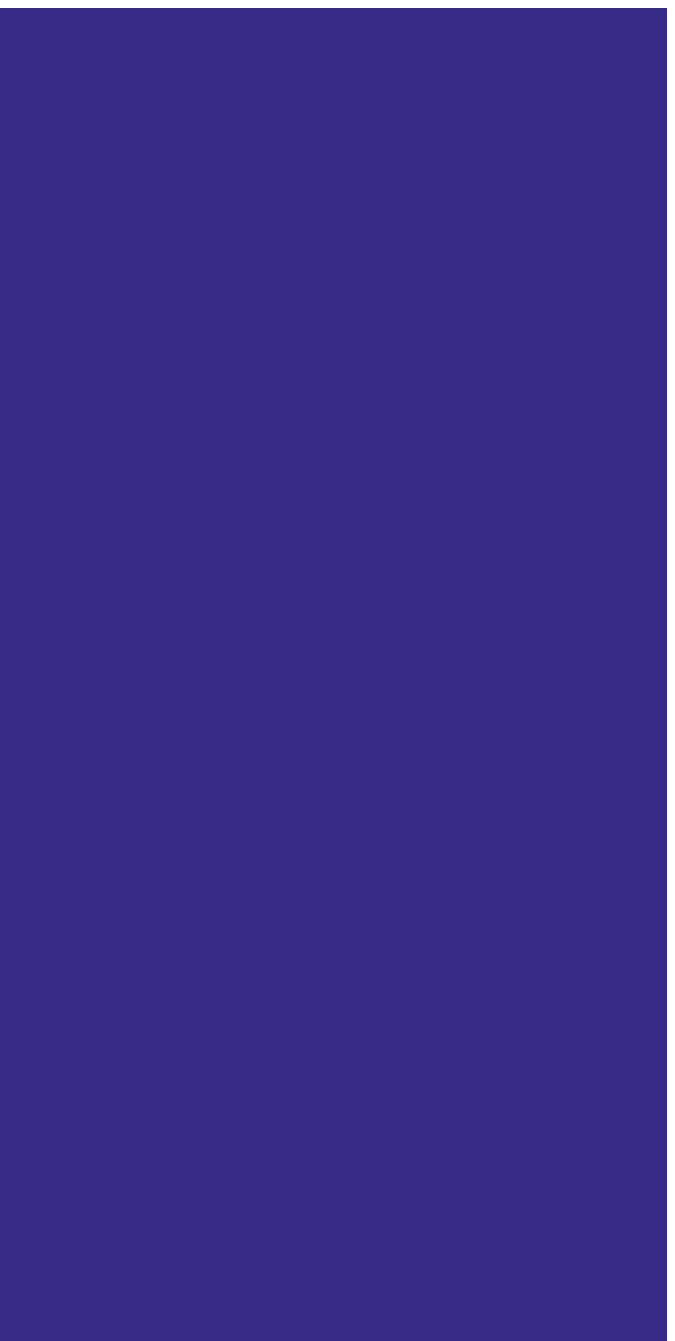}}
\subfigure[] {\includegraphics[width=\myimgsize in]{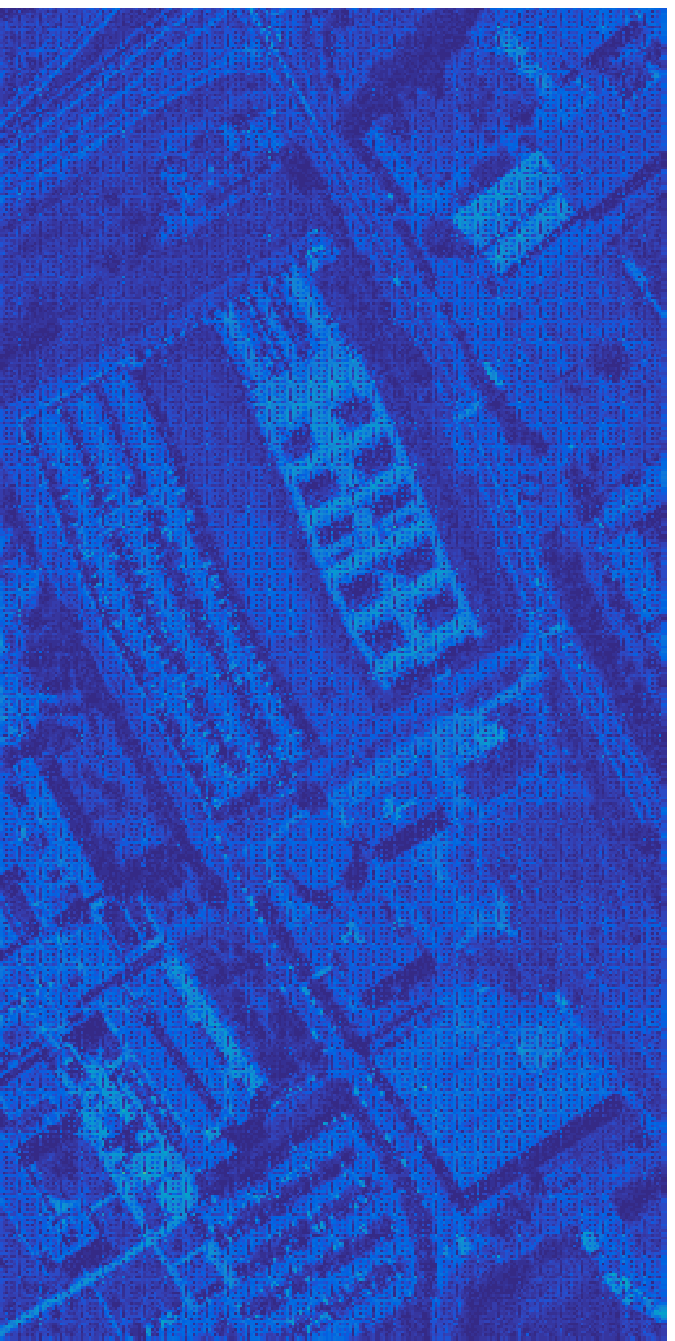}}
\subfigure[] {\includegraphics[width=\myimgsize in]{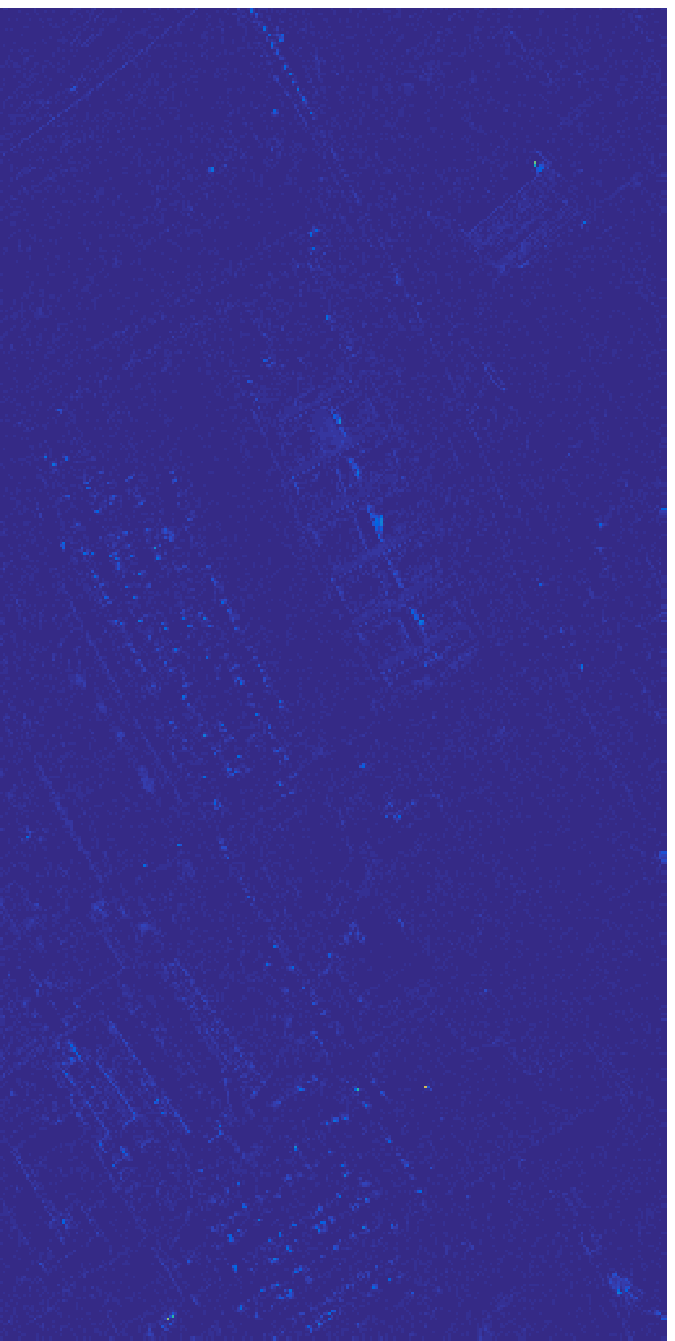}}
\subfigure[] {\includegraphics[width=\myimgsize in]{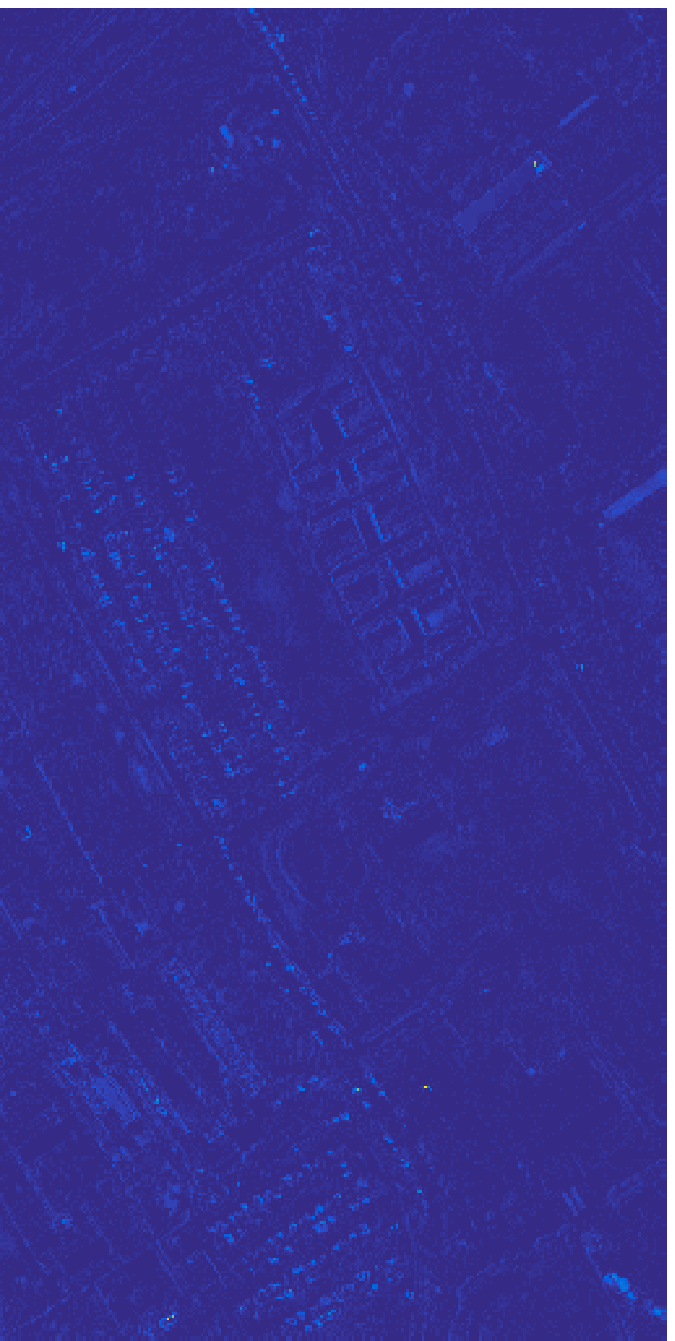}}
\subfigure[] {\includegraphics[width=\myimgsize in]{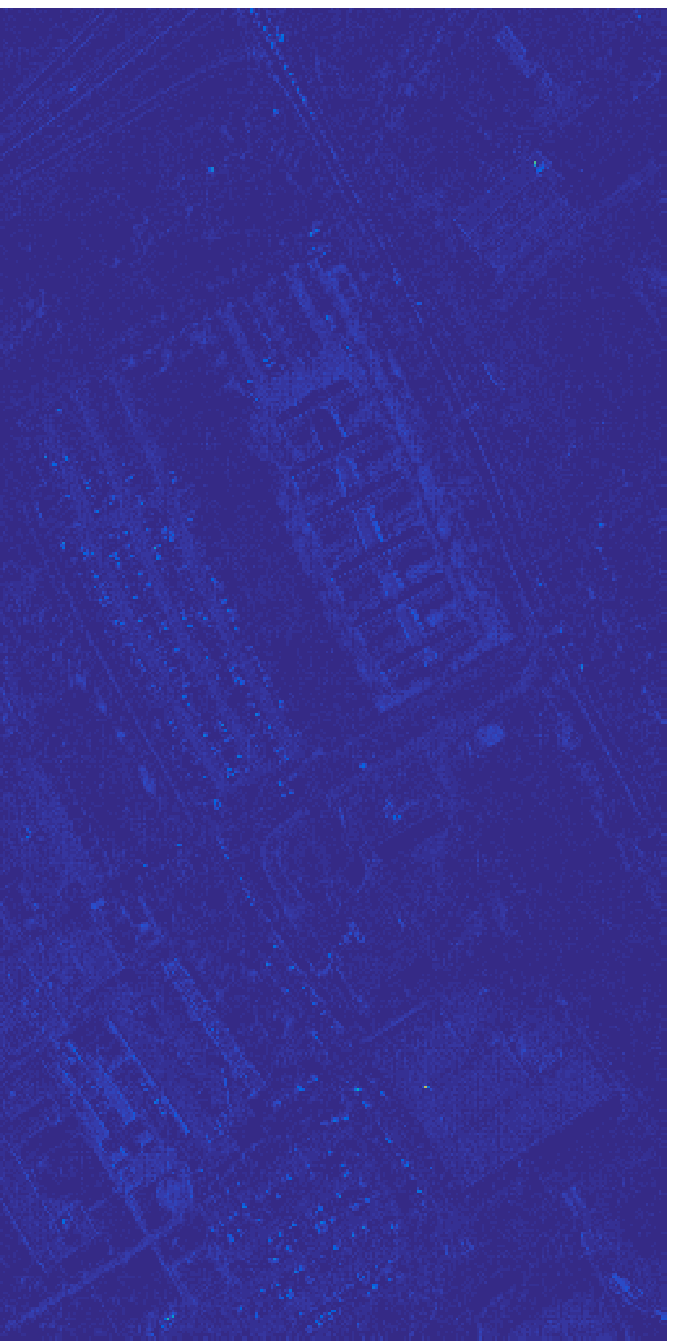}}
\subfigure[] {\includegraphics[width=\myimgsize in]{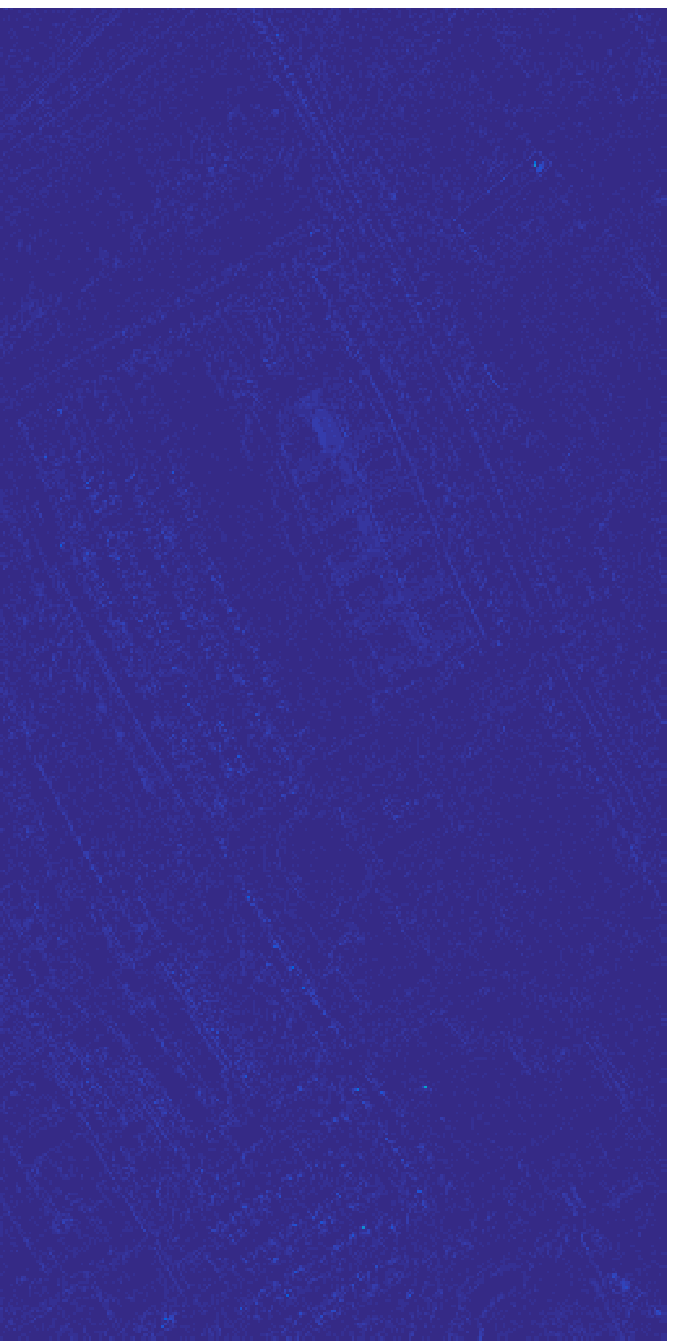}}
\subfigure[] {\includegraphics[width=\myimgsize in]{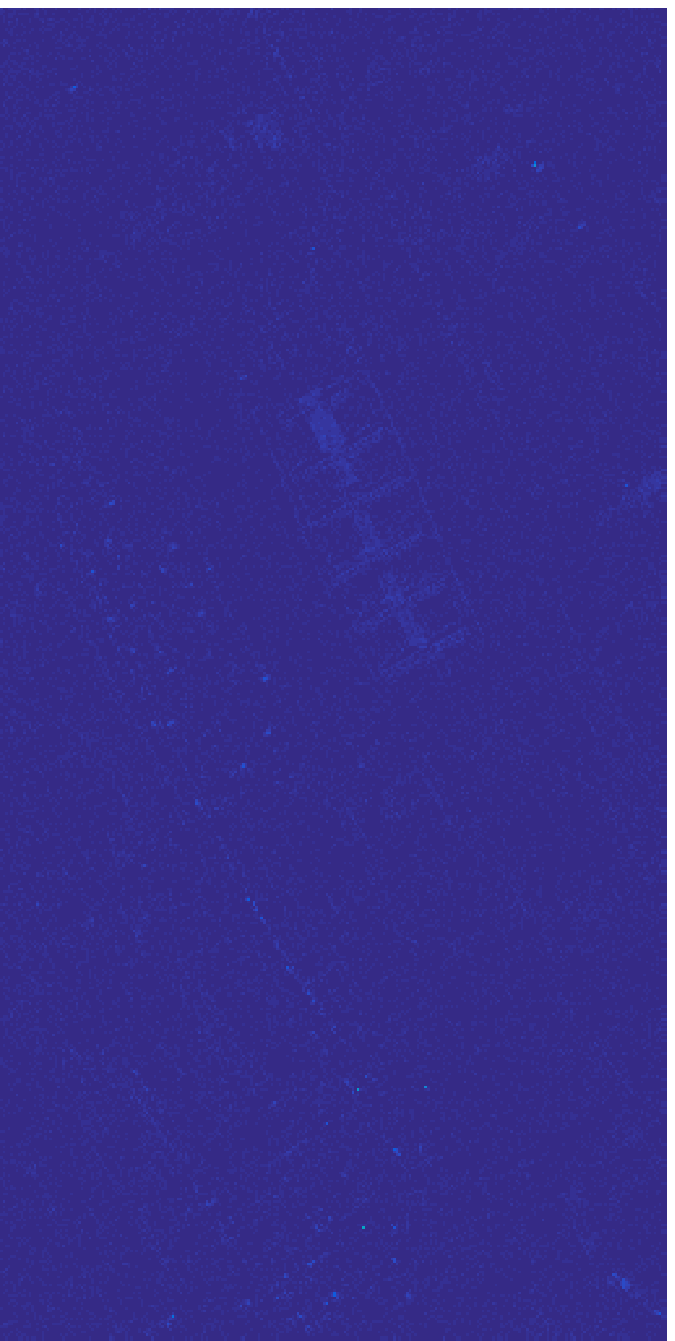}}
\subfigure[] {\includegraphics[width=\myimgsize in]{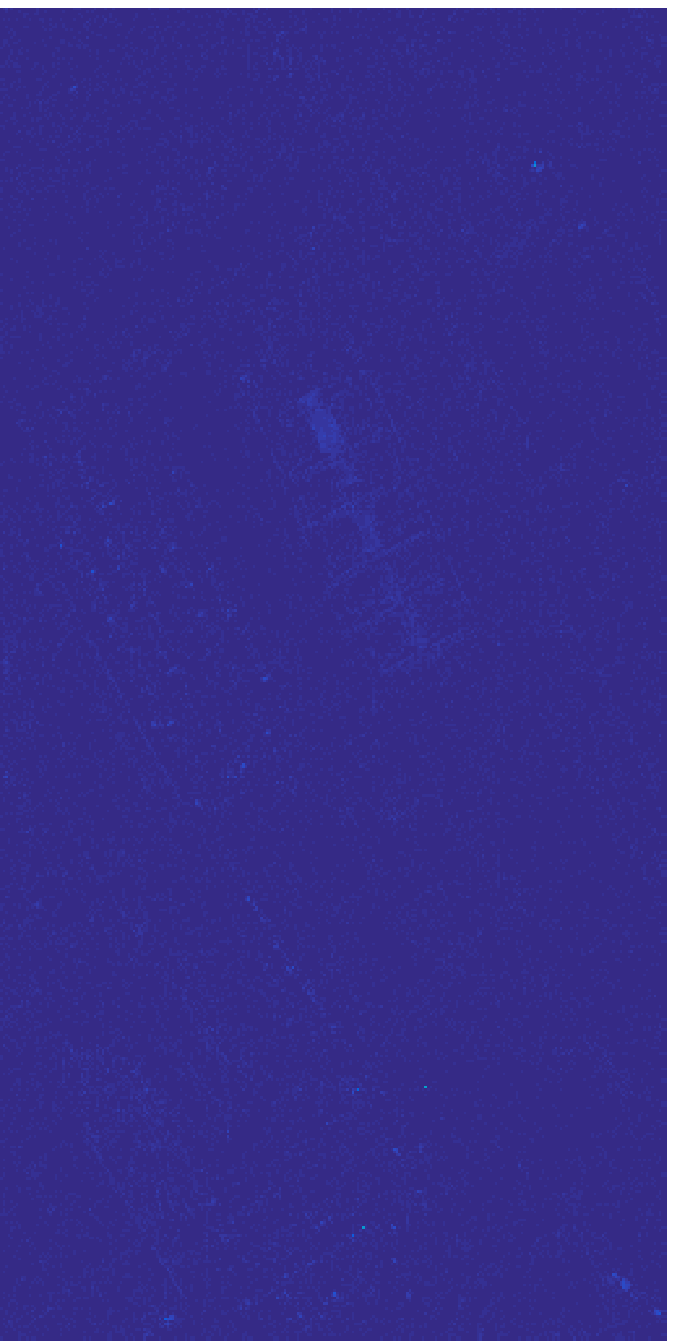}}
\subfigure[] {\includegraphics[width=\myimgsize in]{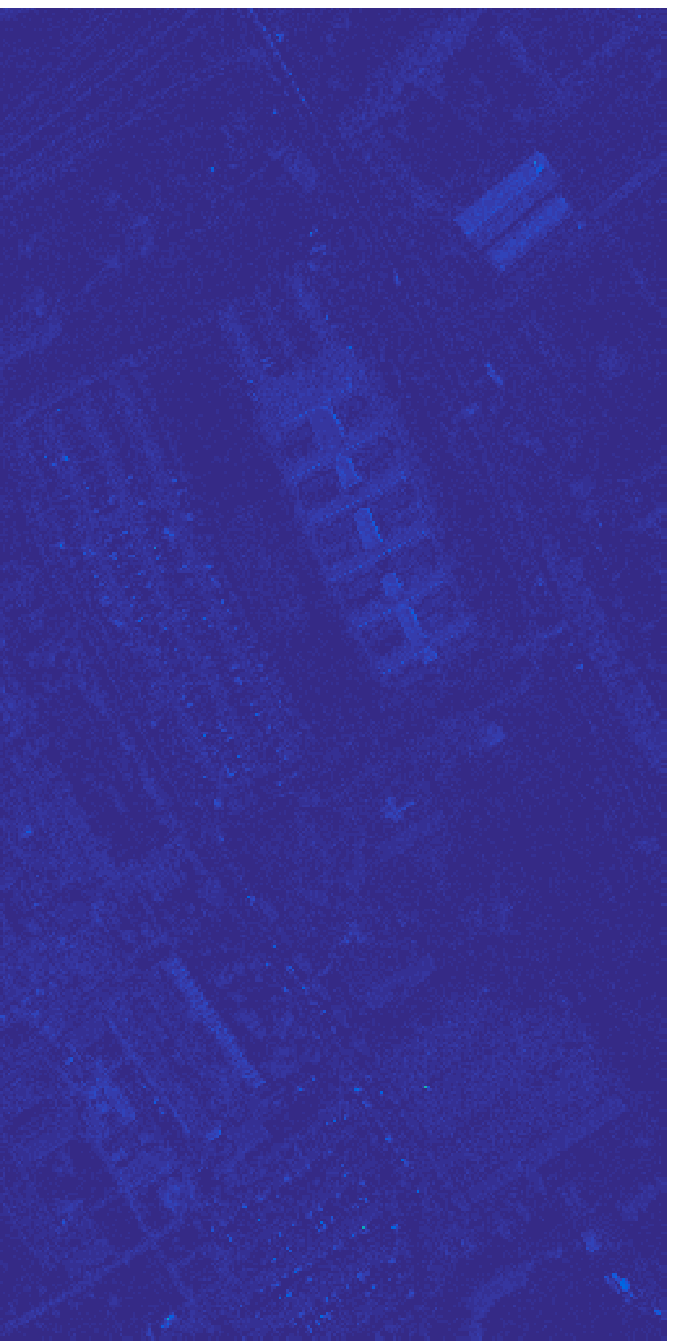}}
\subfigure[] {\includegraphics[width=\myimgsize in]{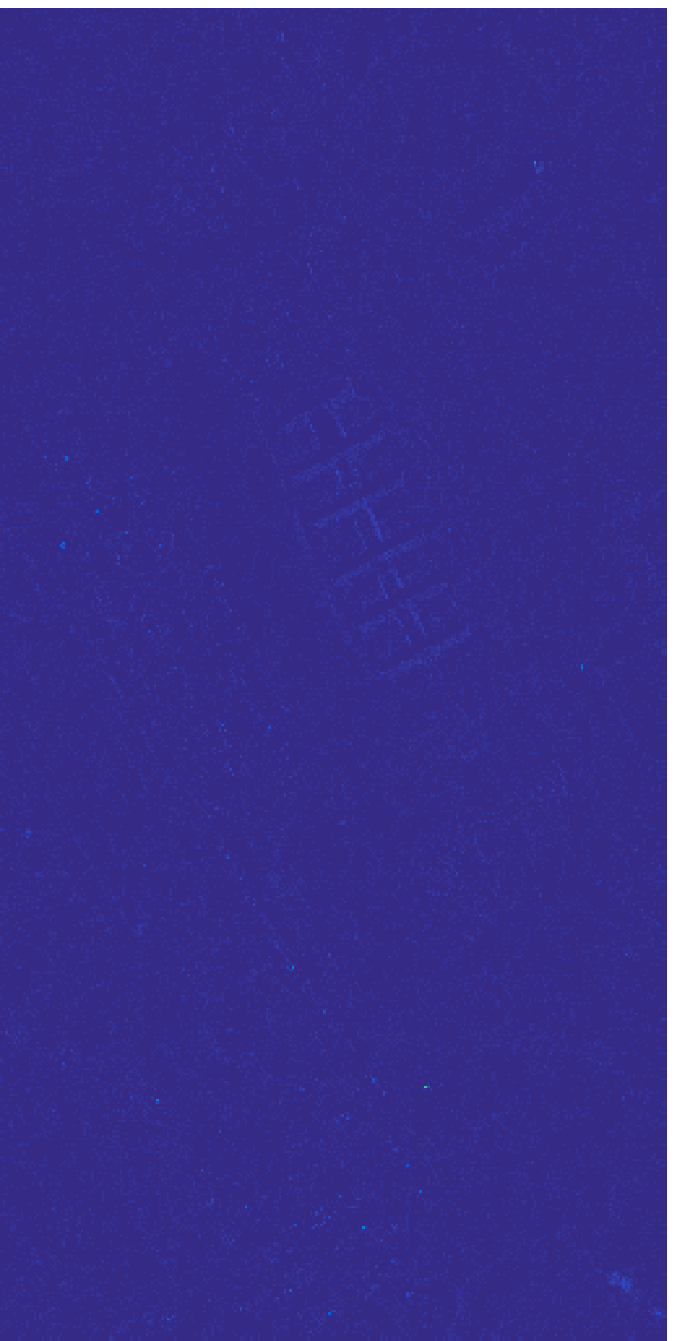}}
\subfigure[] {\includegraphics[width=\myimgsize in]{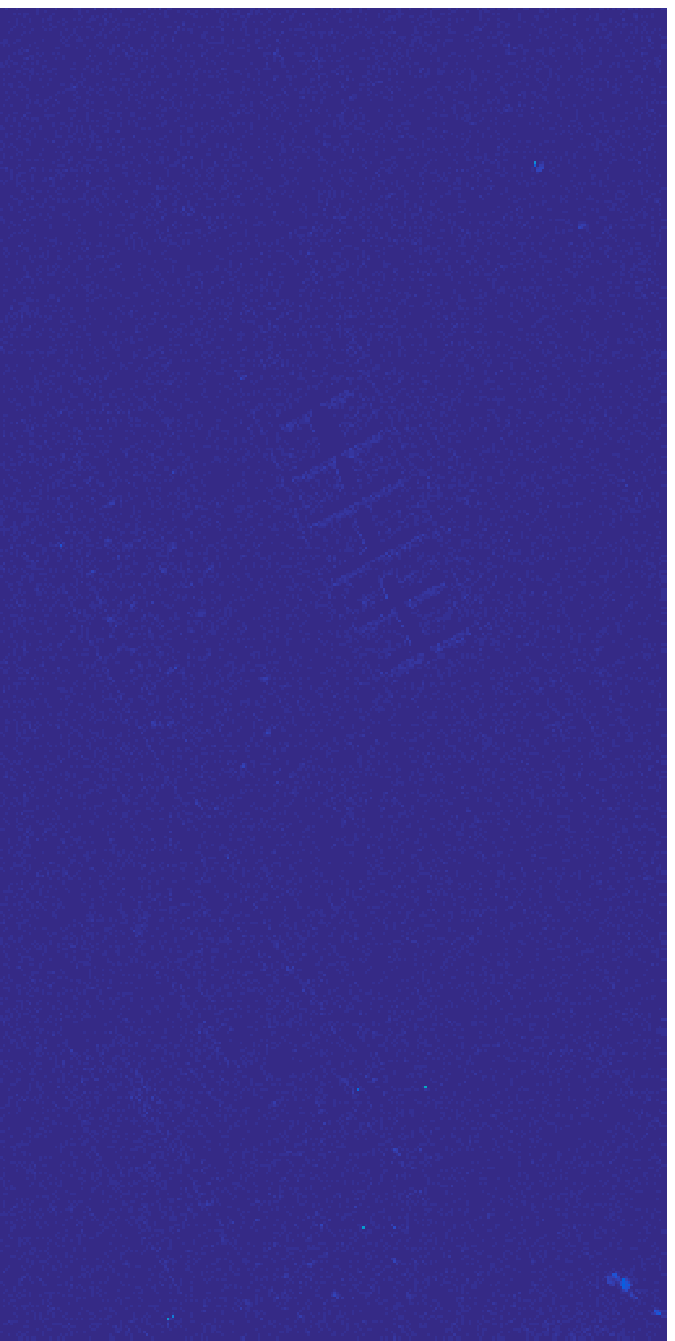}}
\caption{Images (with a meaningful region marked and zoomed in 3 times for easy observation) and error maps at band 30 of HSI super-resolution results when applied to the PaviaU dataset.
(a) Reference image. (b) SLYV. (c) CNMF. (d) CSU.
(e) NSSR. (f) HySure. (g) NPTSR.
(h) CNNFUS. (i) uSDN. (j) HyCoNet. (k) MIAE.}
\label{fig_pavia}
\end{figure*}

\begin{table*}[!t]
\caption{Quality measures for the KSC dataset using different methods (the best values are highlighted)}
\label{tab_ksc}
\centering
\begin{tabular}
{c|c|c|c|c|c|c|c|c|c|c}
\hline\hline
Method & SLYV & CNMF & CSU & NSSR & HySure & NPTSR & CNNFUS & uSDN & HyCoNet & MIAE \\
\hline\hline
      RMSE &0.1574 &0.0454 &0.0465 &0.0513 &0.0453 &0.0450 &0.0534 &0.0504 &0.0441 &\textbf{0.0427} \\
      PSNR &19.33 &32.70 &31.17 &30.77 &32.95 &33.30 &30.95 &30.06 &33.49 &\textbf{34.05} \\
       SAM &23.23 &7.78 &8.02 &8.65 &7.64 &7.29 &9.01 &9.14 &7.22 &\textbf{7.00} \\
     ERGAS &8.753 &3.497 &3.405 &3.738 &3.336 &3.328 &3.954 &3.717 &3.262 &\textbf{3.129} \\
      UIQI &0.506 &0.870 &0.855 &0.836 &0.881 &\textbf{0.887} &0.843 &0.857 &0.878 &0.882 \\
\hline\hline
\end{tabular}
\end{table*}

\begin{figure*}[!t]
\centering
{\includegraphics[width=\myimgsize in]{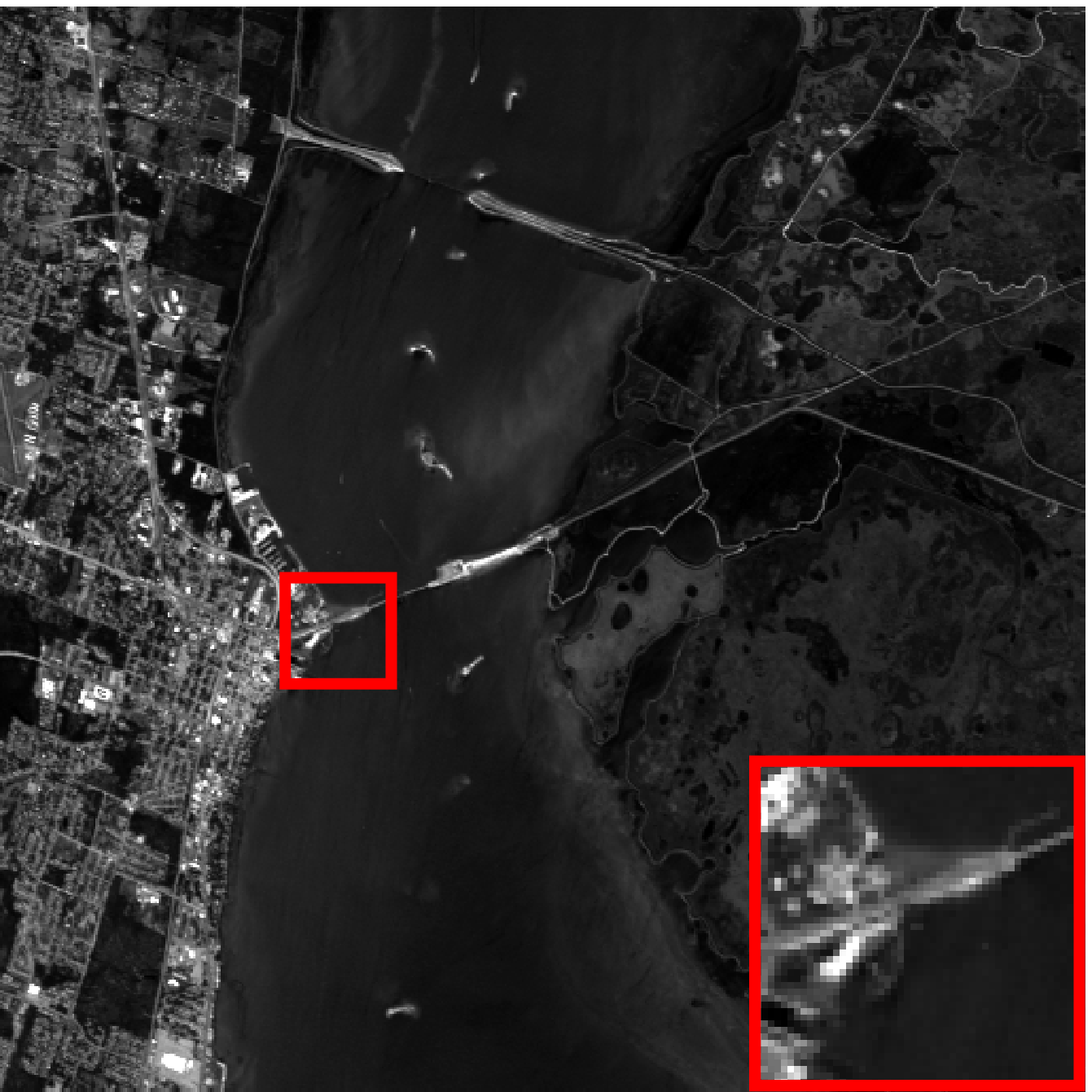}}
{\includegraphics[width=\myimgsize in]{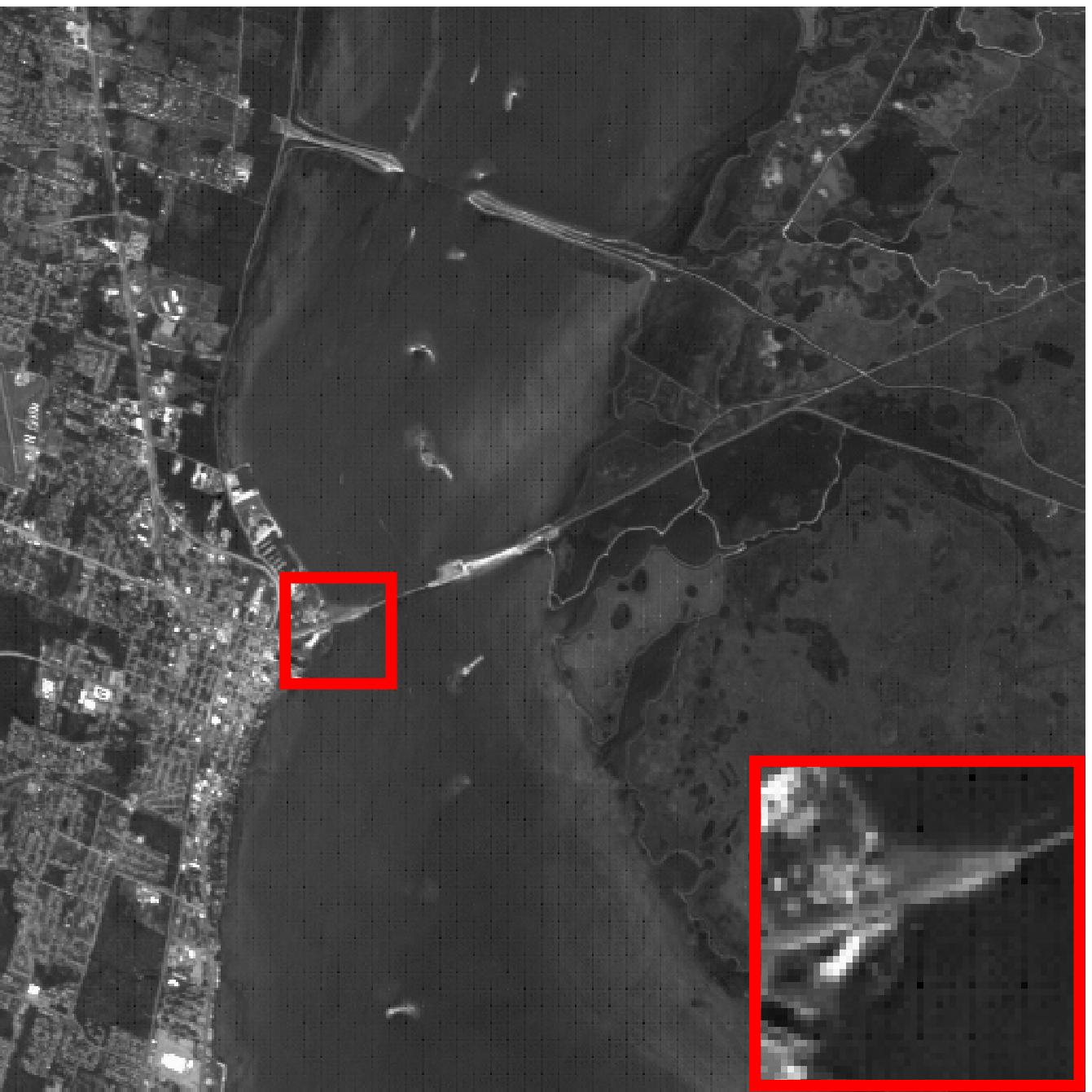}}
{\includegraphics[width=\myimgsize in]{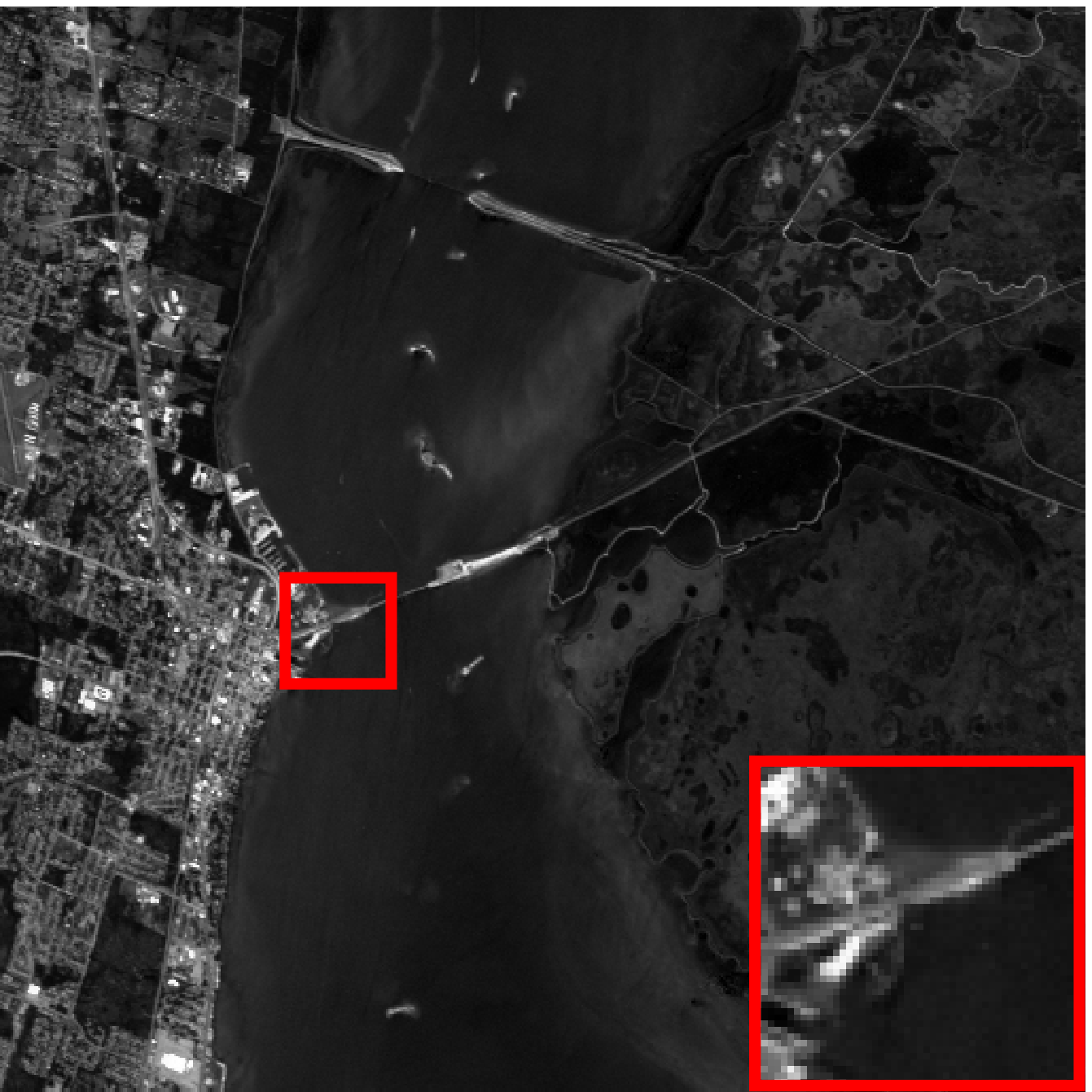}}
{\includegraphics[width=\myimgsize in]{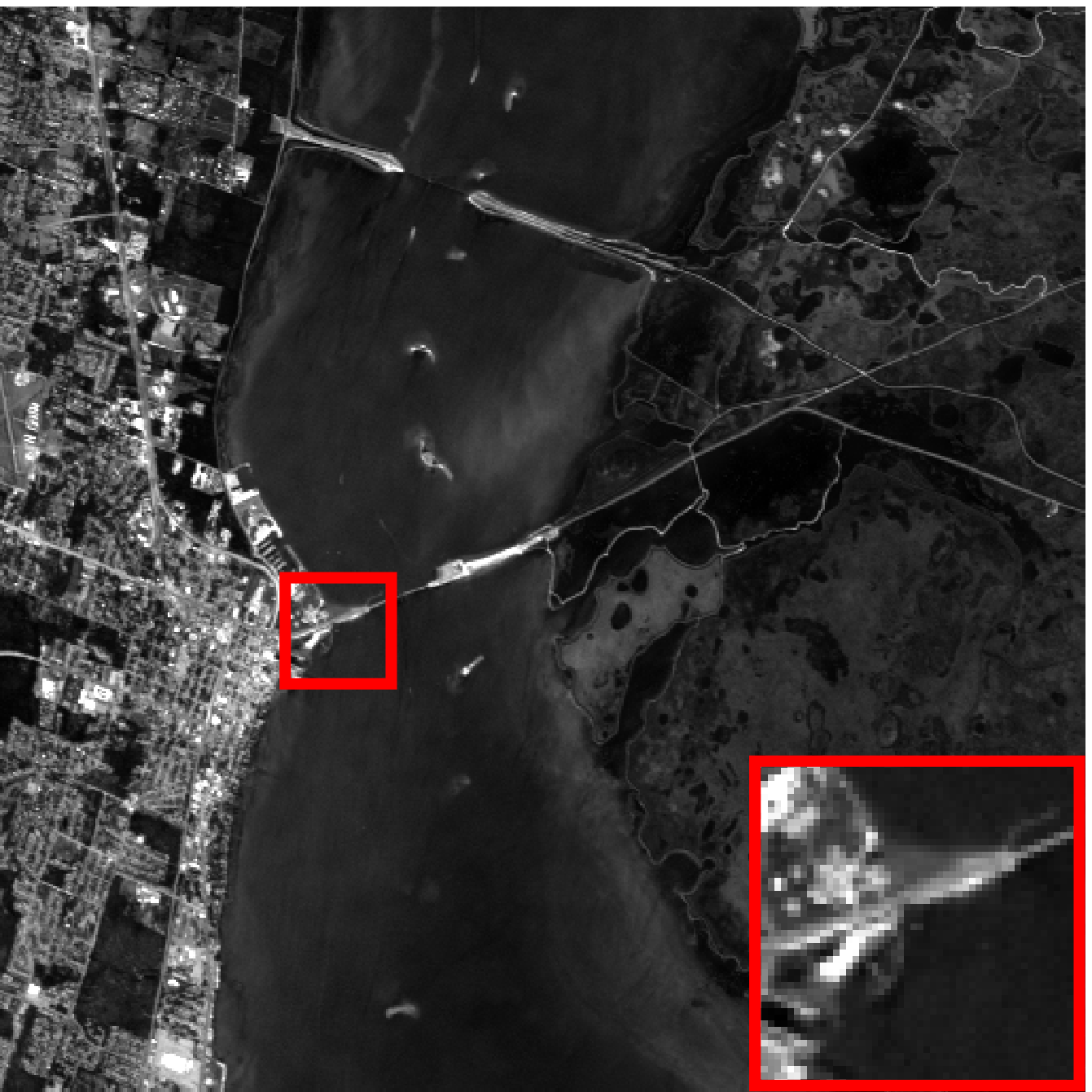}}
{\includegraphics[width=\myimgsize in]{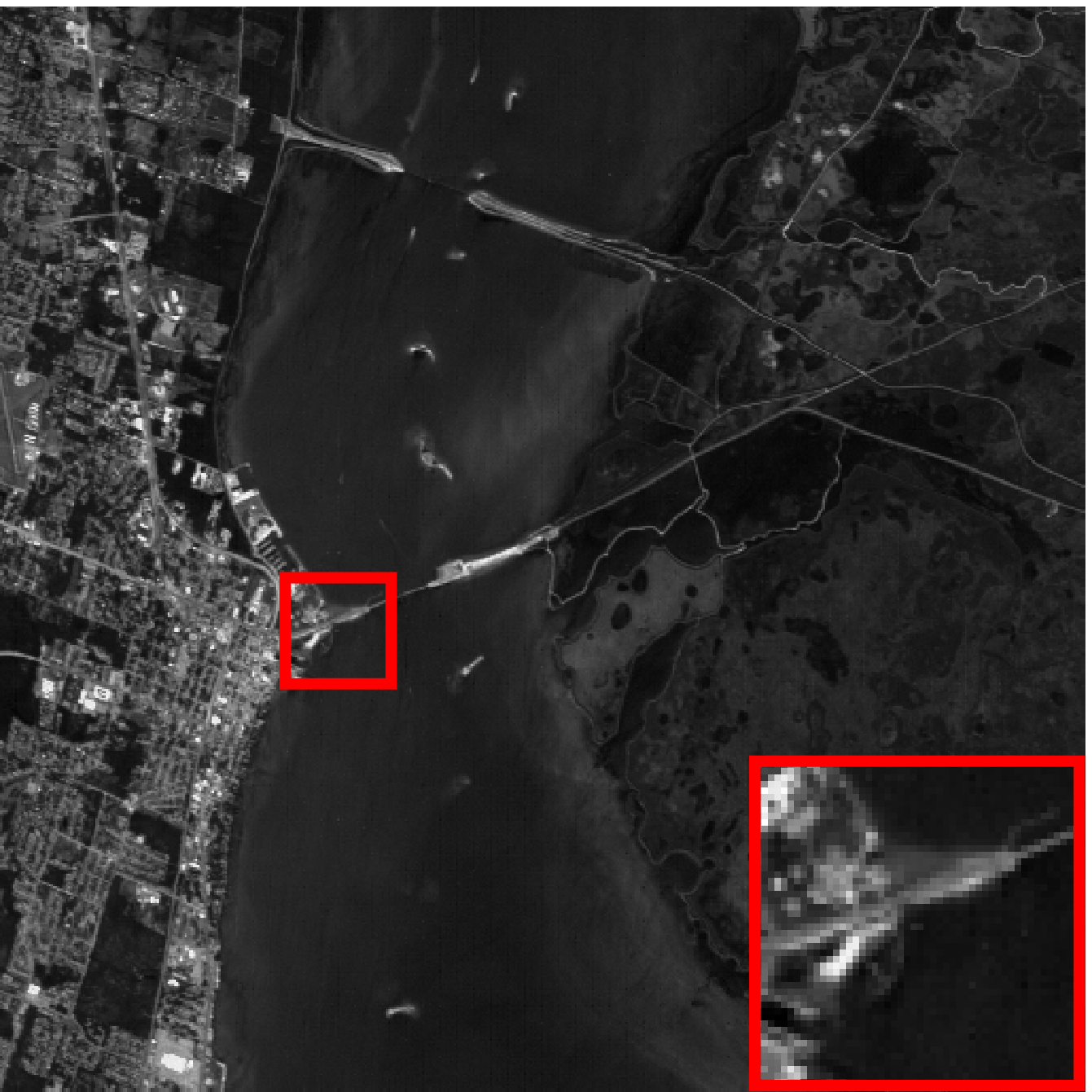}}
{\includegraphics[width=\myimgsize in]{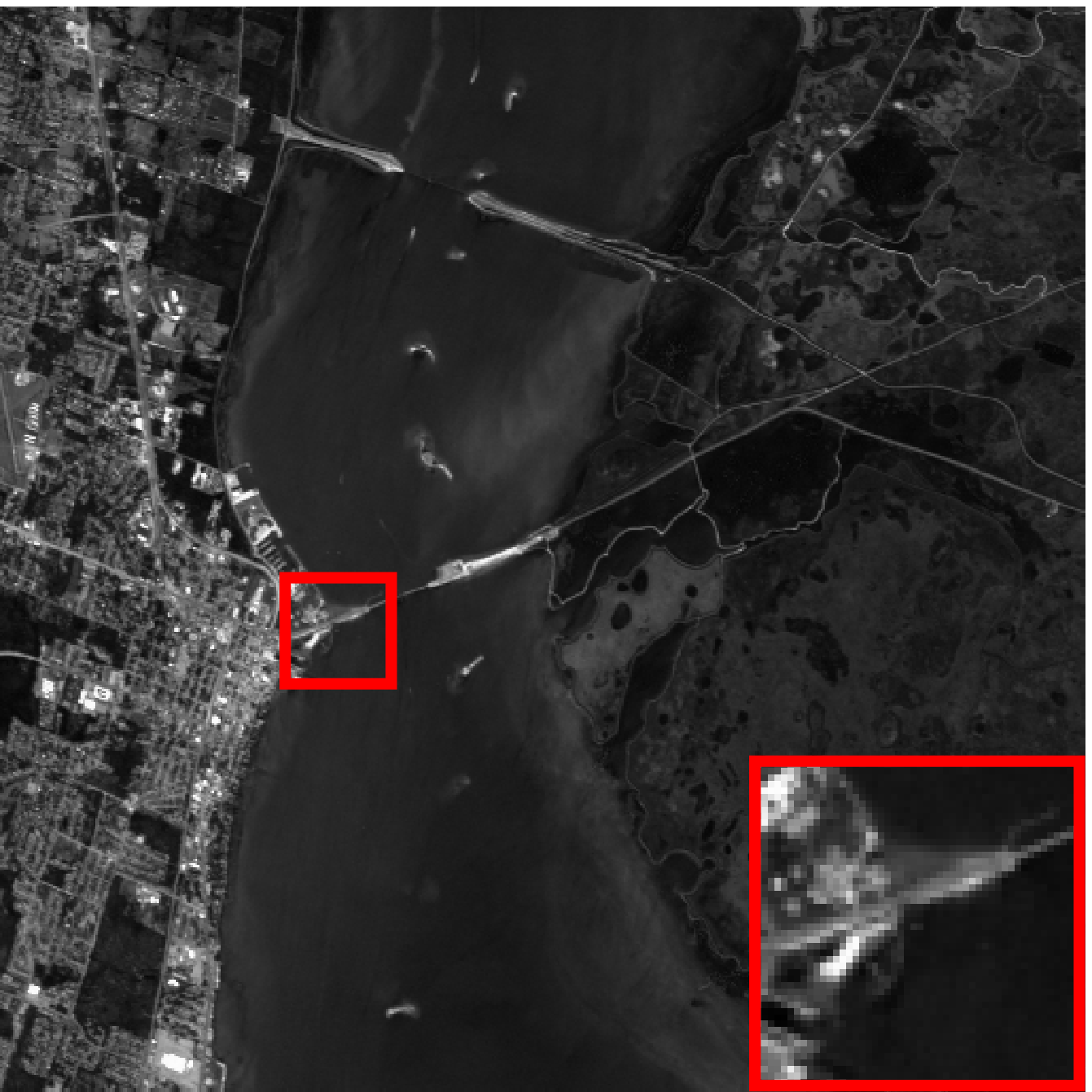}}
{\includegraphics[width=\myimgsize in]{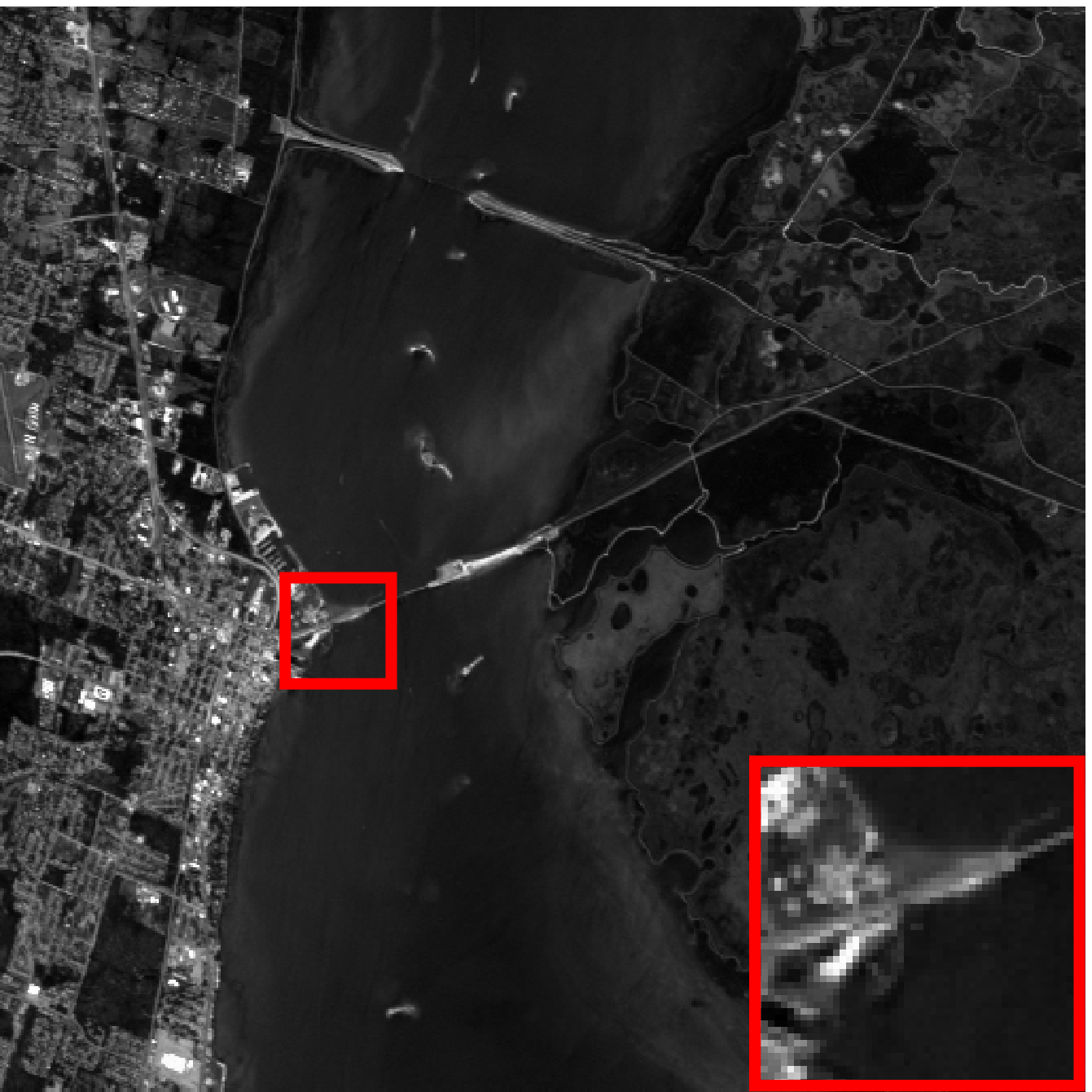}}
{\includegraphics[width=\myimgsize in]{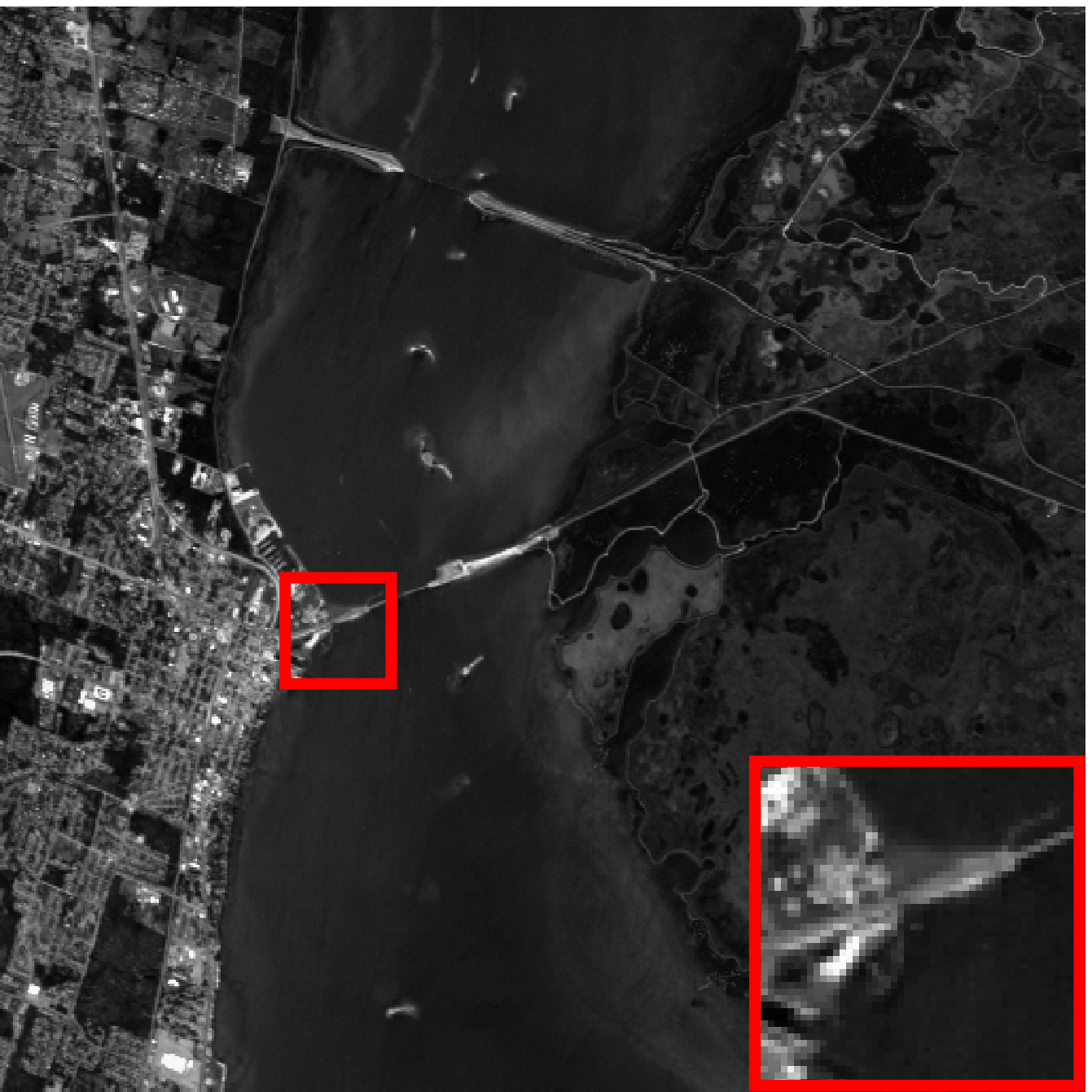}}
{\includegraphics[width=\myimgsize in]{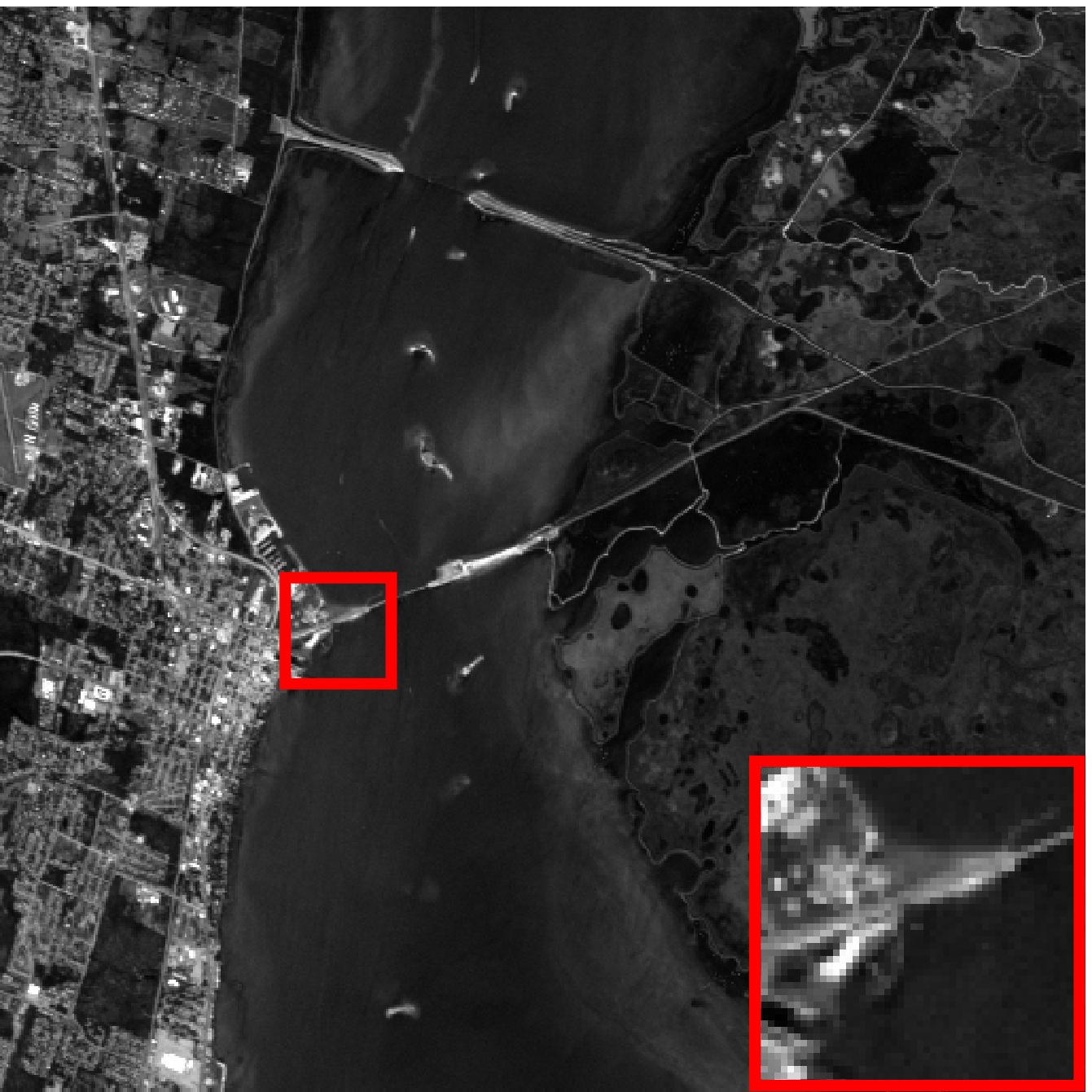}}
{\includegraphics[width=\myimgsize in]{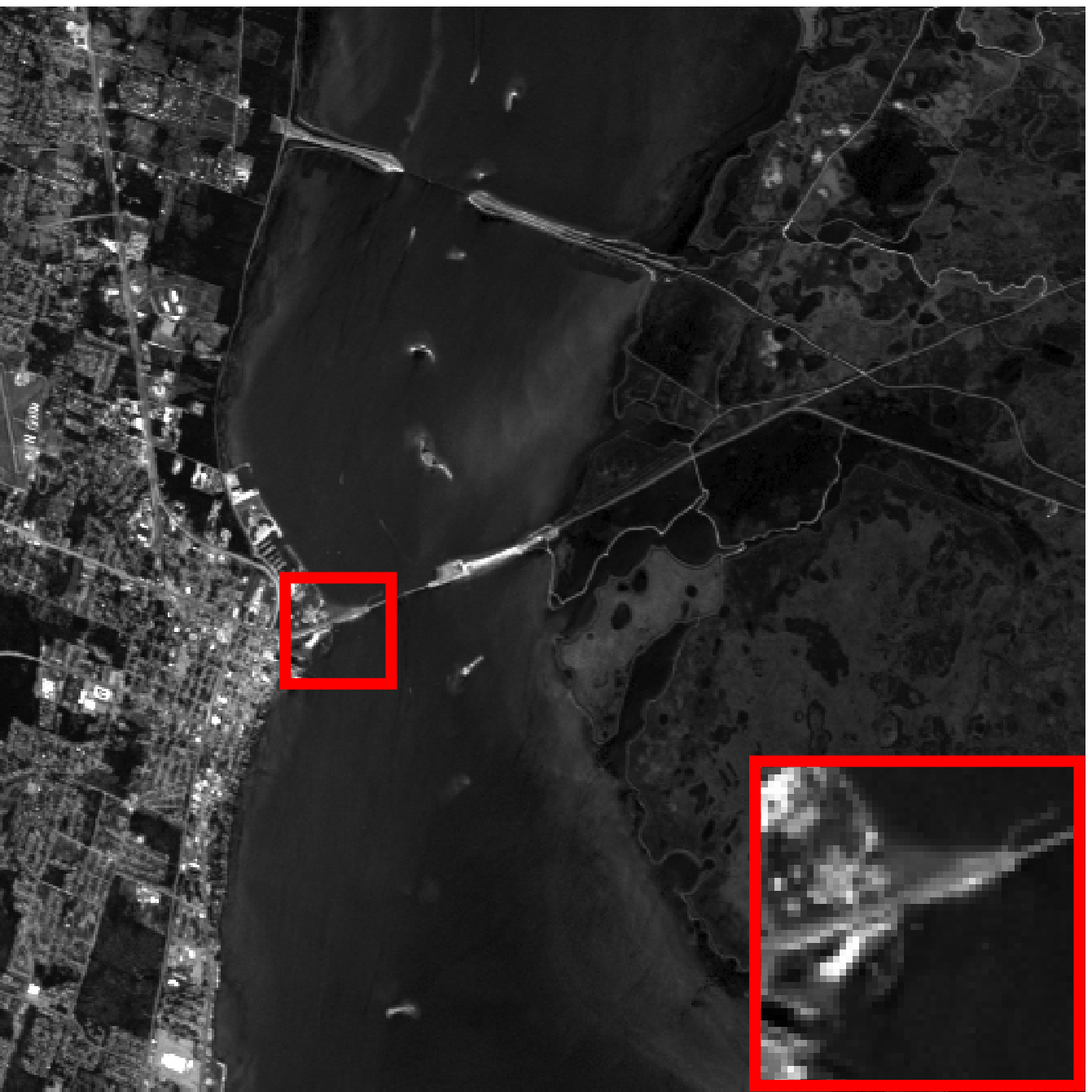}}
{\includegraphics[width=\myimgsize in]{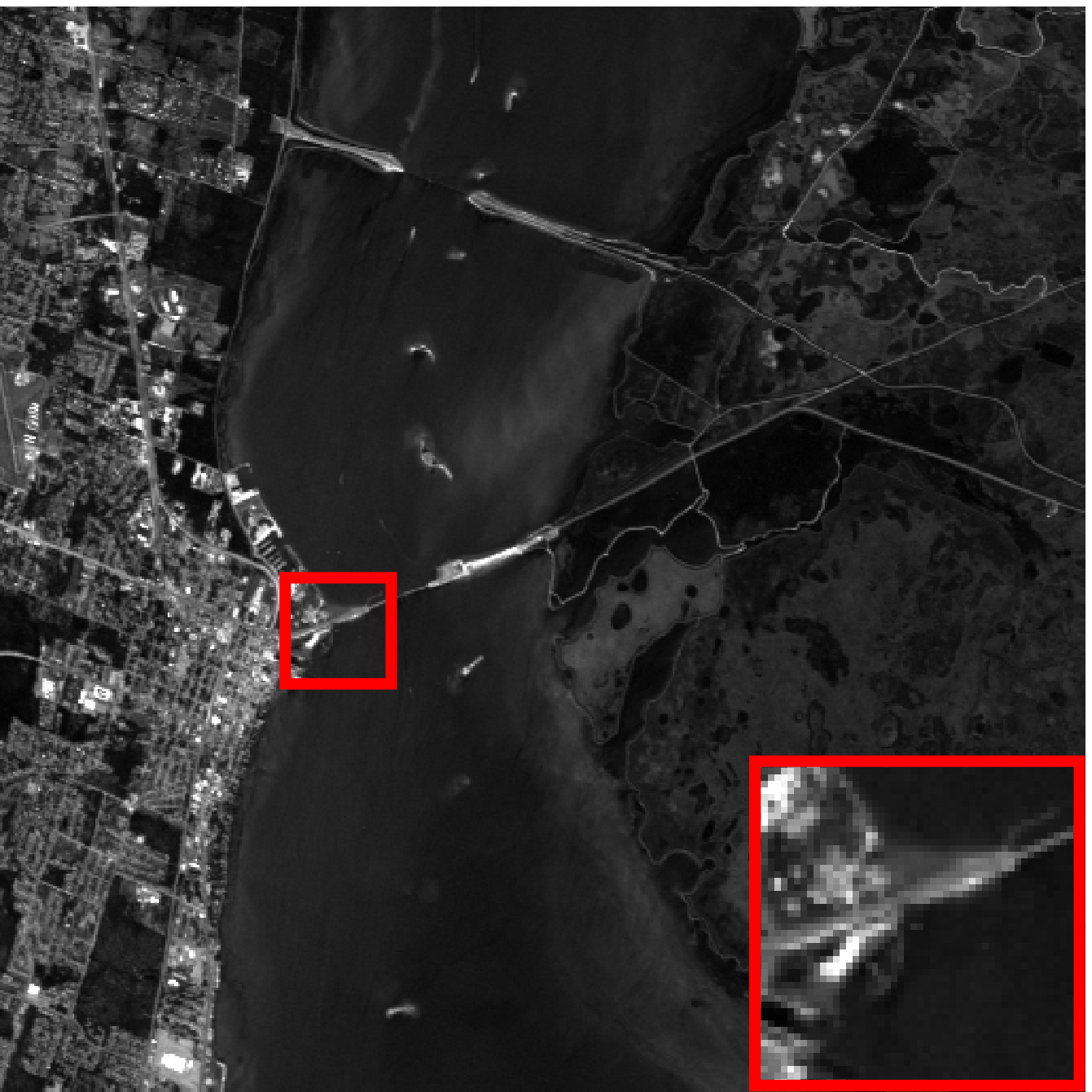}} \\
\subfigure[] {\includegraphics[width=\myimgsize in]{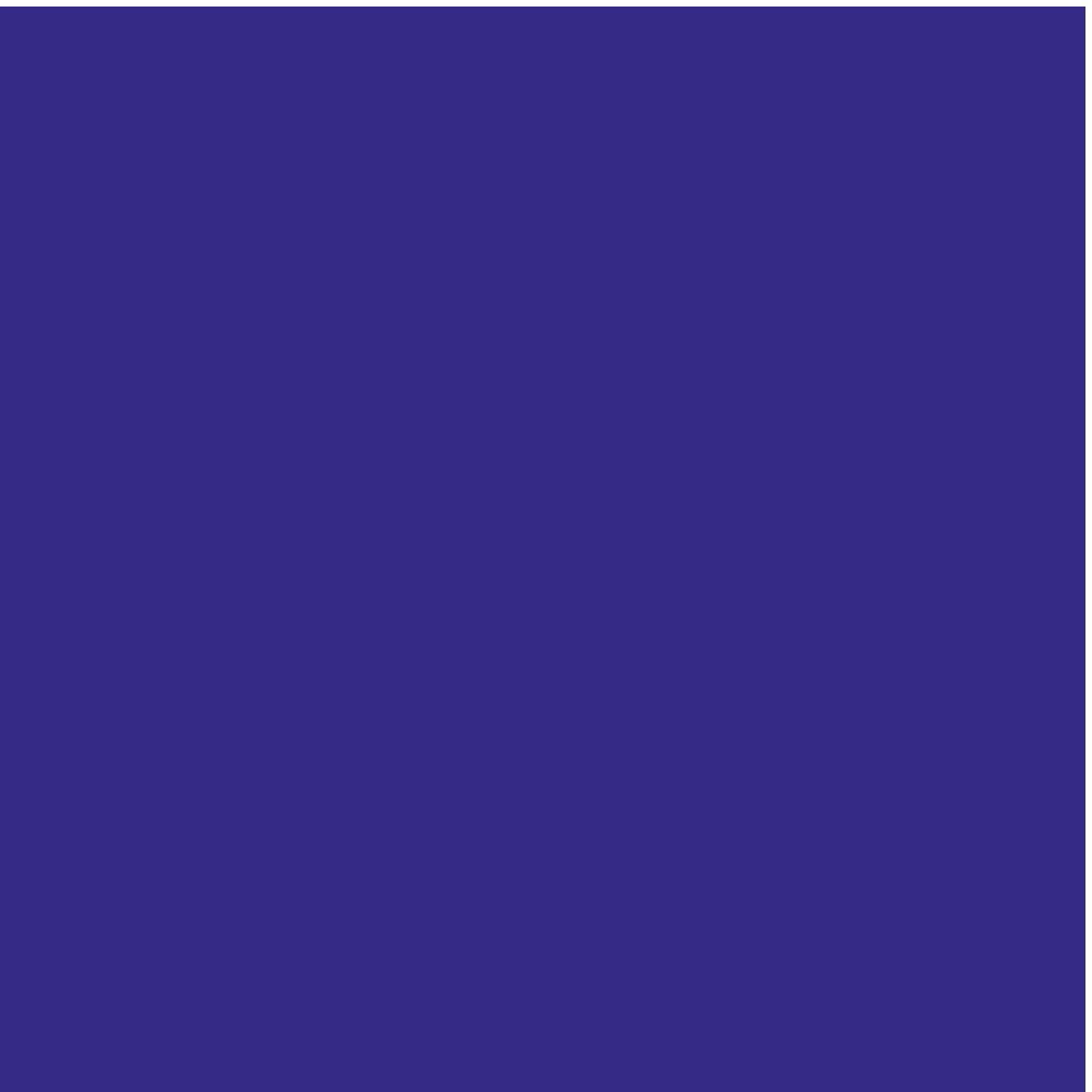}}
\subfigure[] {\includegraphics[width=\myimgsize in]{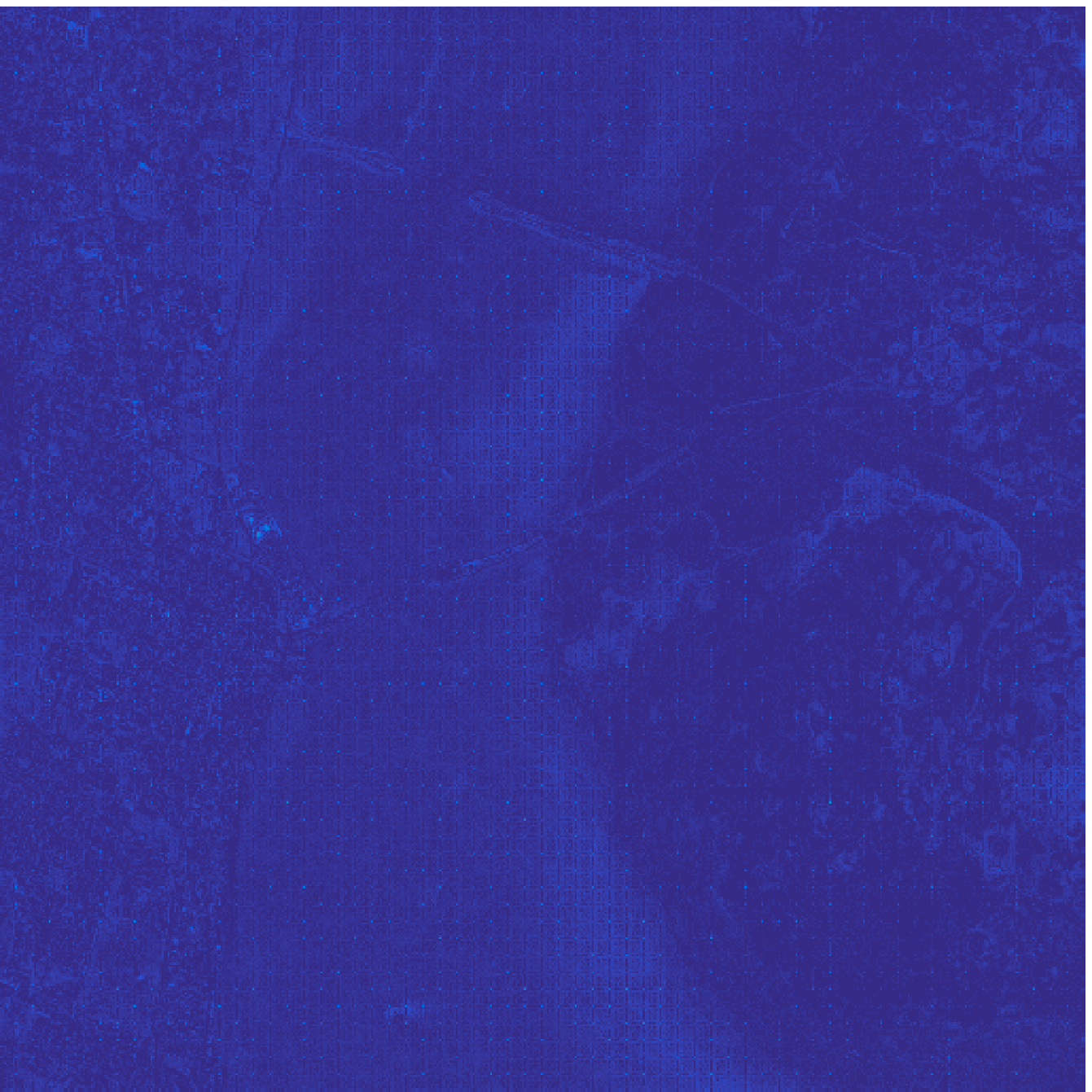}}
\subfigure[] {\includegraphics[width=\myimgsize in]{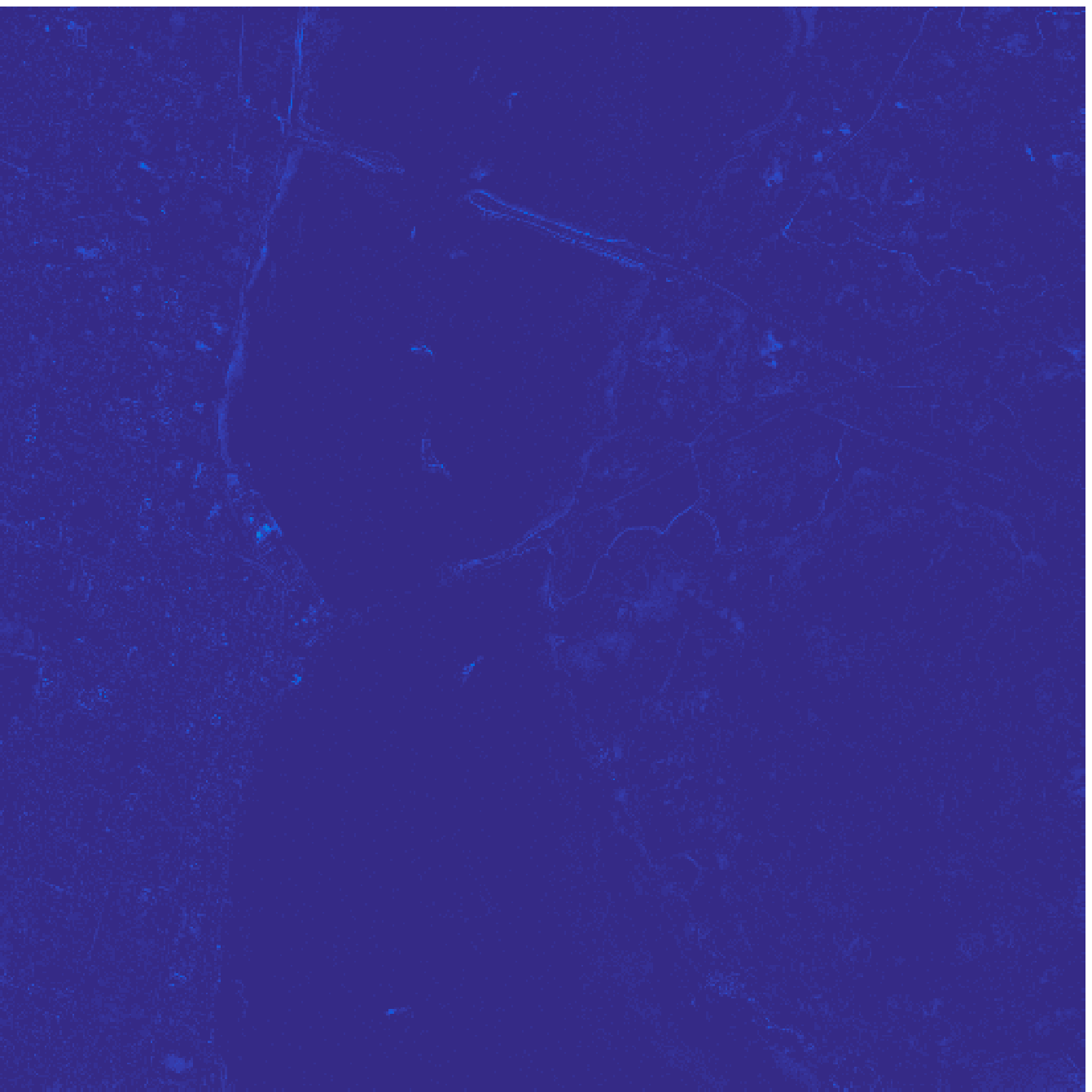}}
\subfigure[] {\includegraphics[width=\myimgsize in]{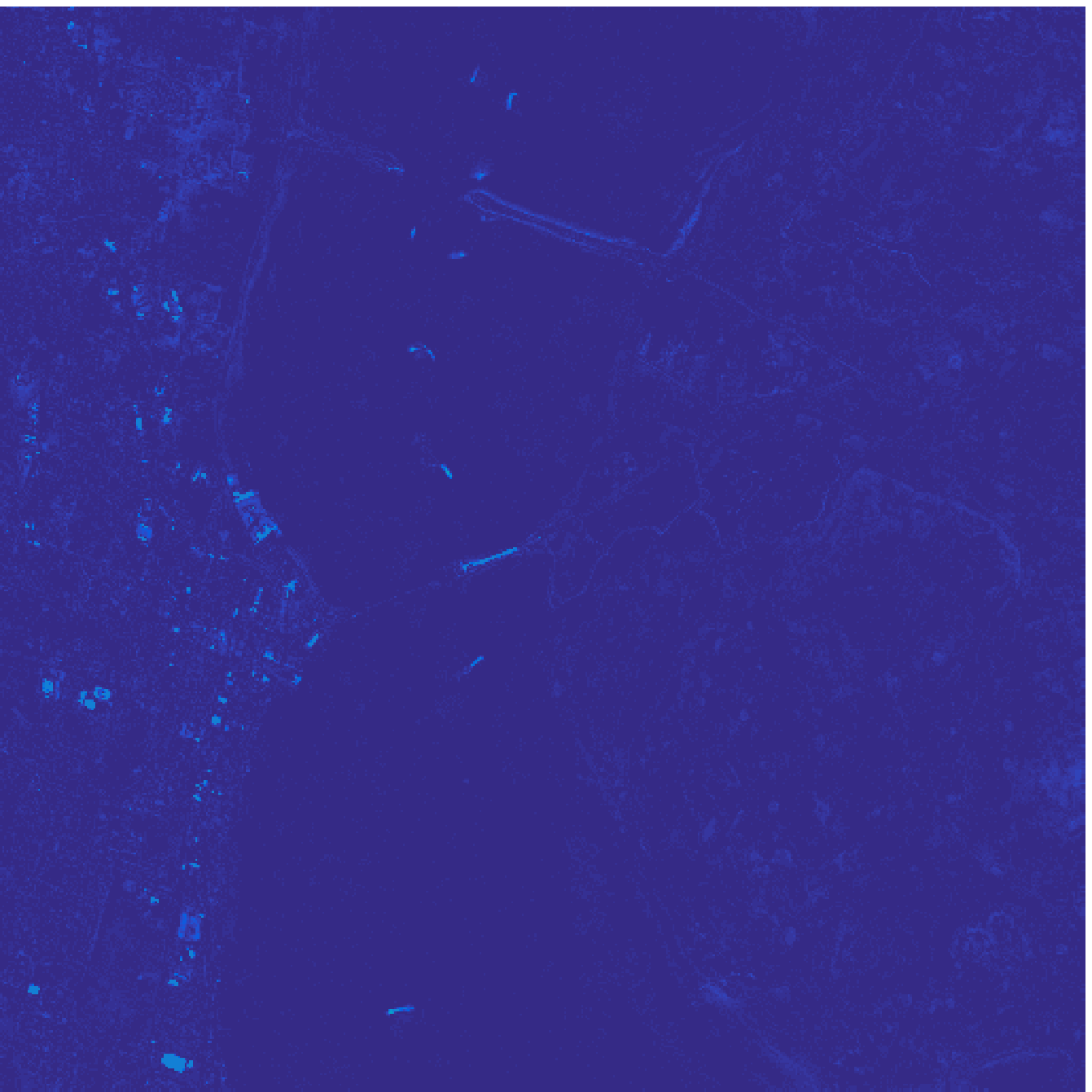}}
\subfigure[] {\includegraphics[width=\myimgsize in]{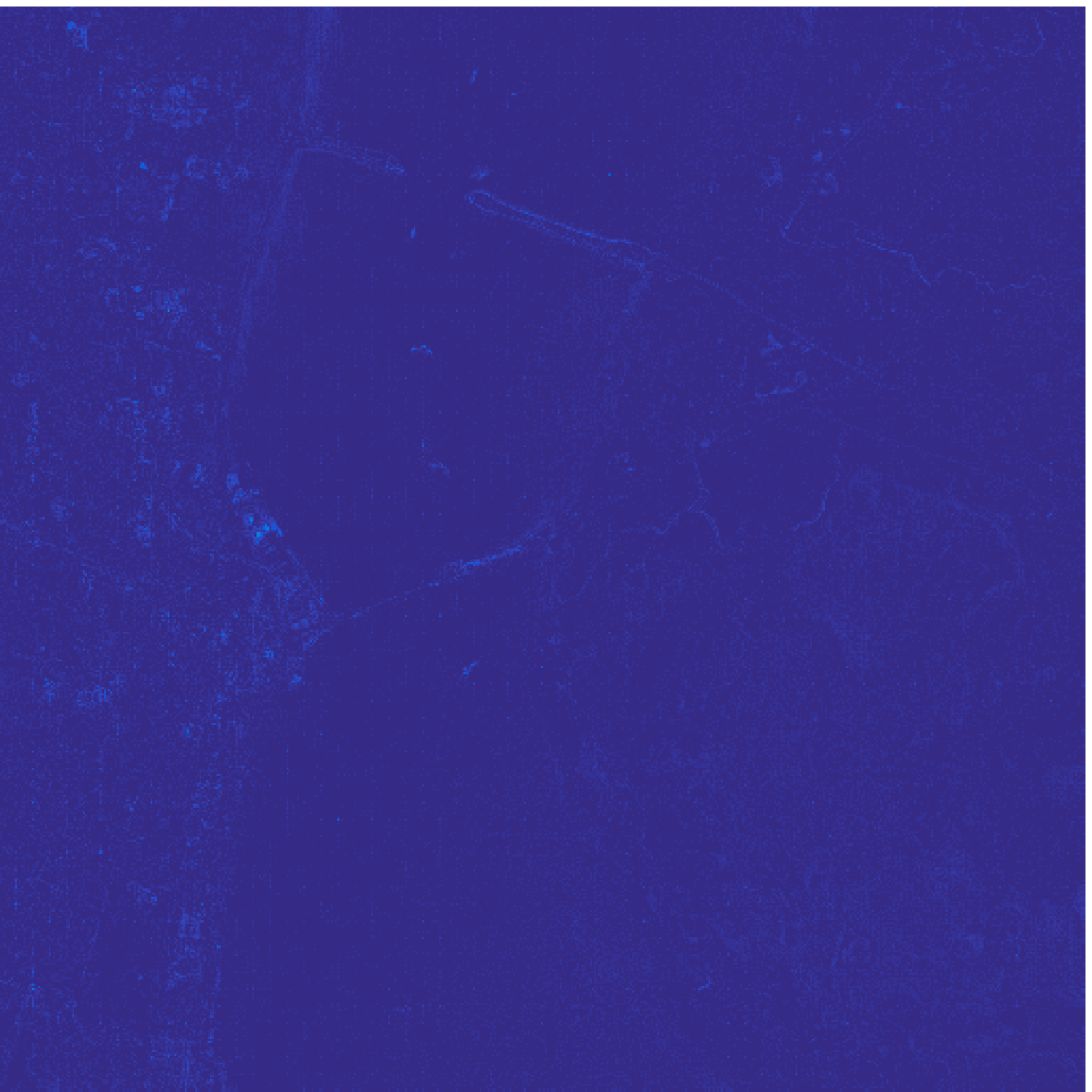}}
\subfigure[] {\includegraphics[width=\myimgsize in]{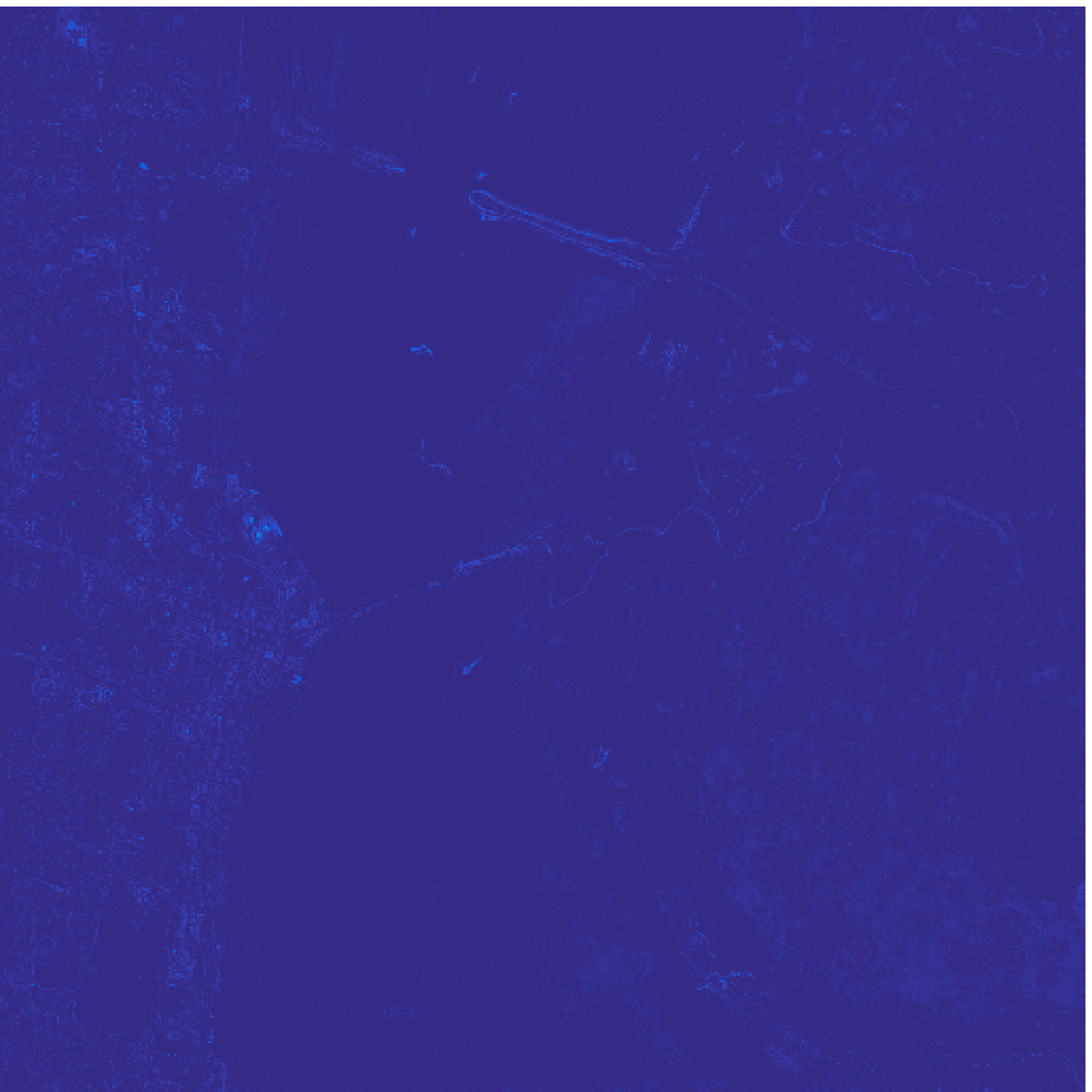}}
\subfigure[] {\includegraphics[width=\myimgsize in]{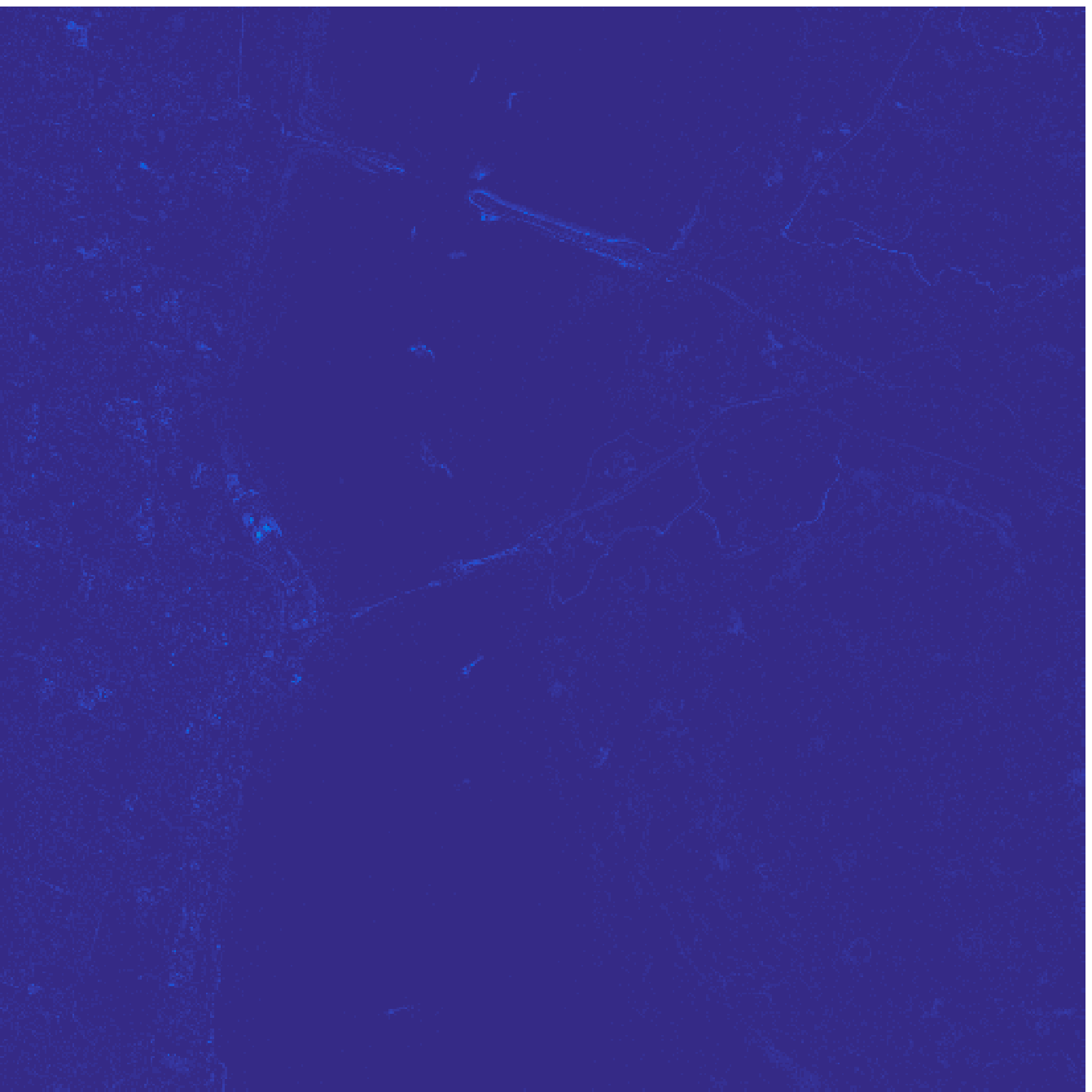}}
\subfigure[] {\includegraphics[width=\myimgsize in]{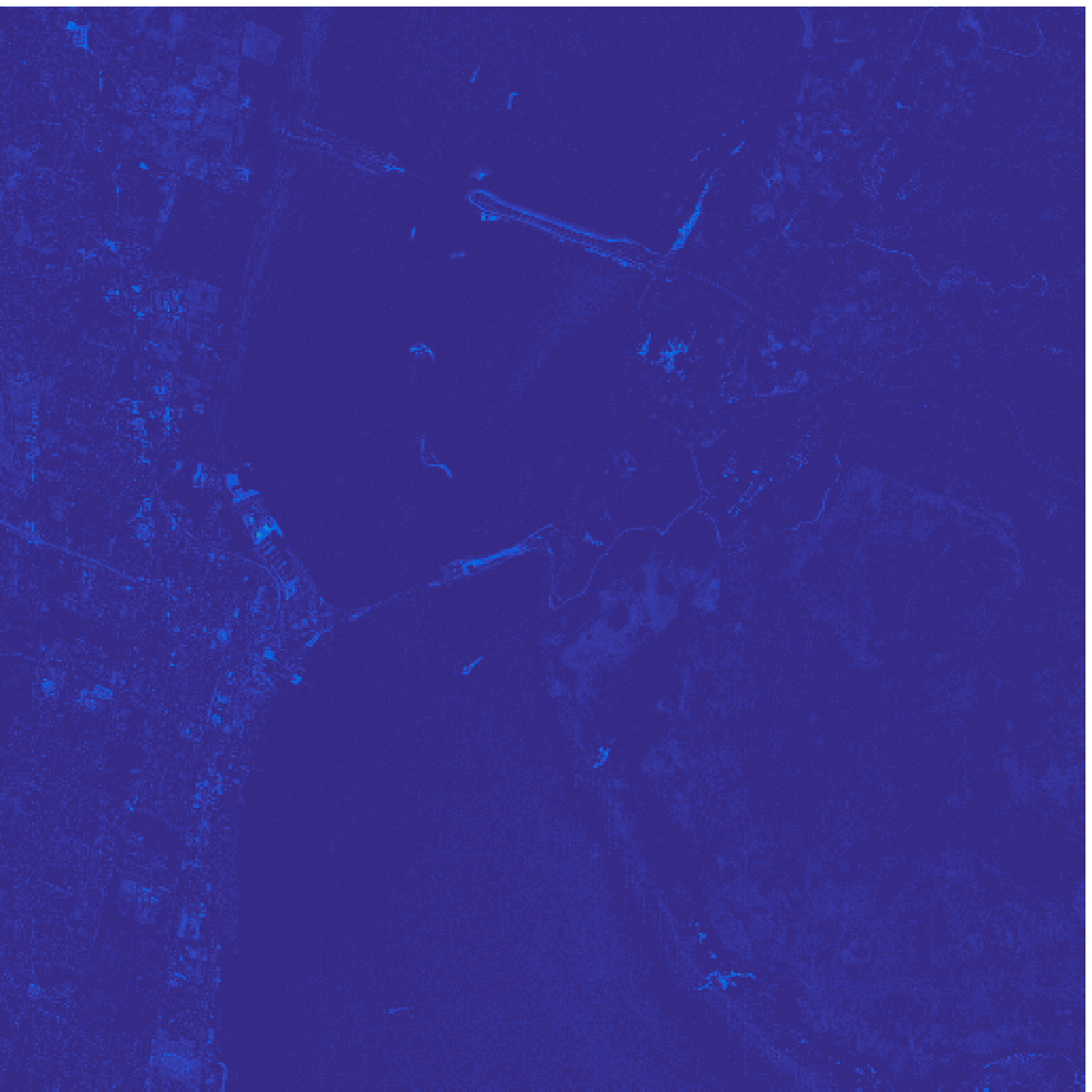}}
\subfigure[] {\includegraphics[width=\myimgsize in]{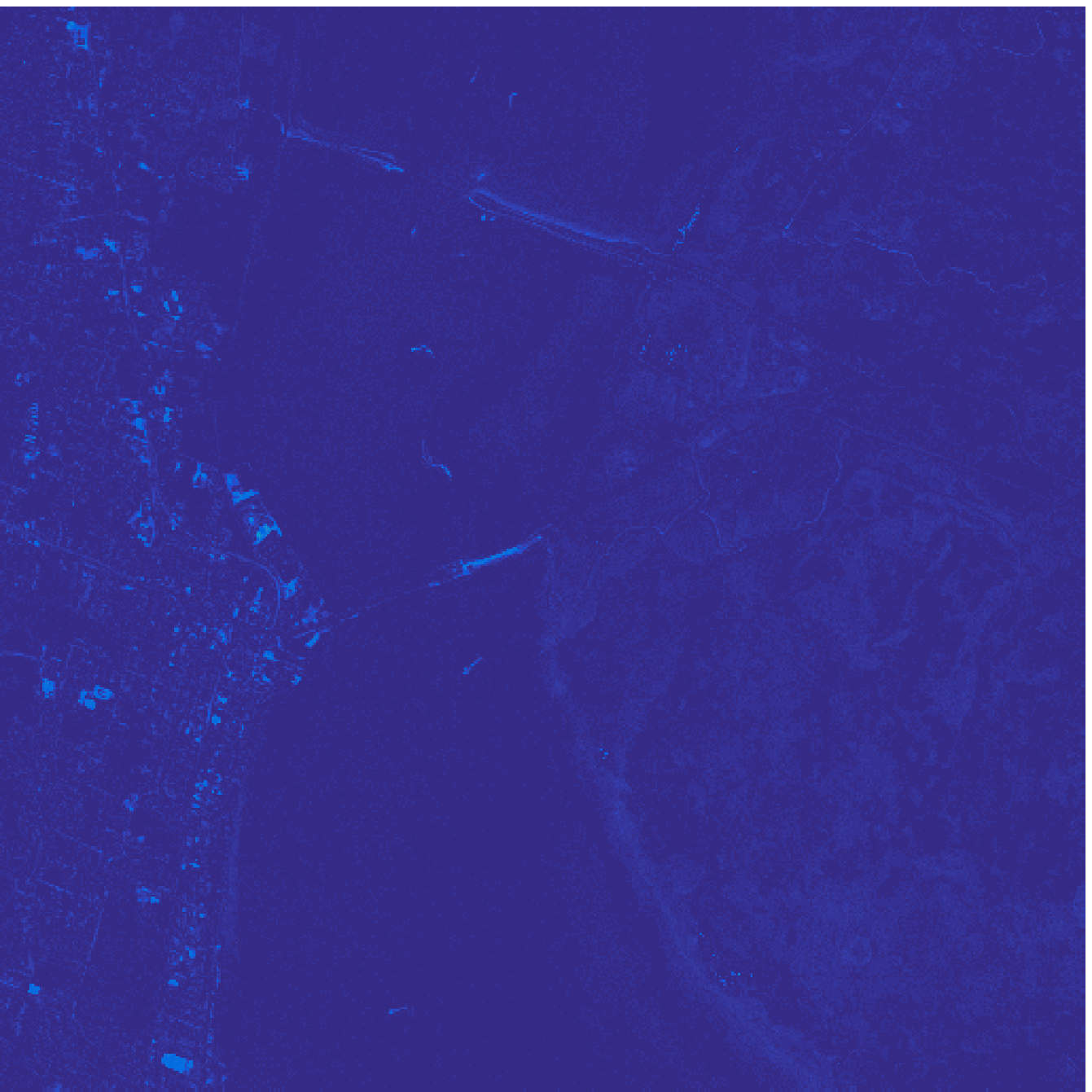}}
\subfigure[] {\includegraphics[width=\myimgsize in]{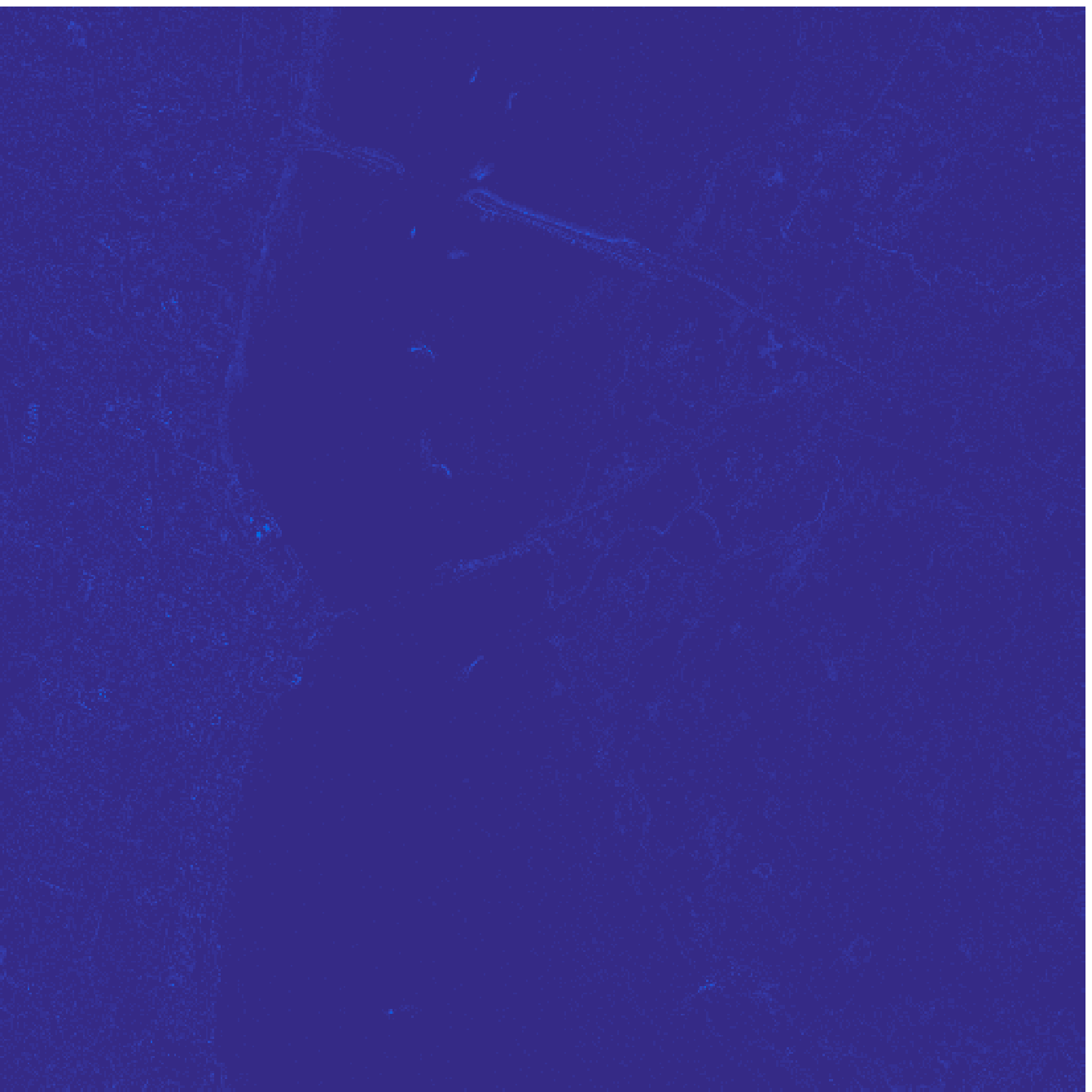}}
\subfigure[] {\includegraphics[width=\myimgsize in]{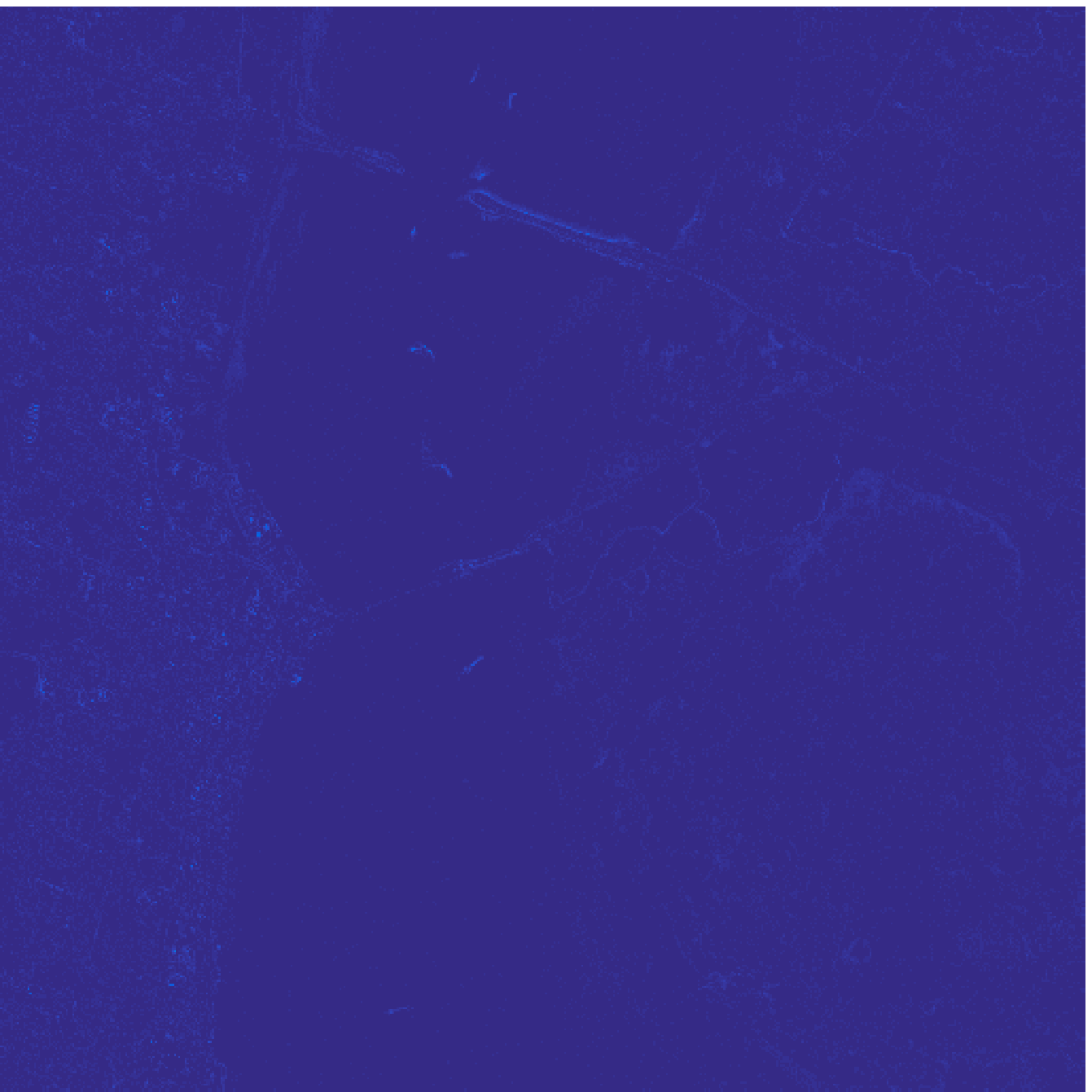}}
\caption{Images (with a meaningful region marked and zoomed in 3 times for easy observation) and error maps at band 30 of HSI super-resolution results when applied to the KSC dataset.
(a) Reference image. (b) SLYV. (c) CNMF. (d) CSU.
(e) NSSR. (f) HySure. (g) NPTSR.
(h) CNNFUS. (i) uSDN. (j) HyCoNet. (k) MIAE.}
\label{fig_ksc}
\end{figure*}

\begin{table*}[!t]
\caption{Quality measures for the DC dataset using different methods (the best values are highlighted)}
\label{tab_dc}
\centering
\begin{tabular}
{c|c|c|c|c|c|c|c|c|c|c}
\hline\hline
Method & SLYV & CNMF & CSU & NSSR & HySure & NPTSR & CNNFUS & uSDN & HyCoNet & MIAE \\
\hline\hline
      RMSE &0.1685 &0.0283 &0.0220 &0.0366 &0.0275 &0.0261 &0.0332 &0.0278 &0.0197 &\textbf{0.0145} \\
      PSNR &18.53 &32.61 &32.48 &29.90 &32.15 &32.77 &32.39 &29.24 &31.63 &\textbf{35.90} \\
       SAM &26.56 &3.82 &2.82 &5.33 &3.48 &3.54 &3.70 &3.53 &\textbf{1.83} &1.84 \\
     ERGAS &37.085 &15.182 &14.406 &14.447 &13.797 &13.886 &14.300 &17.345 &\textbf{10.943} &14.224 \\
      UIQI &0.469 &0.930 &0.929 &0.905 &0.930 &0.944 &0.954 &0.856 &0.914 &\textbf{0.975} \\
\hline\hline
\end{tabular}
\end{table*}

\begin{figure*}[!t]
\centering
{\includegraphics[width=\myimgsize in]{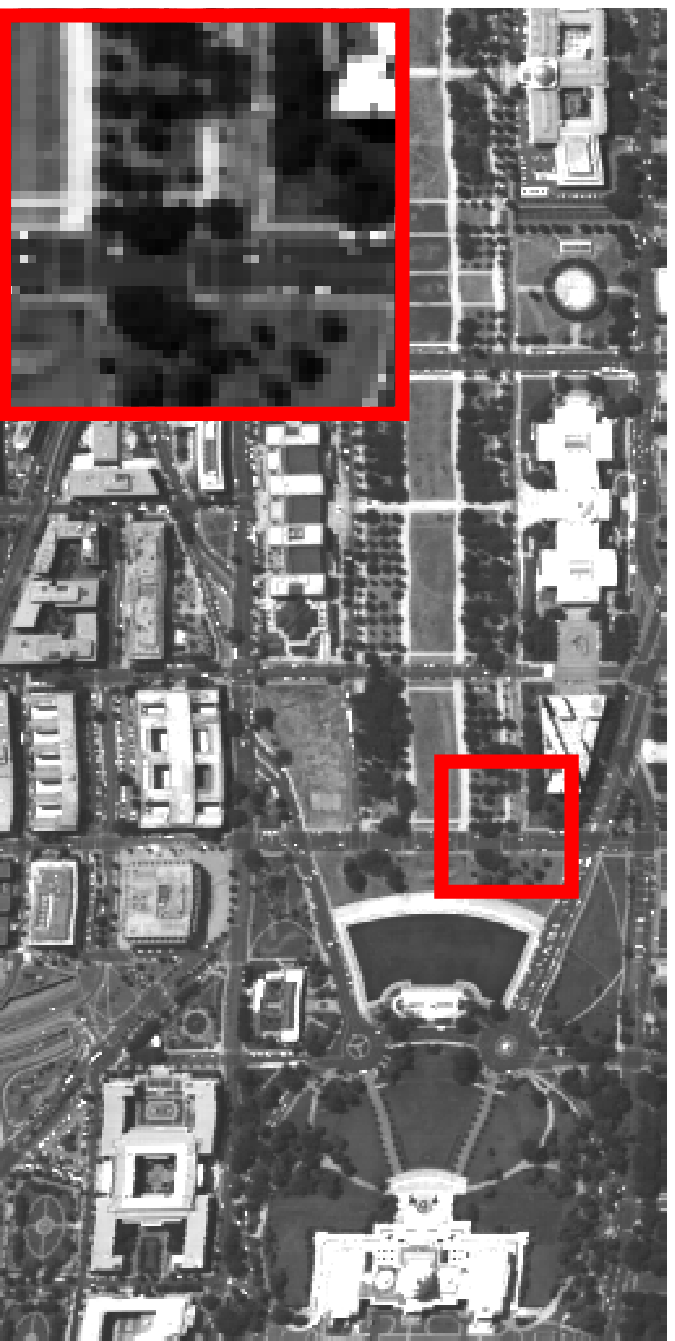}}
{\includegraphics[width=\myimgsize in]{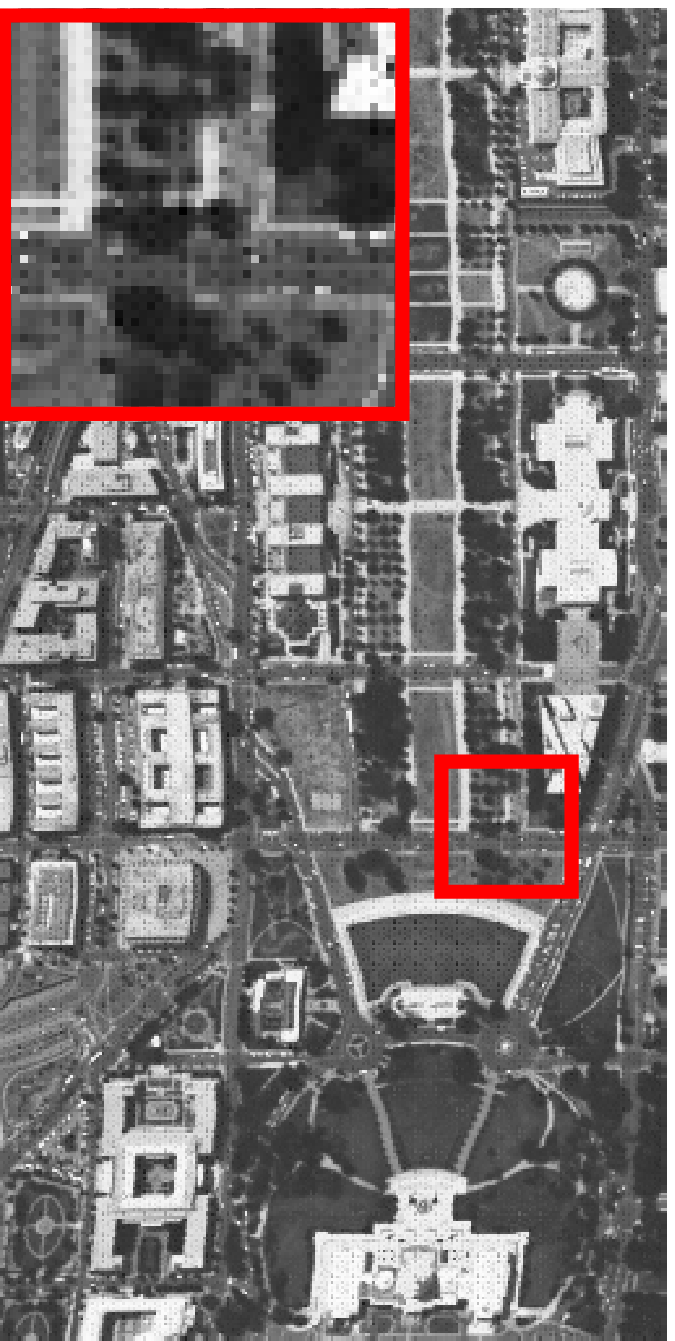}}
{\includegraphics[width=\myimgsize in]{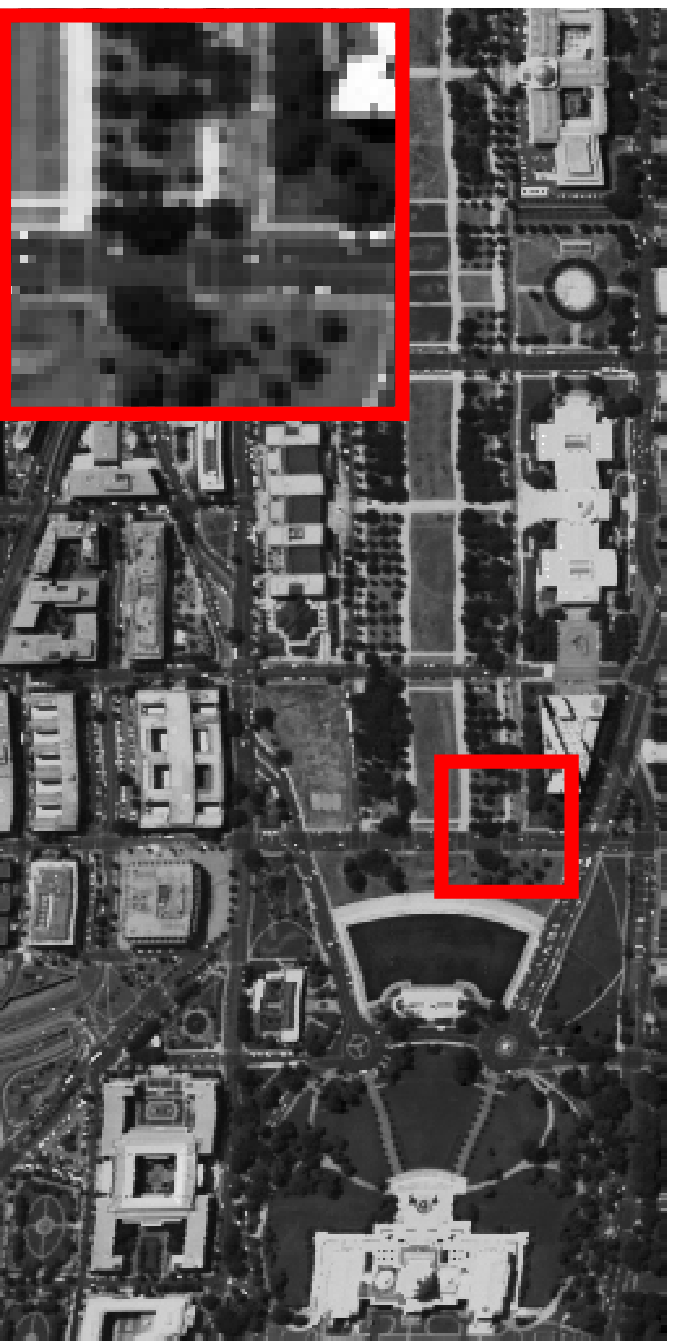}}
{\includegraphics[width=\myimgsize in]{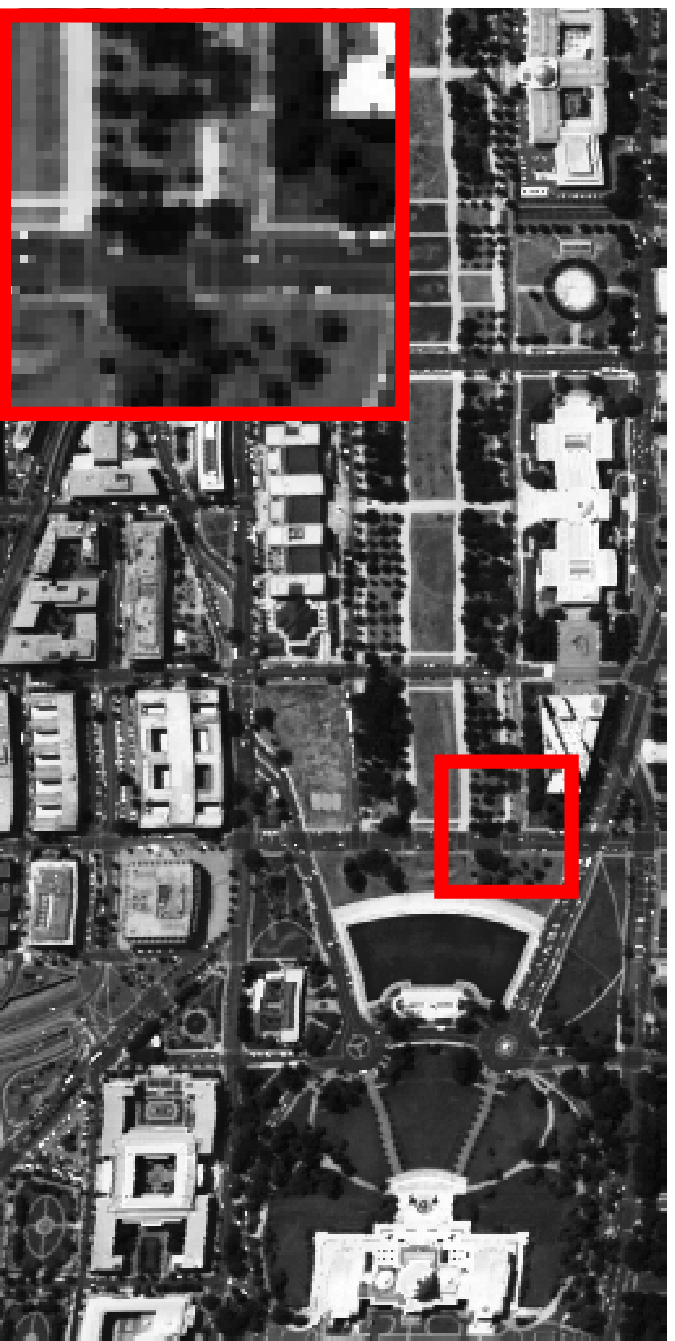}}
{\includegraphics[width=\myimgsize in]{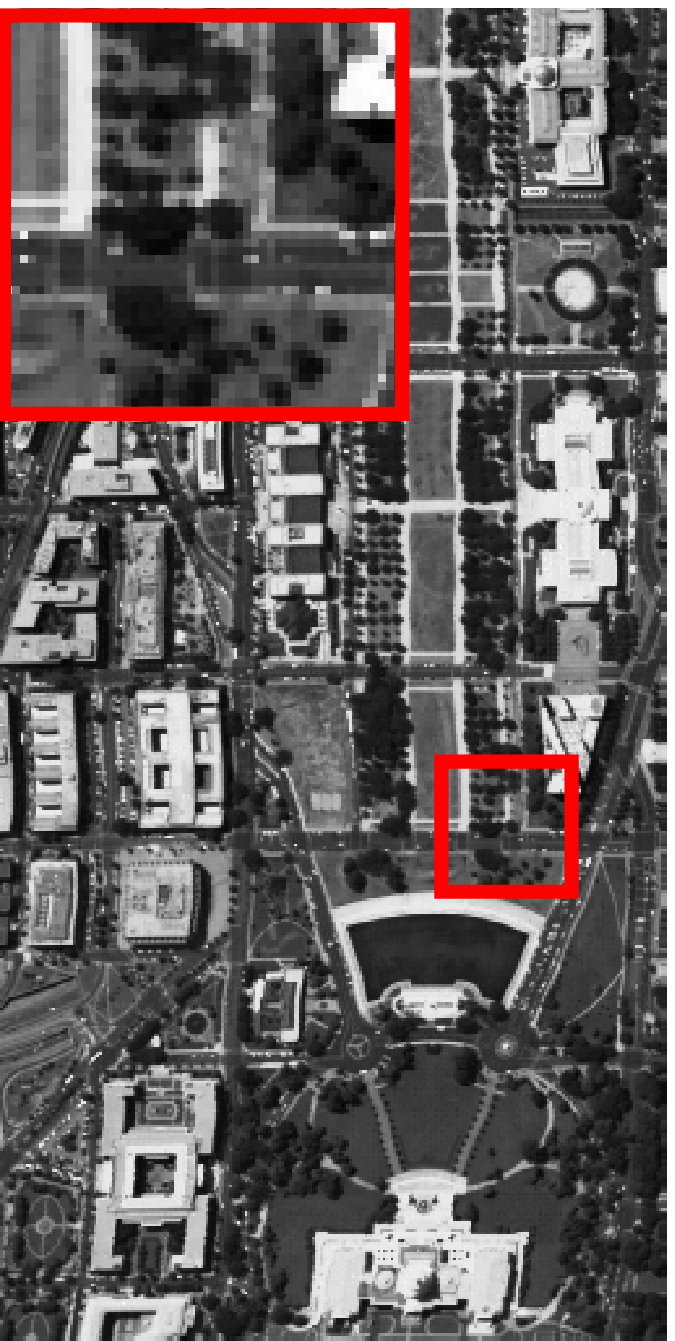}}
{\includegraphics[width=\myimgsize in]{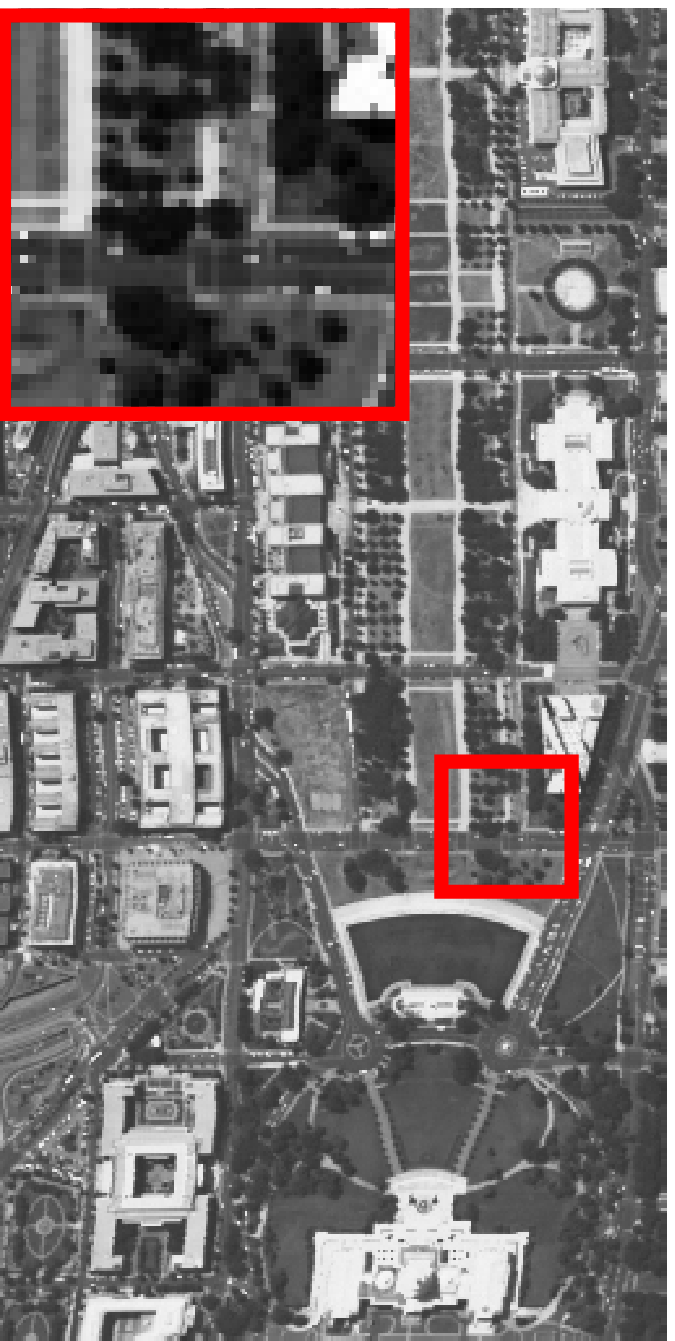}}
{\includegraphics[width=\myimgsize in]{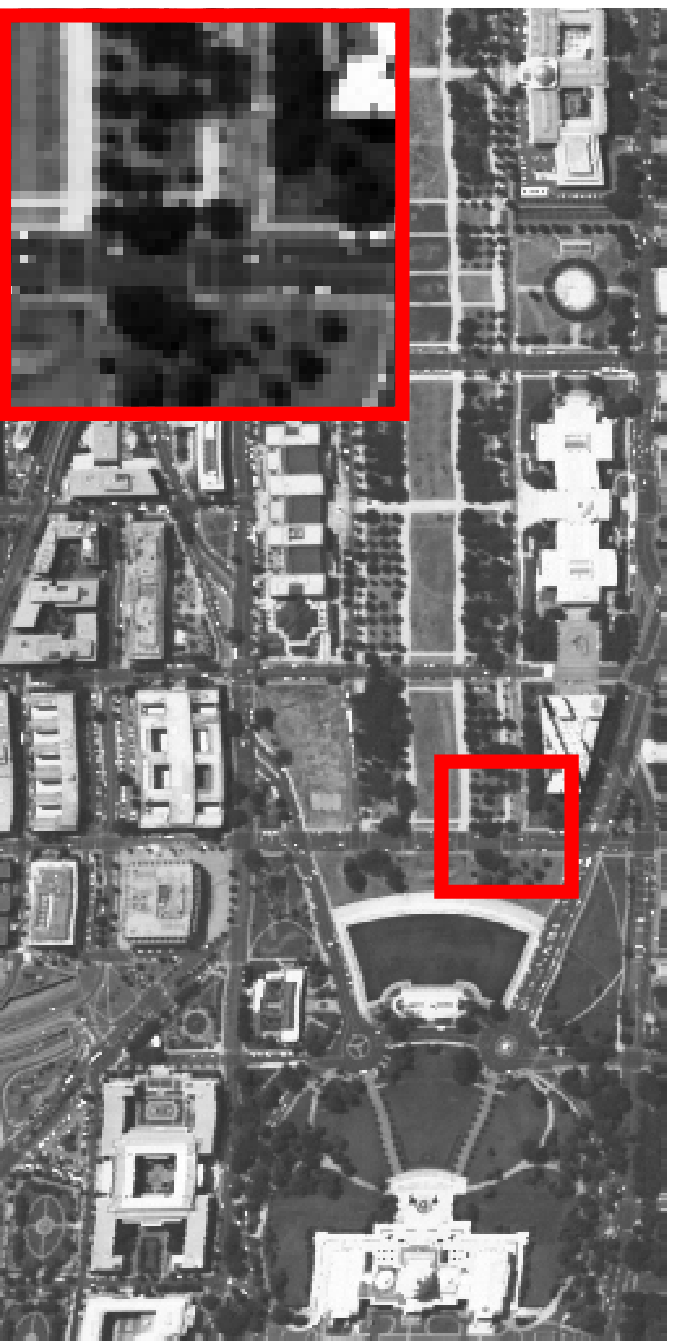}}
{\includegraphics[width=\myimgsize in]{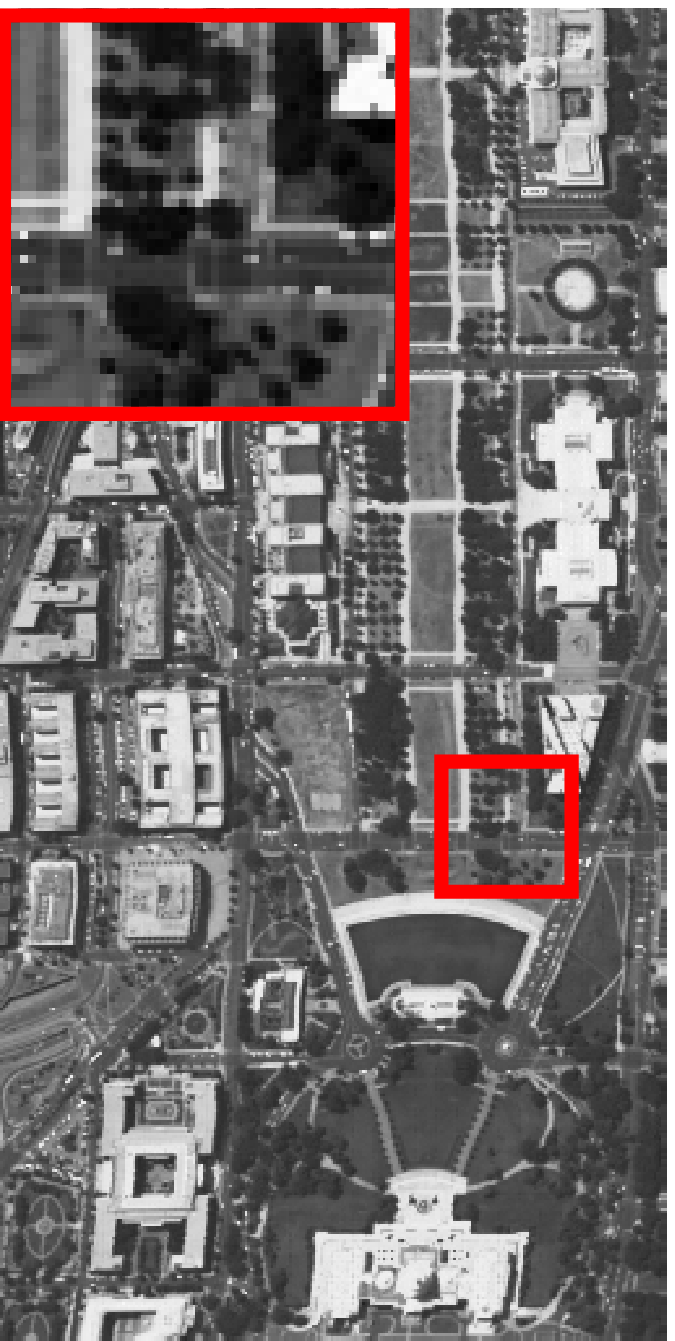}}
{\includegraphics[width=\myimgsize in]{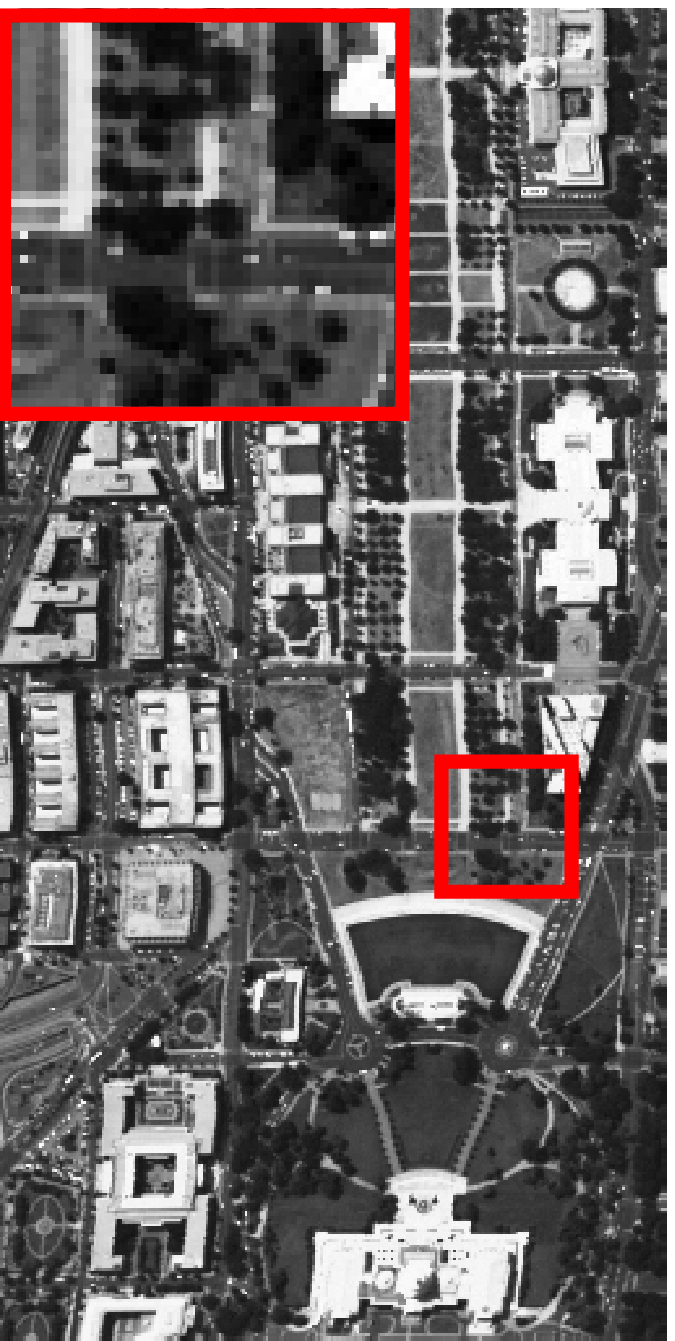}}
{\includegraphics[width=\myimgsize in]{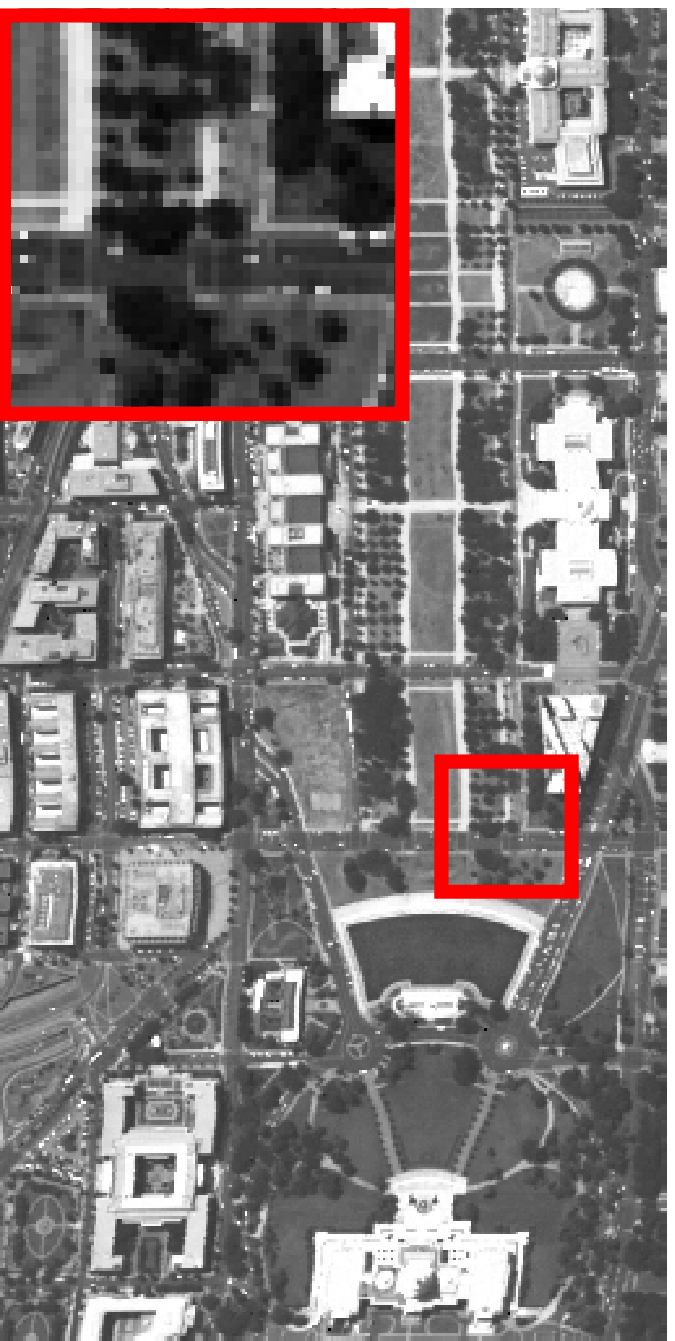}}
{\includegraphics[width=\myimgsize in]{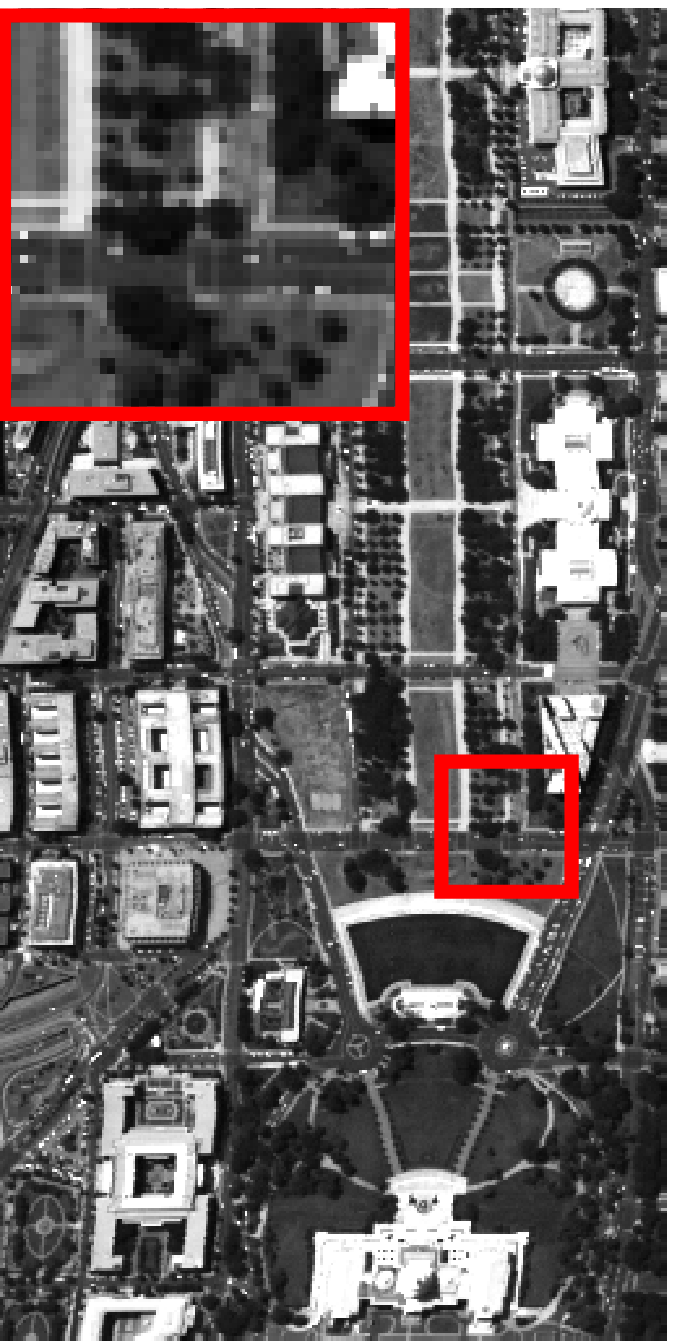}} \\
\subfigure[] {\includegraphics[width=\myimgsize in]{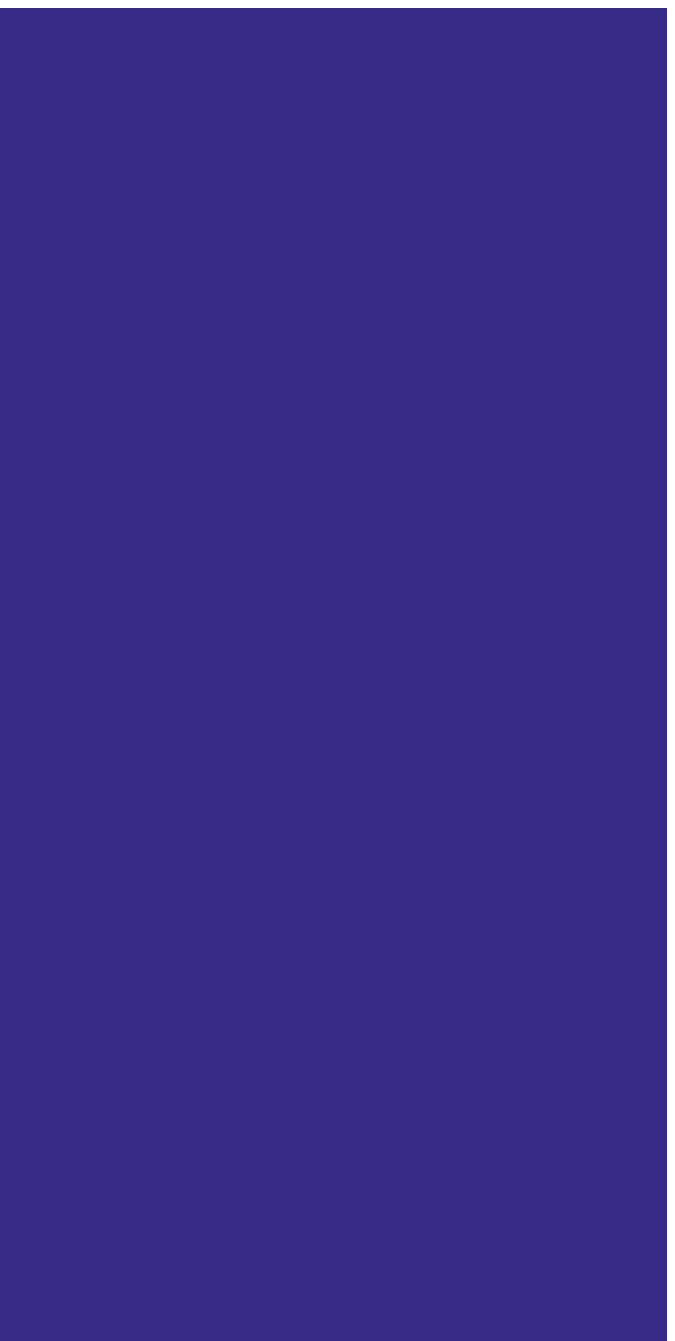}}
\subfigure[] {\includegraphics[width=\myimgsize in]{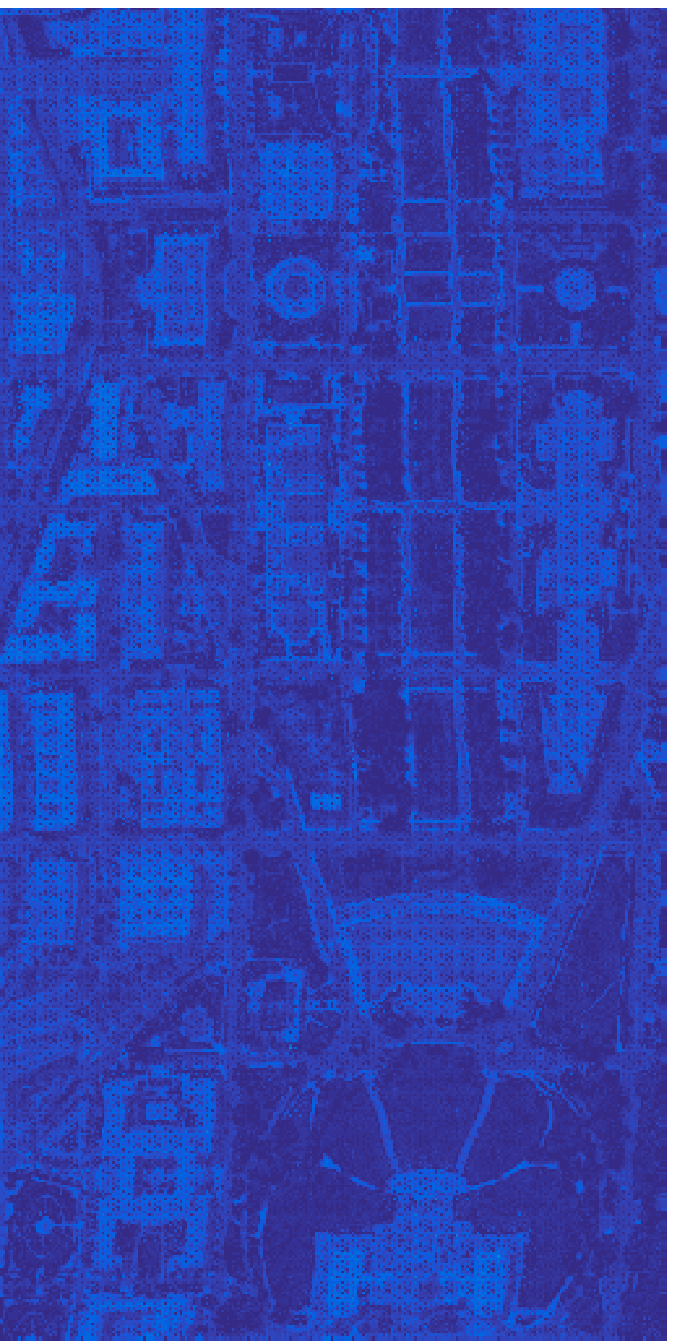}}
\subfigure[] {\includegraphics[width=\myimgsize in]{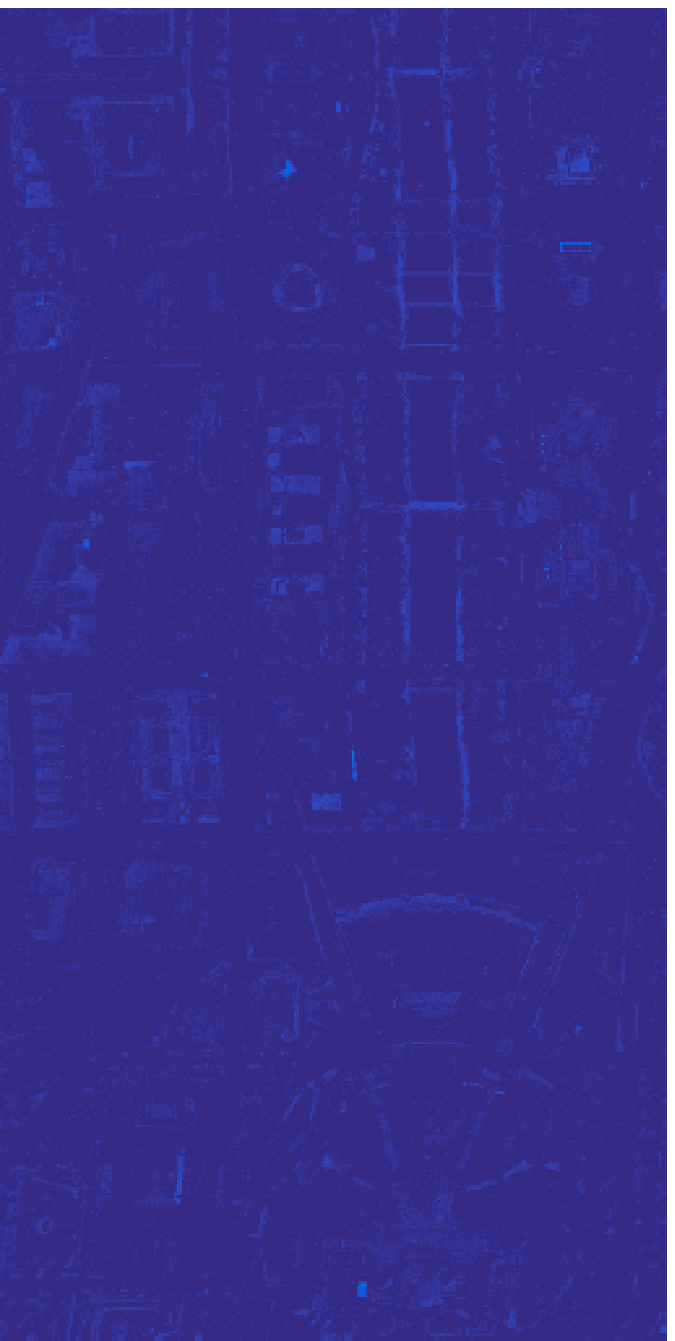}}
\subfigure[] {\includegraphics[width=\myimgsize in]{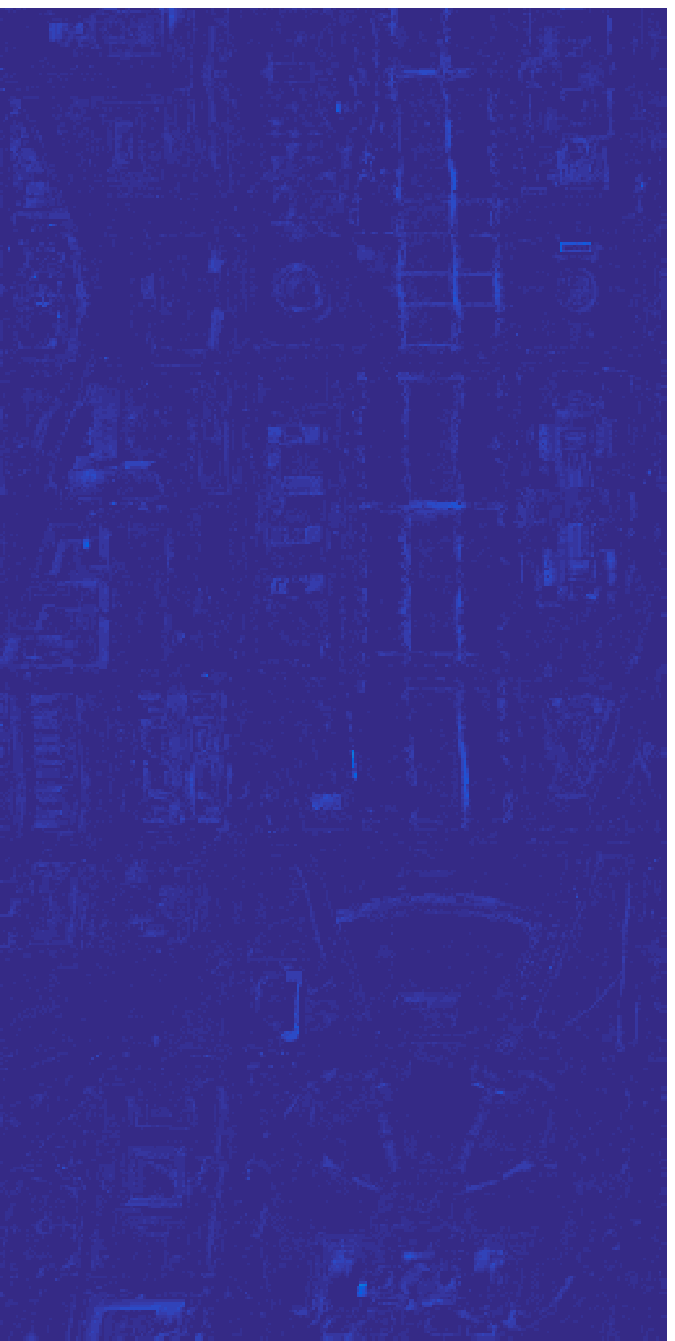}}
\subfigure[] {\includegraphics[width=\myimgsize in]{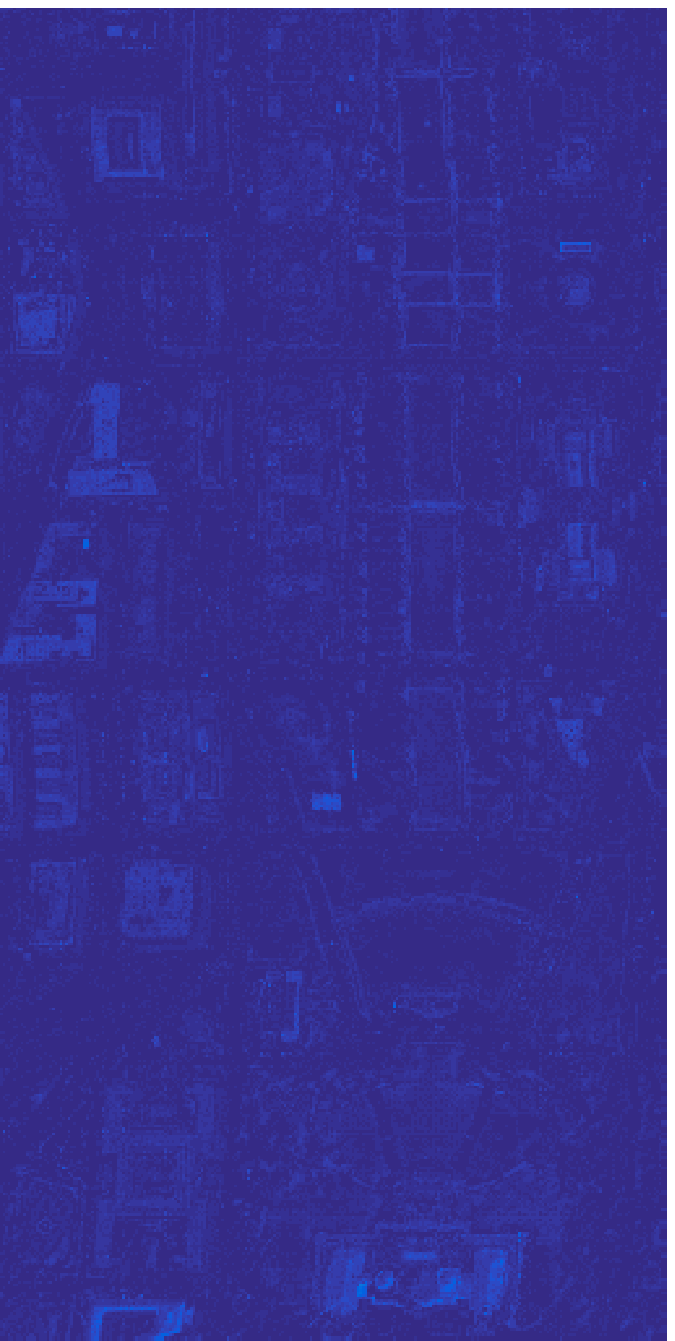}}
\subfigure[] {\includegraphics[width=\myimgsize in]{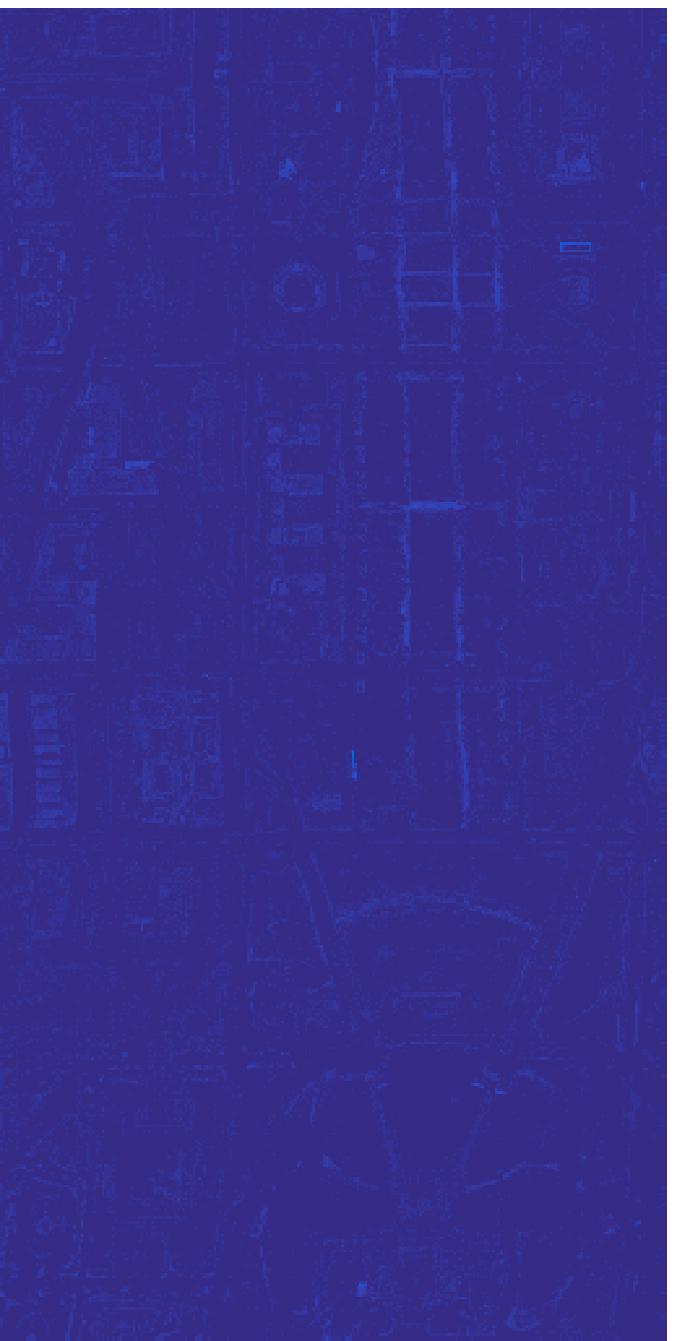}}
\subfigure[] {\includegraphics[width=\myimgsize in]{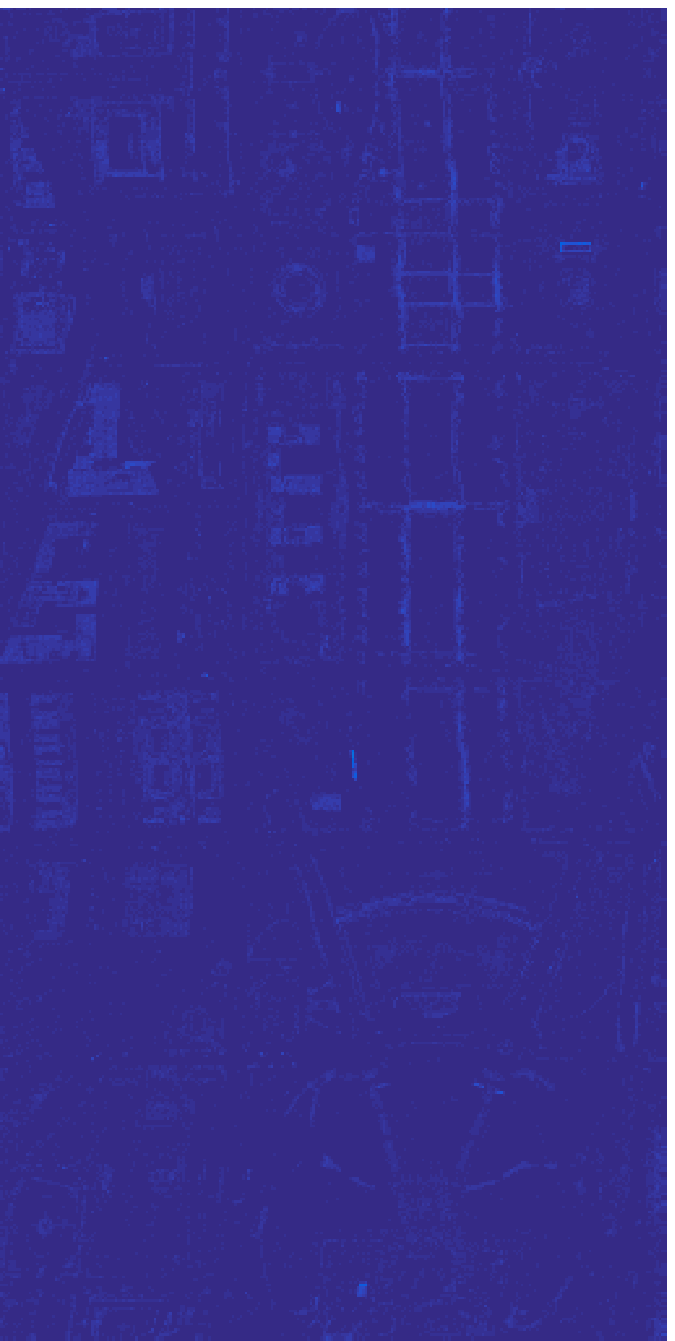}}
\subfigure[] {\includegraphics[width=\myimgsize in]{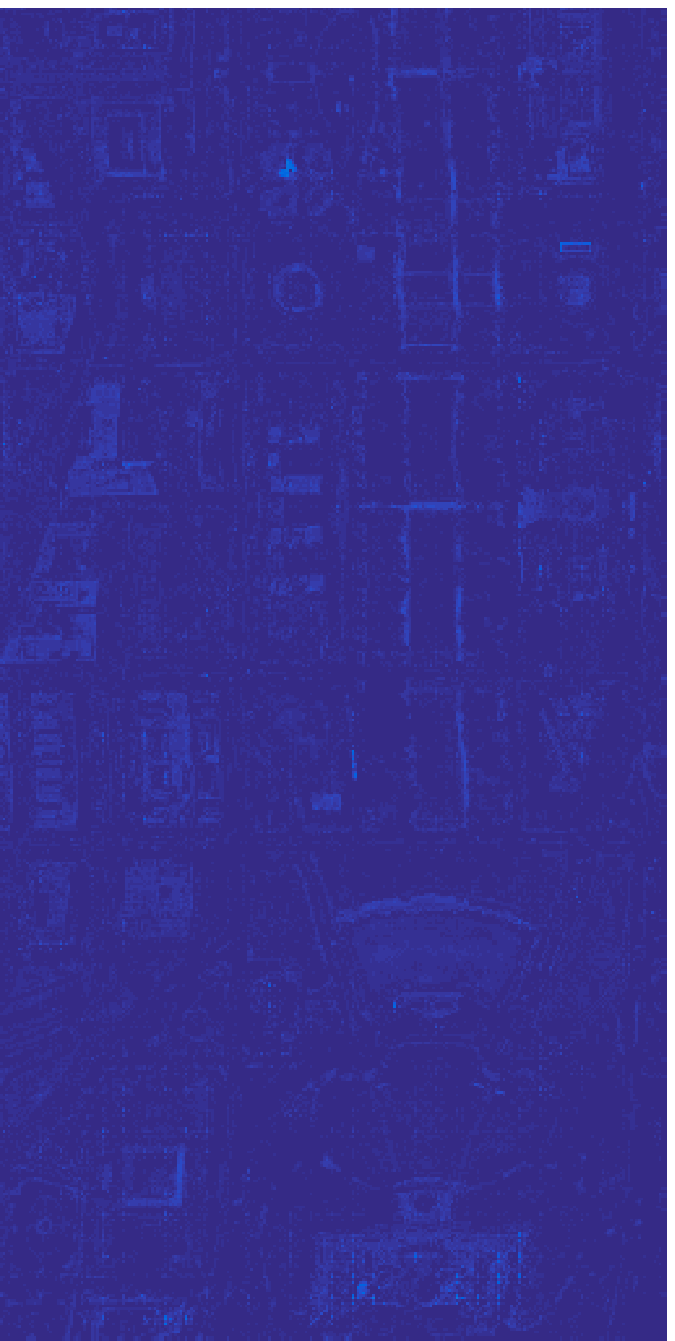}}
\subfigure[] {\includegraphics[width=\myimgsize in]{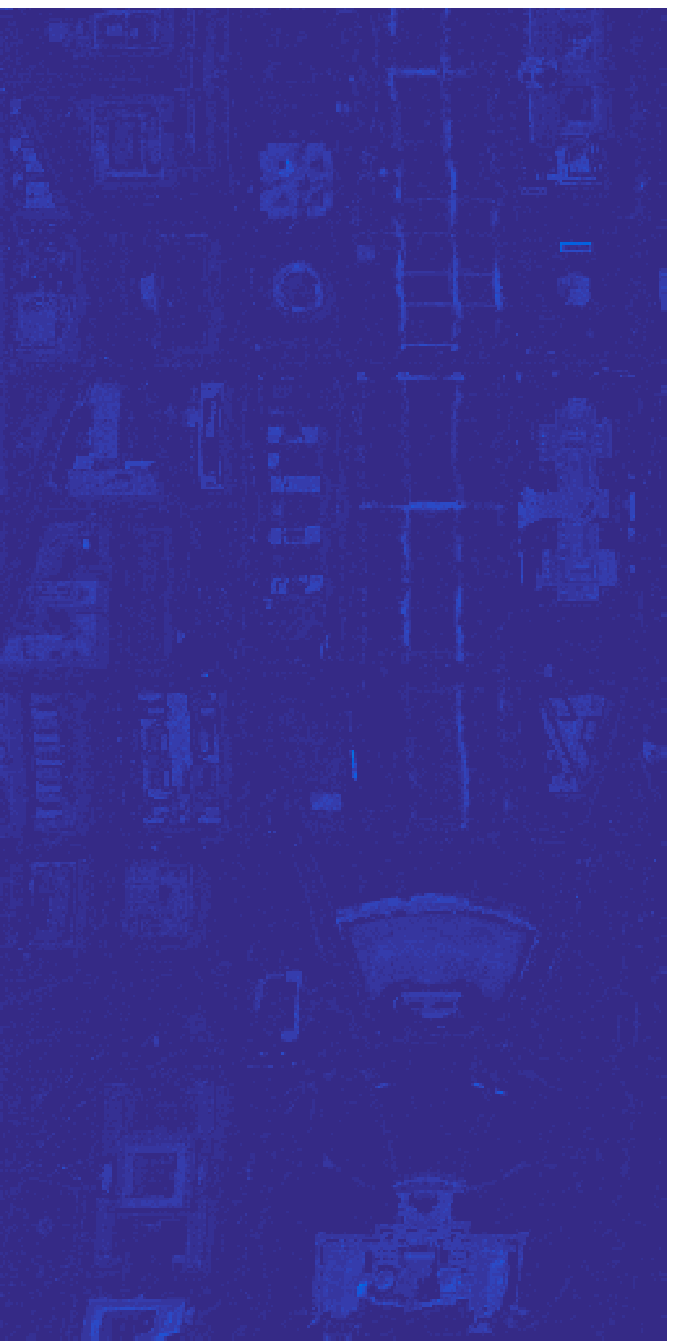}}
\subfigure[] {\includegraphics[width=\myimgsize in]{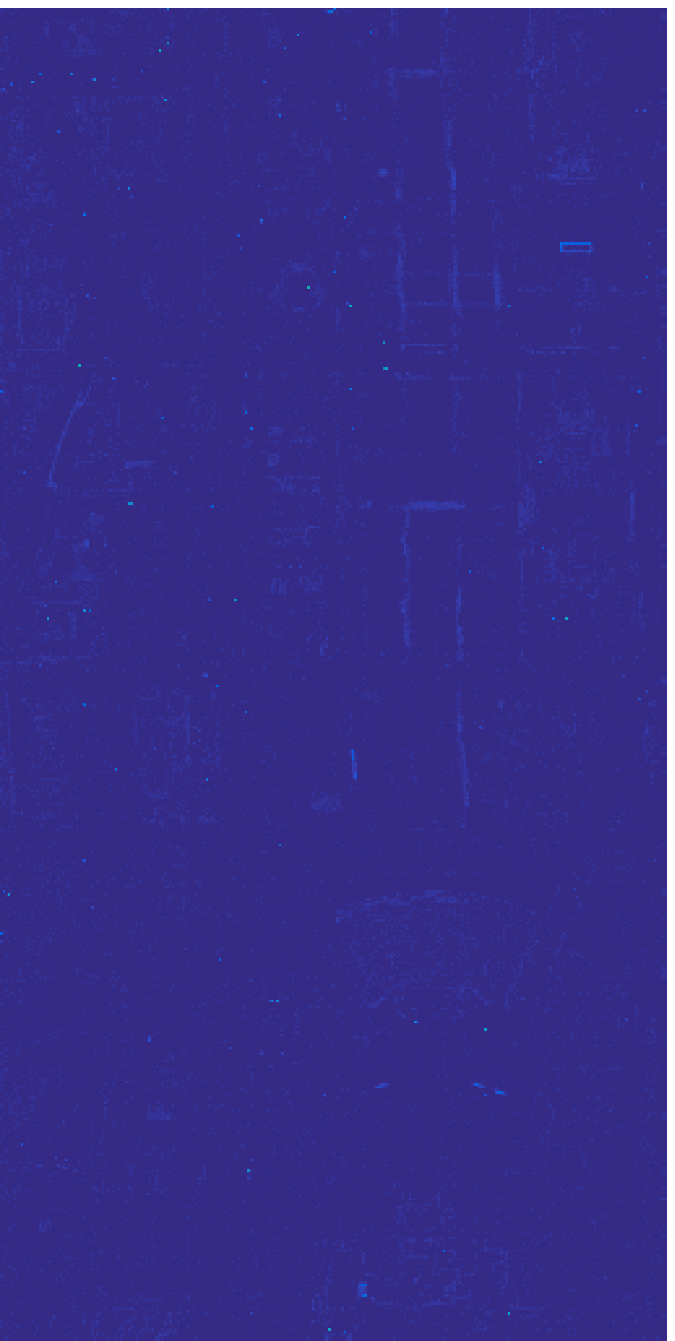}}
\subfigure[] {\includegraphics[width=\myimgsize in]{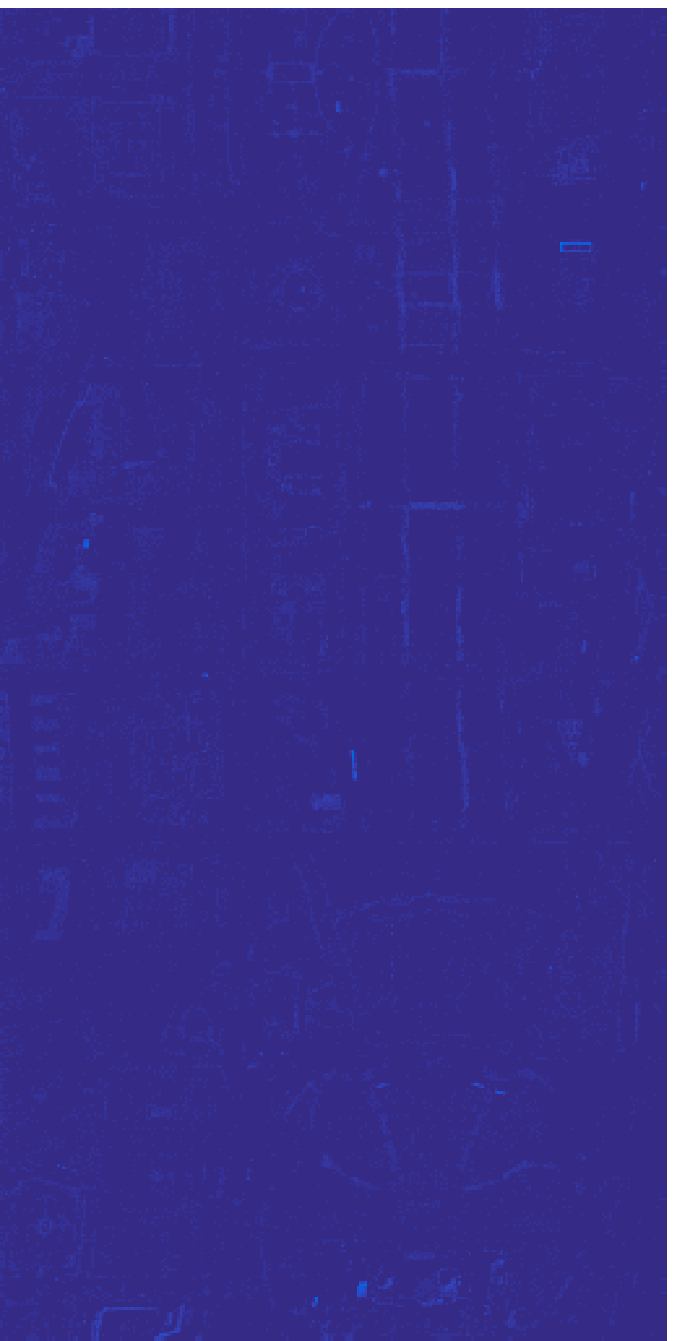}}
\caption{Images (with a meaningful region marked and zoomed in 3 times for easy observation) and error maps at band 30 of HSI super-resolution results when applied to the DC dataset.
(a) Reference image. (b) SLYV. (c) CNMF. (d) CSU.
(e) NSSR. (f) HySure. (g) NPTSR.
(h) CNNFUS. (i) uSDN. (j) HyCoNet. (k) MIAE.}
\label{fig_dc}
\end{figure*}

\begin{figure*}[!t]
\centering
\subfigure[] {\includegraphics[width=0.3 \textwidth]{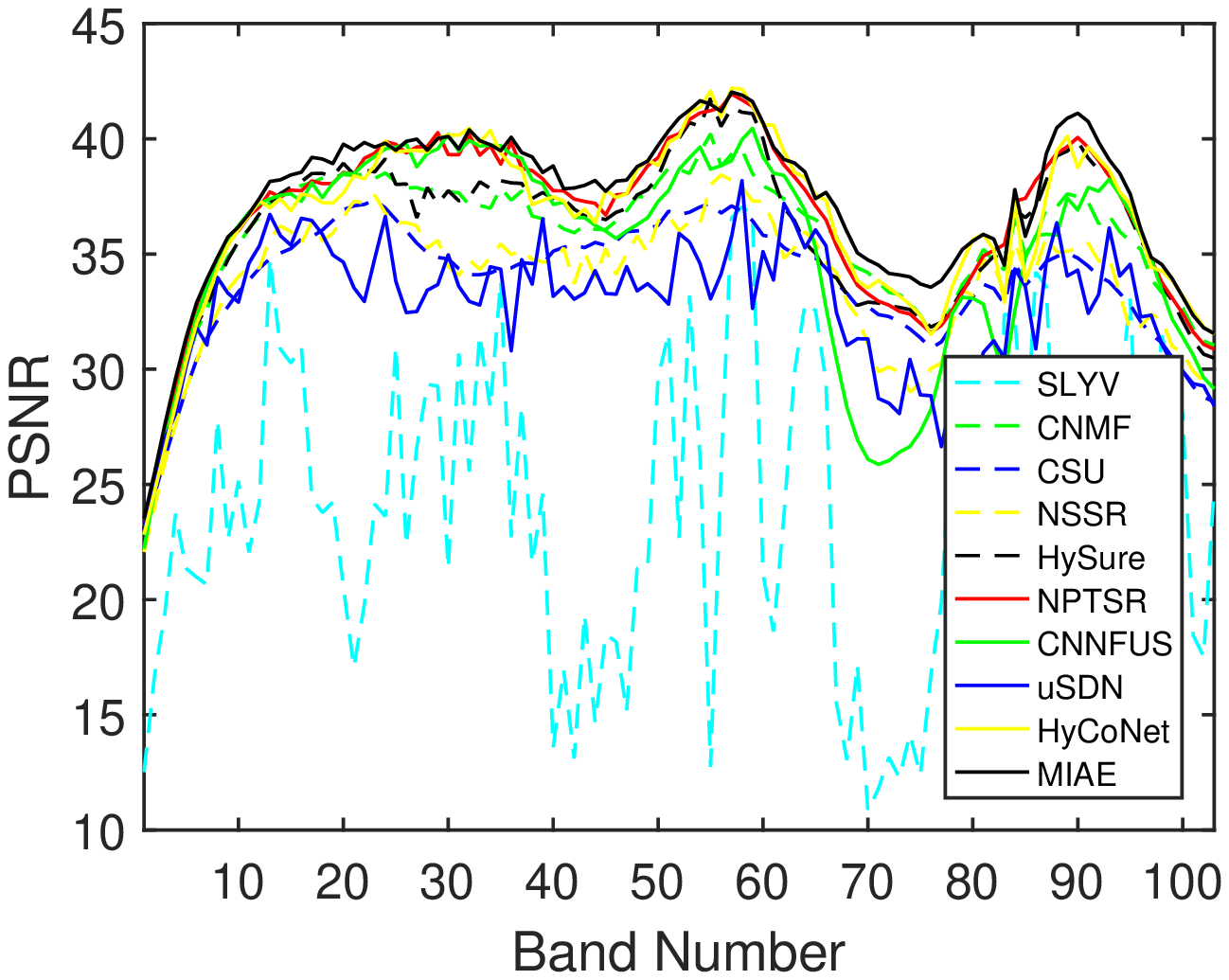}}~
\subfigure[] {\includegraphics[width=0.3 \textwidth]{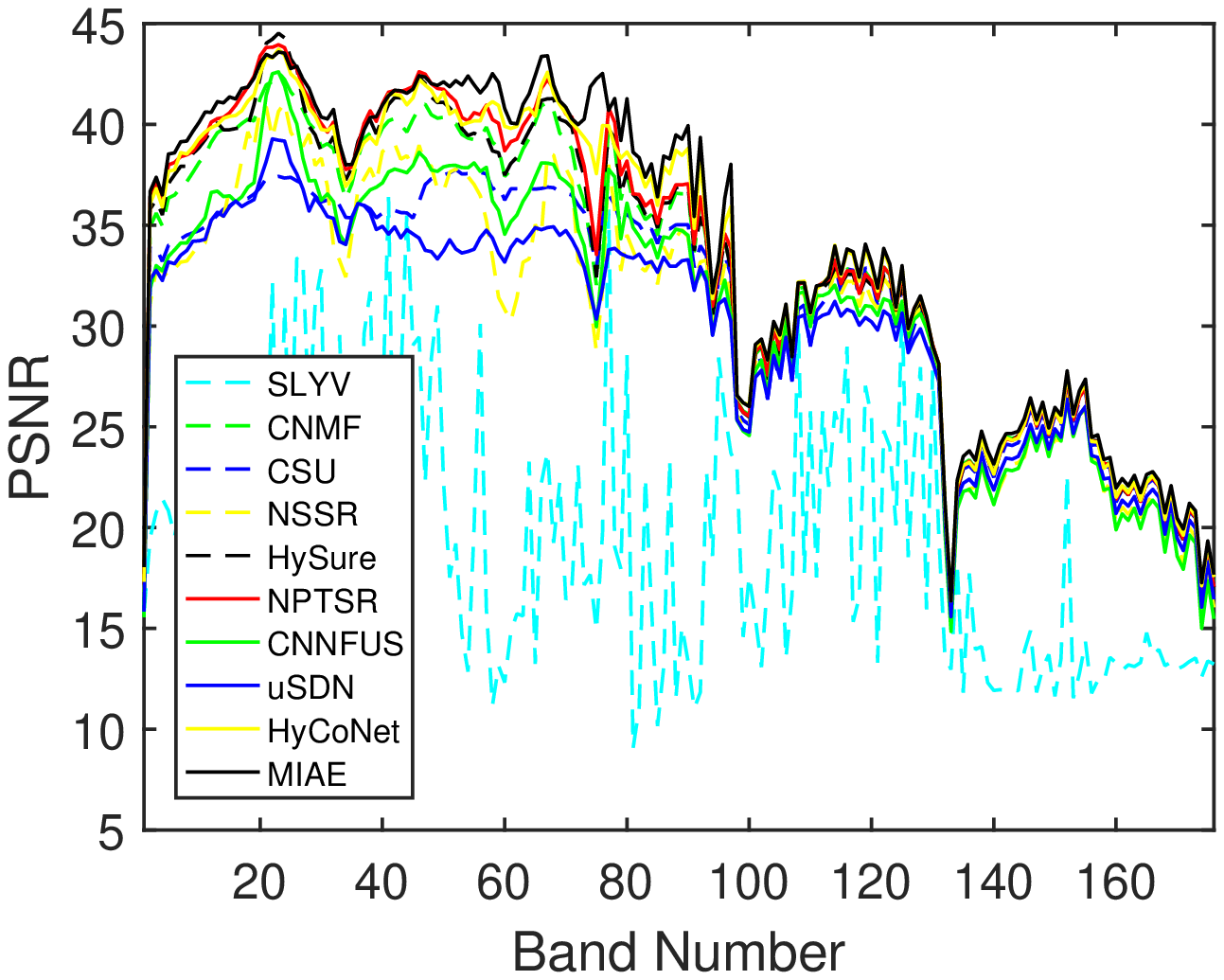}}~
\subfigure[] {\includegraphics[width=0.3 \textwidth]{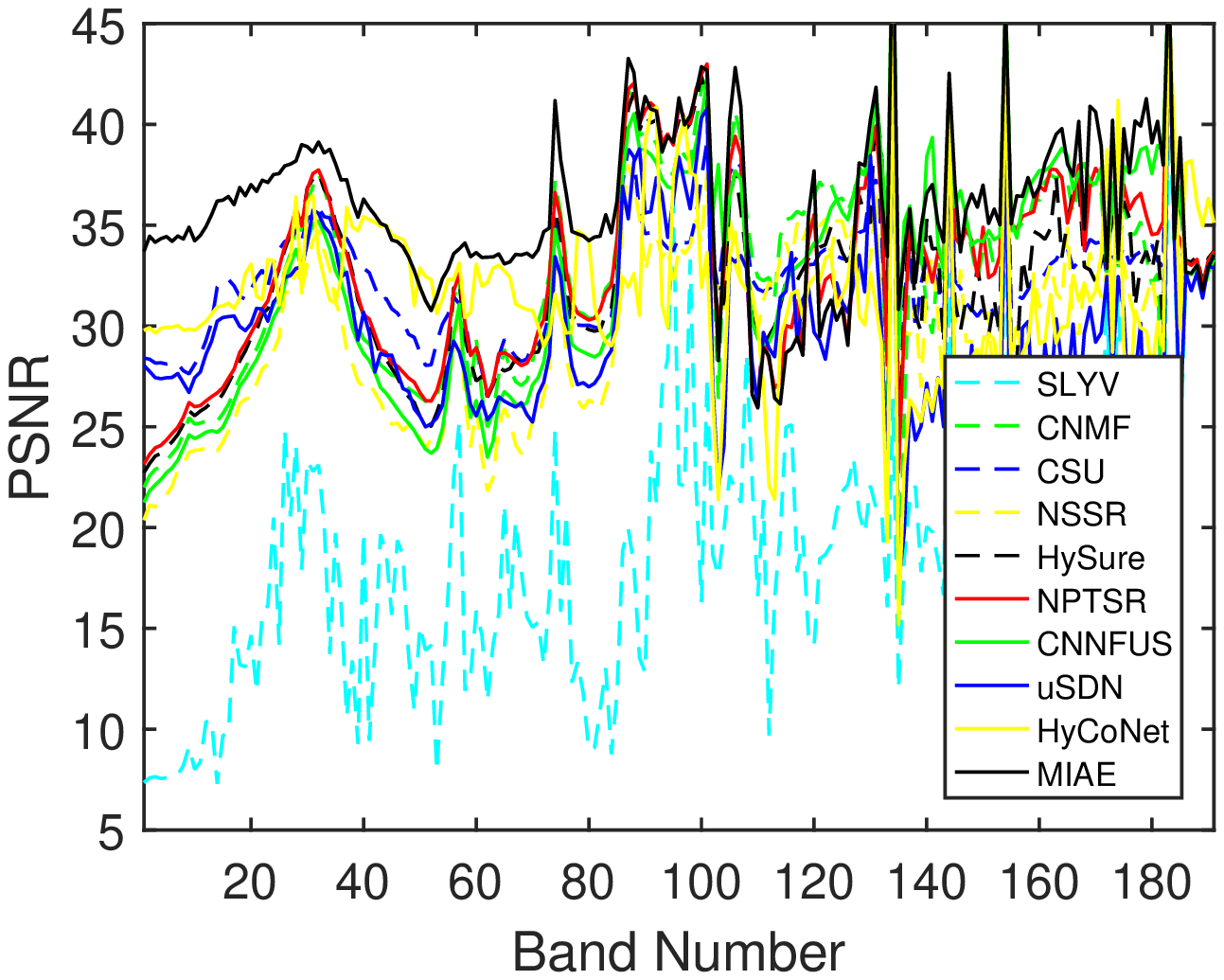}}
\caption{PSNR as a function of spectral band. (a) PaviaU dataset. (b) KSC dataset. (c) DC dataset.}
\label{fig_psnr}
\end{figure*}

\begin{figure*}[!t]
\centering
\subfigure[] {\includegraphics[width=0.3 \textwidth]{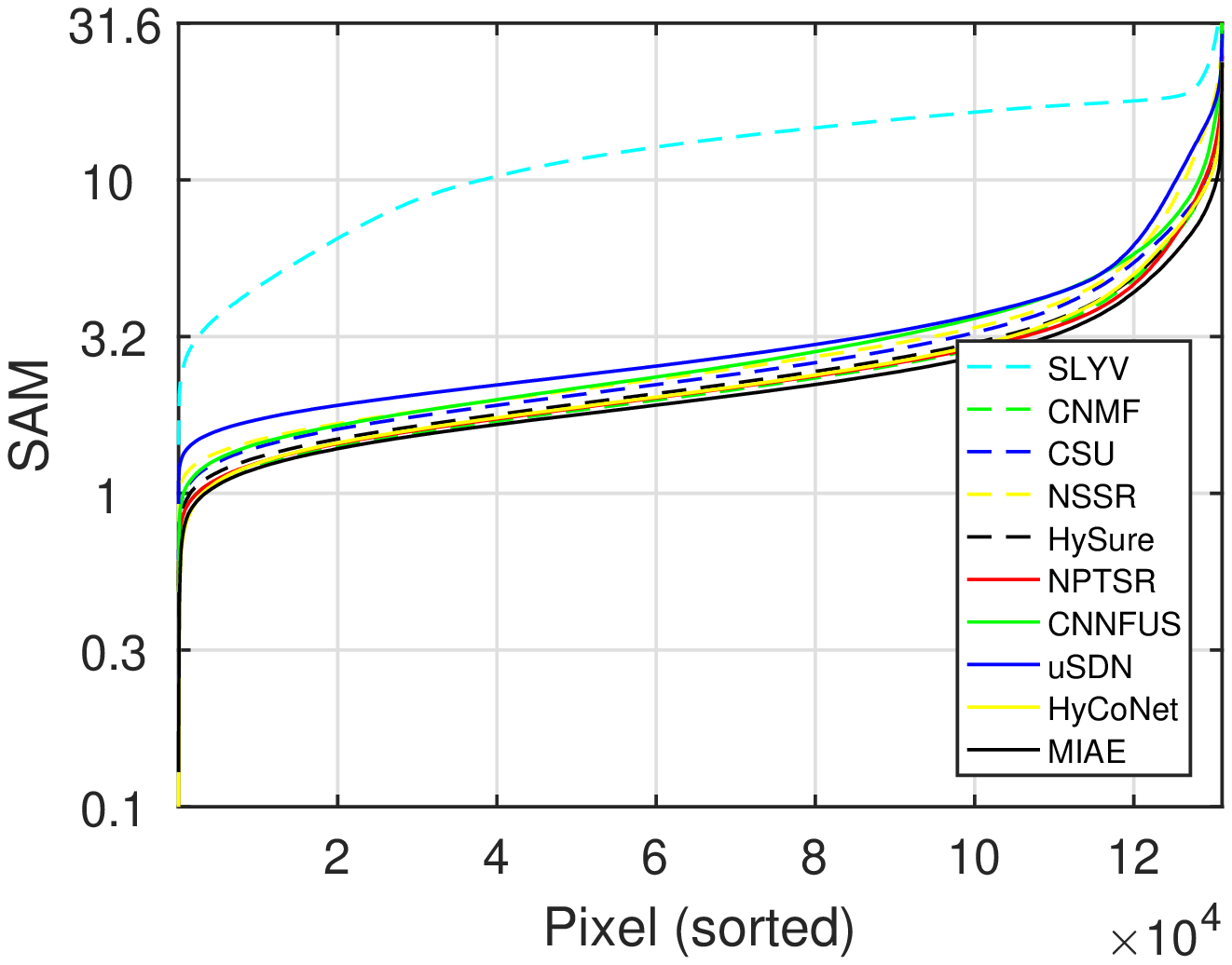}}
\subfigure[] {\includegraphics[width=0.3 \textwidth]{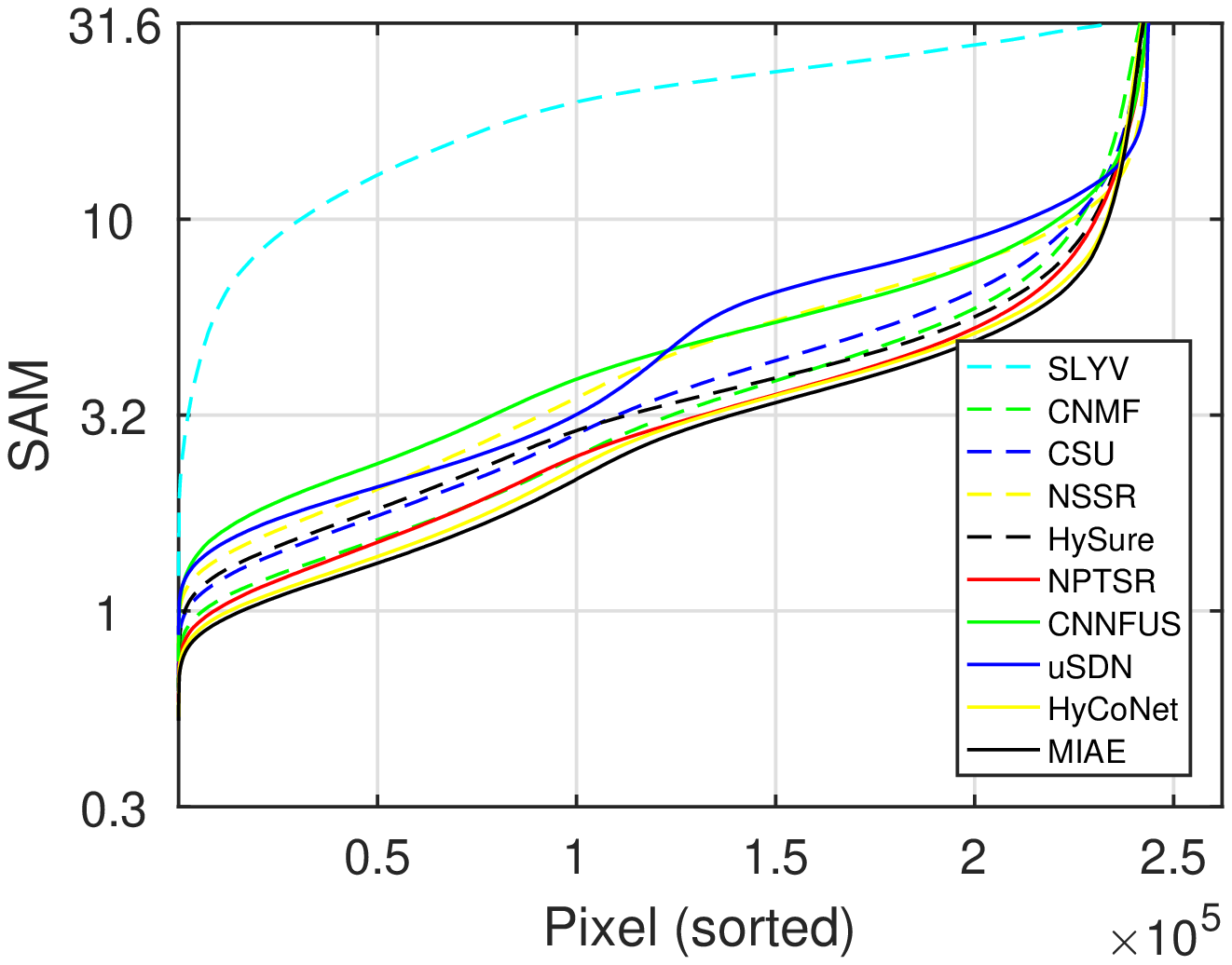}}
\subfigure[] {\includegraphics[width=0.3 \textwidth]{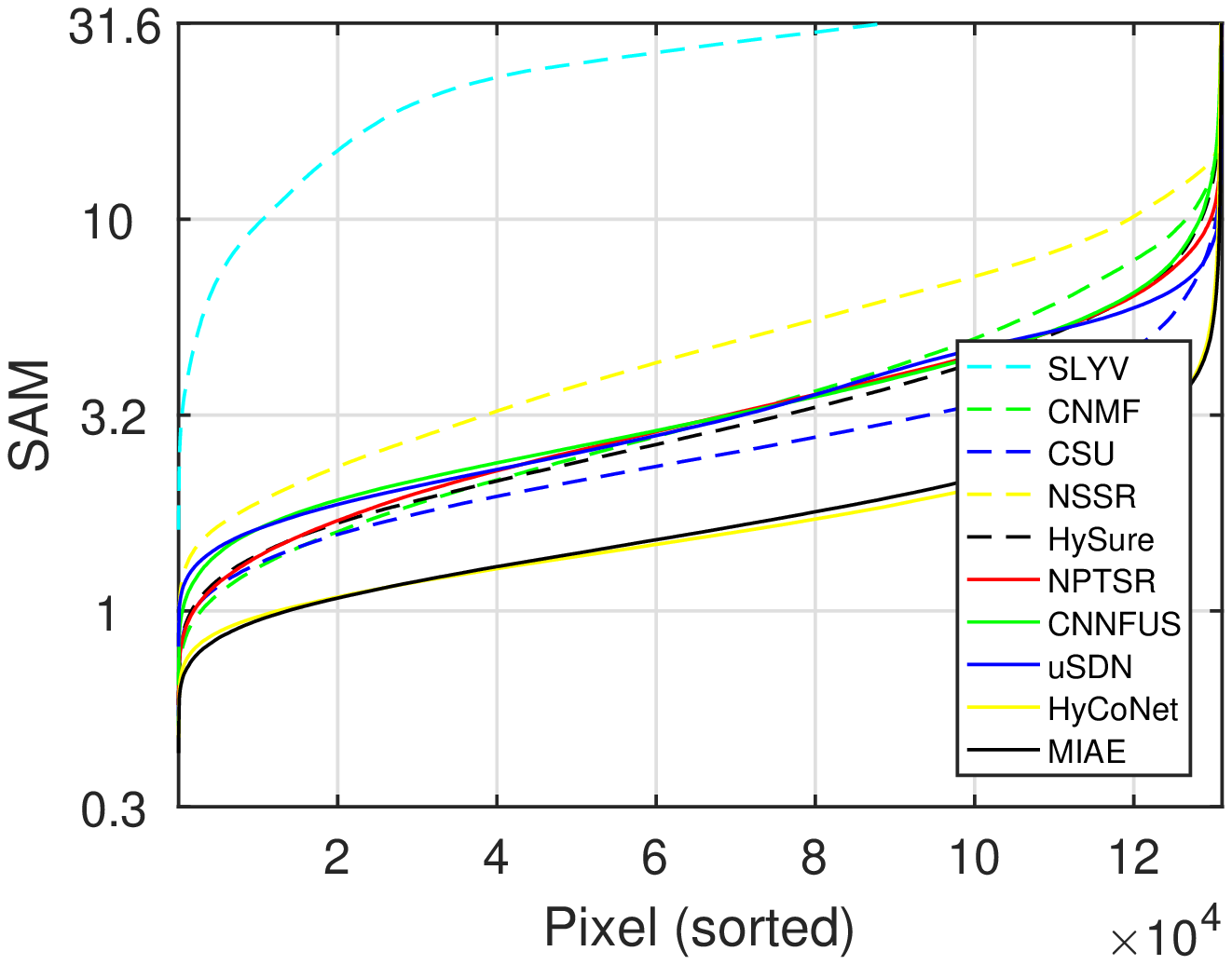}}
\caption{SAM (plotted in a $\log_{10}(\cdot)$ scale) as a function of sorted pixel. (a) PaviaU dataset. (b) KSC dataset. (c) DC dataset.}
\label{fig_sam}
\end{figure*}

Nine unsupervised methods, which can be divided into model- and deep learning-based approaches, are compared to evaluate the performance of MIAE.
The model-based approaches consist of six methods.
The first method is the baseline one (denoted by SLYV) that solves a Sylvester equation \cite{wei2015fast},
and the next five methods are coupled NMF (CNMF) \cite{yokoya2011coupled}, coupled spectral unmixing (CSU) \cite{lanaras2015hyperspectral}, NSSR \cite{dong2016hyperspectral}, HySure \cite{simoes2015convex}, and NPTSR \cite{Xu19}.
The deep learning-based approaches are CNNFUS \cite{dian2021regularizing} and three autoencoder-based methods, i.e., uSDN \cite{uSDN}, HyCoNet \cite{HyCoNet} and also MIAE.
The free parameters of the compared methods are tuned to be optimal with the test datasets, and the default training strategies are used for the deep learning-based approaches.
All of the compared methods are blind, where the estimated ${\bf B}$ and ${\bf R}$ are used.
For those methods that do not involve blind estimation procedures, ${\bf B}$ and ${\bf R}$ are estimated by the proposed blind estimation network.

The five quantitative results of the compared methods for the PaviaU dataset are shown in Table \ref{tab_pavia} with the best values marked in bold.
It can be seen that, all methods outperform the baseline method SLYV, and the proposed MIAE method gives the best quantitative results followed by NPTSR and HyCoNet. Both the model- and deep learning-based approaches can yield good results.
Fig. \ref{fig_pavia} illustrates the reference image and the fusion results of the compared methods in form of the 30th band gray and error images. Visually, it can be observed that the baseline method SLYV has severe spatial distortion and all other methods outperform it. MIAE and HyCoNet perform better than the other methods in terms of both zoomed region and error map.
Fig. \ref{fig_psnr} (a) shows the PSNR as a function of spectral band for the compared methods.
It can be seen that the proposed MIAE method performs best in almost all bands followed by HyCoNet and NPTSR.
Fig. \ref{fig_sam} (a) shows the SAM between the reference image and the fusion results for each pixel using the compared methods, with the pixels sorted by ascending error.
As illustrated in this figure, MIAE consistently outperforms the others at the pixel level.

Table \ref{tab_ksc} reports the five quality measures of the compared methods for the KSC dataset.
From this table, we can see that the baseline method SLYV performs the worst, NPTSR gives the best UIQI result, and the proposed MIAE method performs best for the remaining four quality measures. NPTSR and HyCoNet are only inferior to MIAE.
In Fig. \ref{fig_ksc}, we show the reference image and the fusion results of the compared methods in form of the 30th band gray and error images. Visually, it can be observed that the reconstructed results of HyCoNet and MIAE are better than the others, and the baseline method SLYV gives the worst images.
Fig. \ref{fig_psnr} (b) gives PSNR as a function of the spectral band for the compared methods.
MIAE, HyCoNet and NPTSR achieve high results in most bands.
Fig. \ref{fig_sam} (b) gives the SAMs for each pixel between the reference image and the fusion results, with the pixels sorted in order of ascending error.
It can be observed that MIAE is the best followed by HyCoNet and NPTSR.

Table \ref{tab_dc} summarizes the five quality measures of the compared methods for the DC dataset.
From this table, we can see that MIAE gives three best quantitative results and one second best, and HyCoNet gives two best.
The 30th band gray and error images of the reference image and the fusion results of the compared methods are given in Fig. \ref{fig_dc}. Through visual inspection, we can see that MIAE and HyCoNet exhibit good reconstructed results.
PSNR and SAM, as functions of the spectral band and by pixel sorted on error, are shown in Figs. \ref{fig_psnr} (c) and \ref{fig_sam} (c), respectively. It can be seen that MIAE outperforms the others in terms of band-level PSNR, and MIAE and HyCoNet achieve higher results than the others in terms of pixel-level SAM.

\subsubsection{Computational Efficiency}

\begin{table*}[!t]
\caption{Running/Training times (in seconds) of the compared methods}
\label{tab_time}
\centering
\begin{tabular}
{c|c|c|c|c|c|c|c|c|c|c}
\hline\hline
Method & SLYV & CNMF & CSU & NSSR & HySure & NPTSR & CNNFUS & uSDN & HyCoNet & MIAE \\
\hline\hline
PaviaU &1.6 &16.4 &156.9 &181.9 &167.6 &1339.0 &9.4 &504.5 &562.6 &186.5 \\
KSC &5.6 &36.5 &302.7 &499.0 &336.1 &4599.2 &14.1 &827.9 &970.7 &396.4 \\
DC &3.0 &16.5 &172.2 &290.6 &171.1 &2424.3 &8.8 &477.8 &577.9 &239.1 \\
\hline\hline
\end{tabular}
\end{table*}

\begin{table}[!t]
\caption{Number of trainable parameters}
\label{tab_para}
\centering
\tabcolsep = 4.0pt
\begin{tabular}
{c|c|c|c}
\hline\hline
      & uSDN & HyCoNet & MIAE \\
\hline
PaviaU & 37.9K &  377.7K & 87.8K \\
KSC & 48.9K & 389.5K & 99.5K \\
DC & 51.1K & 391.9K & 21.7K \\
\hline\hline
\end{tabular}
\end{table}

All experiments are carried out using a desktop computer with an Intel Core i9-7900X CPU, a GeForce GTX 2080Ti GPU, and 64-GB memory. The first seven methods SLYV, CNMF, CSU, NSSR, HySure, NPTSR and CNNFUS are performed using MATLAB, and the remaining three autoencoder-based methods uSDN, HyCoNet and MIAE are implemented by the PyTorch framework.
Table \ref{tab_time} summarizes the running times of the first seven methods and the training times of the autoencoder-based methods, and the number of trainable parameters for each autoencoder network is reported in Table \ref{tab_para}.
It can be seen that, MIAE takes less time to train the network than the other two autoencoder-based methods, and its trainable parameters are much less than HyCoNet. Ignoring the platform, MIAE is comparable to the model-based approaches.

\subsection{Experiment Results on Real Data}

\begin{figure*}[!t]
\centering
\subfigure[] {\includegraphics[width=\myimgsizer in]{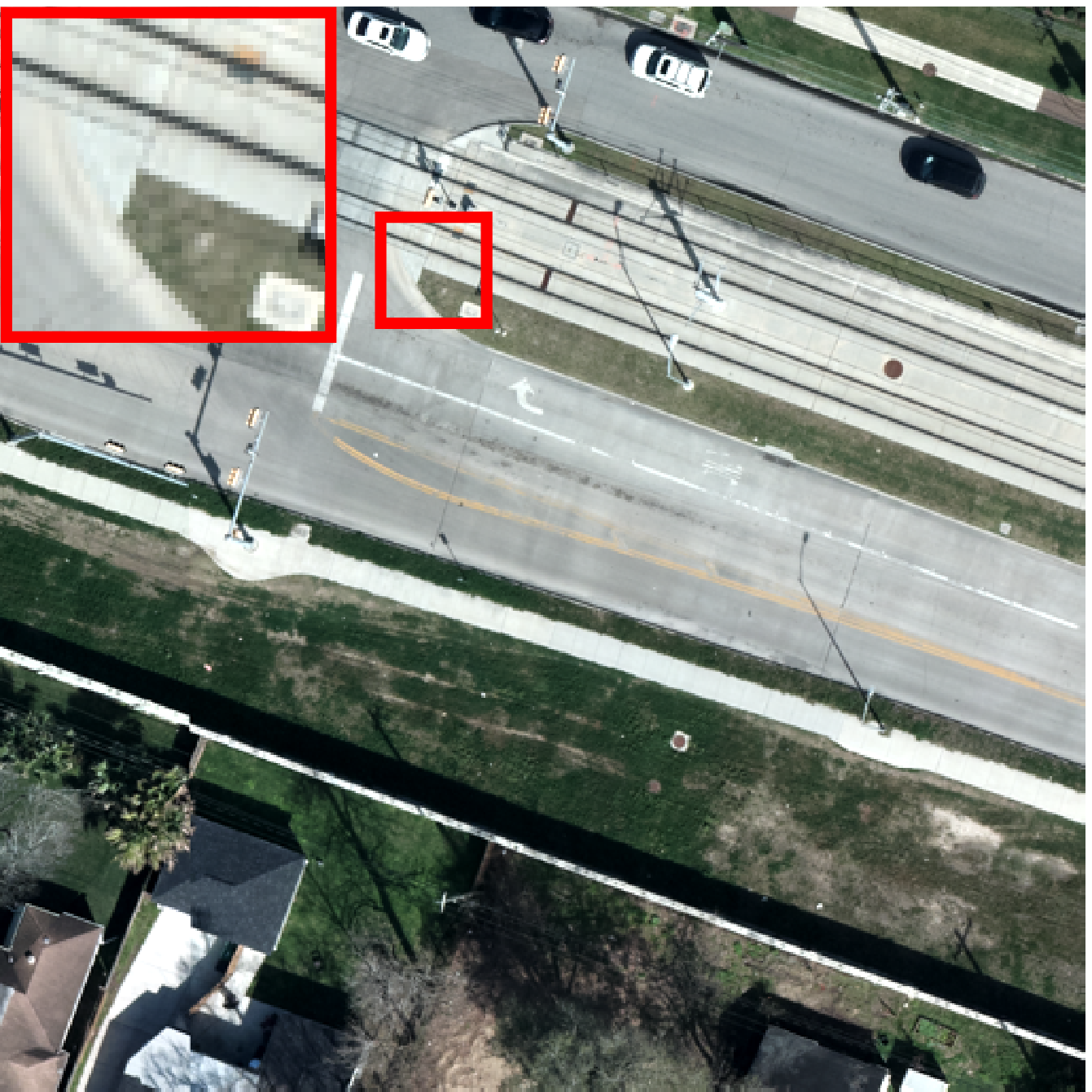}}
\subfigure[] {\includegraphics[width=\myimgsizer in]{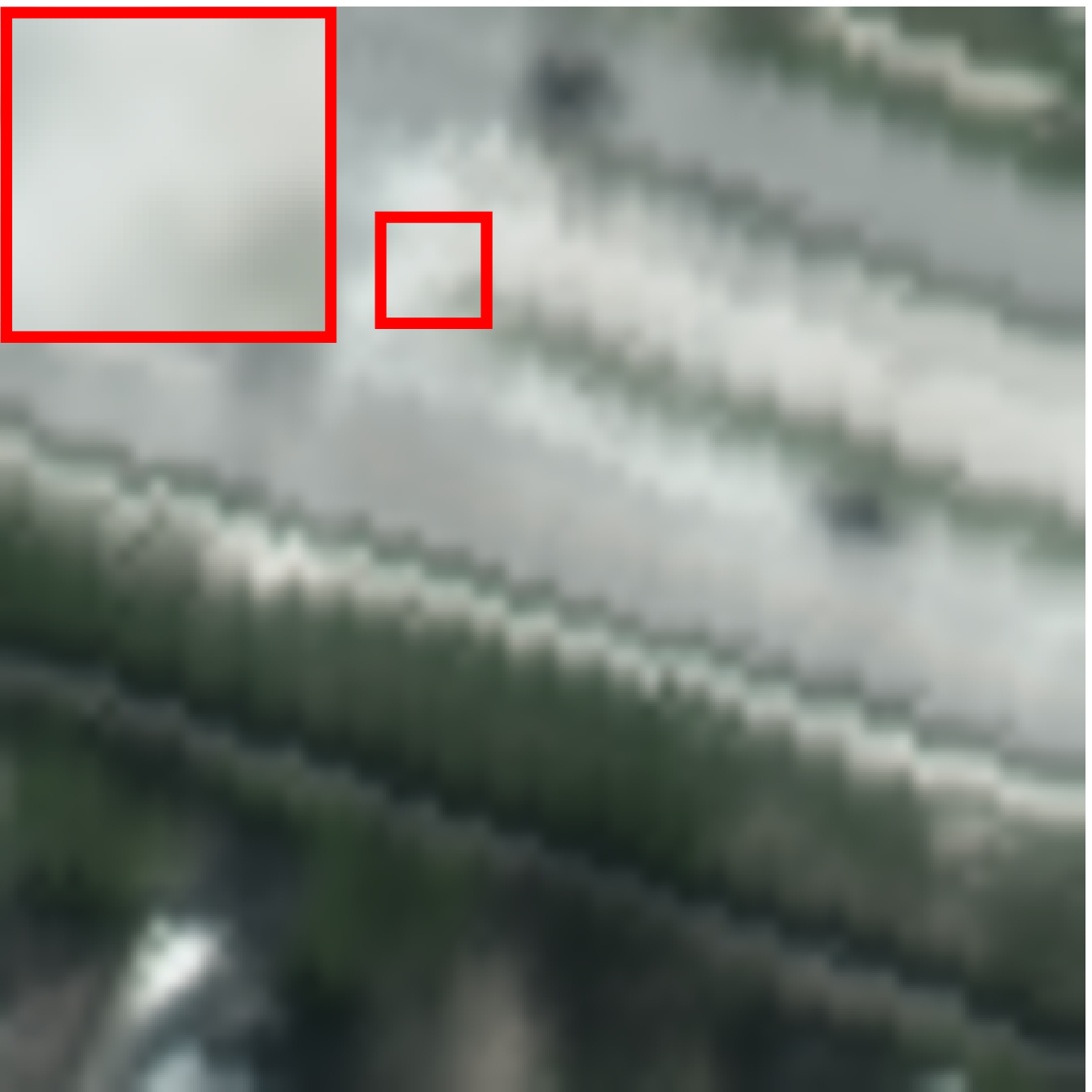}}
\subfigure[] {\includegraphics[width=\myimgsizer in]{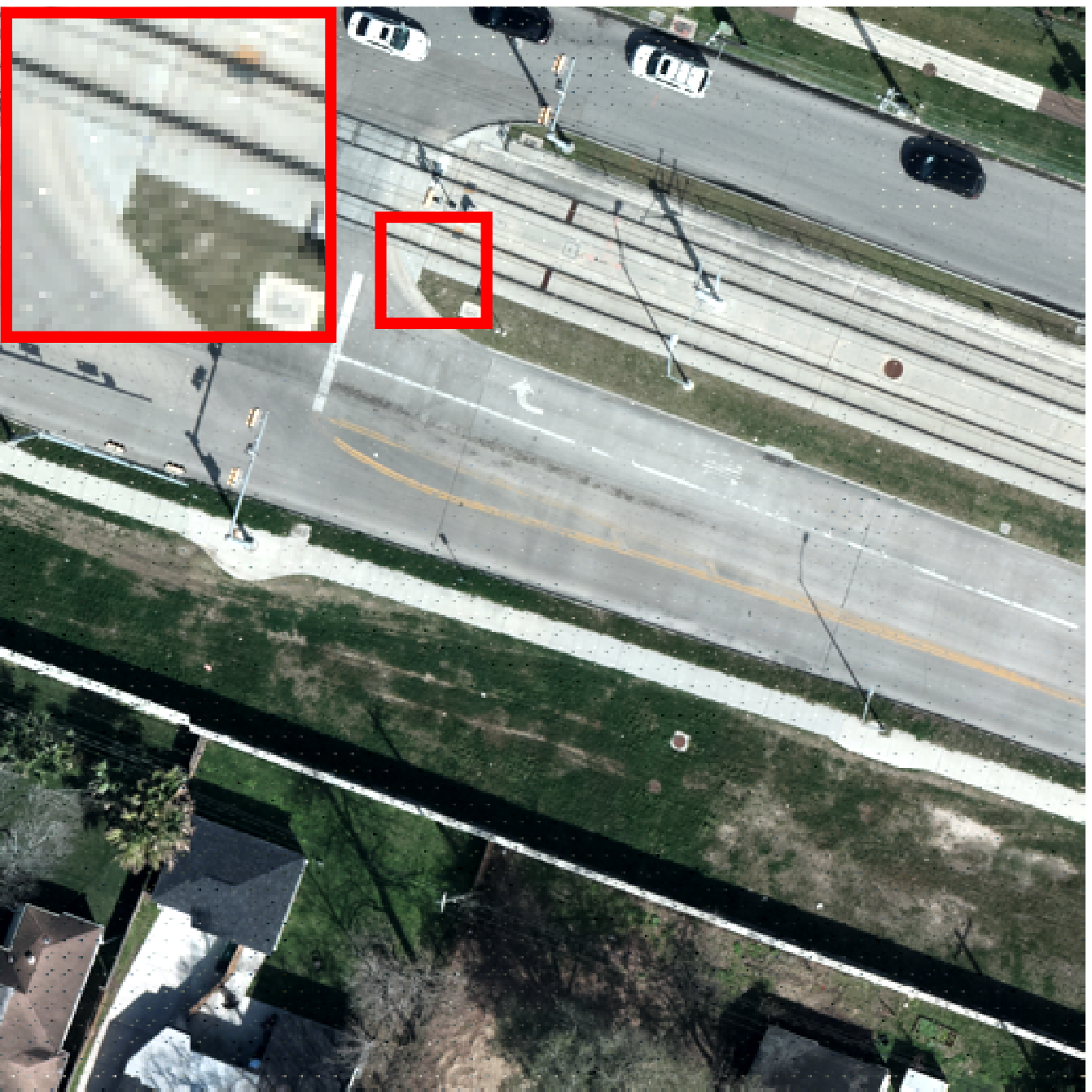}}
\subfigure[] {\includegraphics[width=\myimgsizer in]{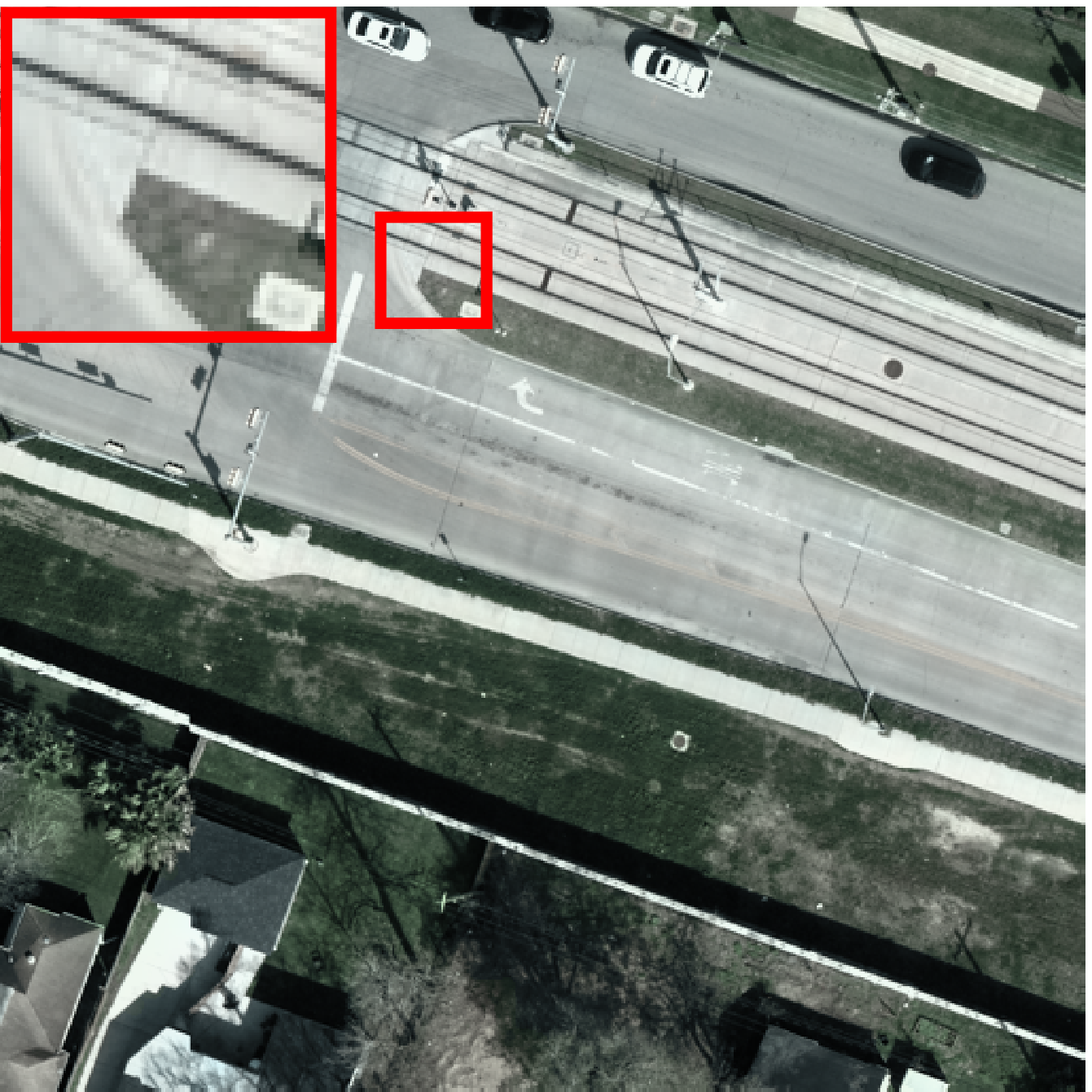}}
\subfigure[] {\includegraphics[width=\myimgsizer in]{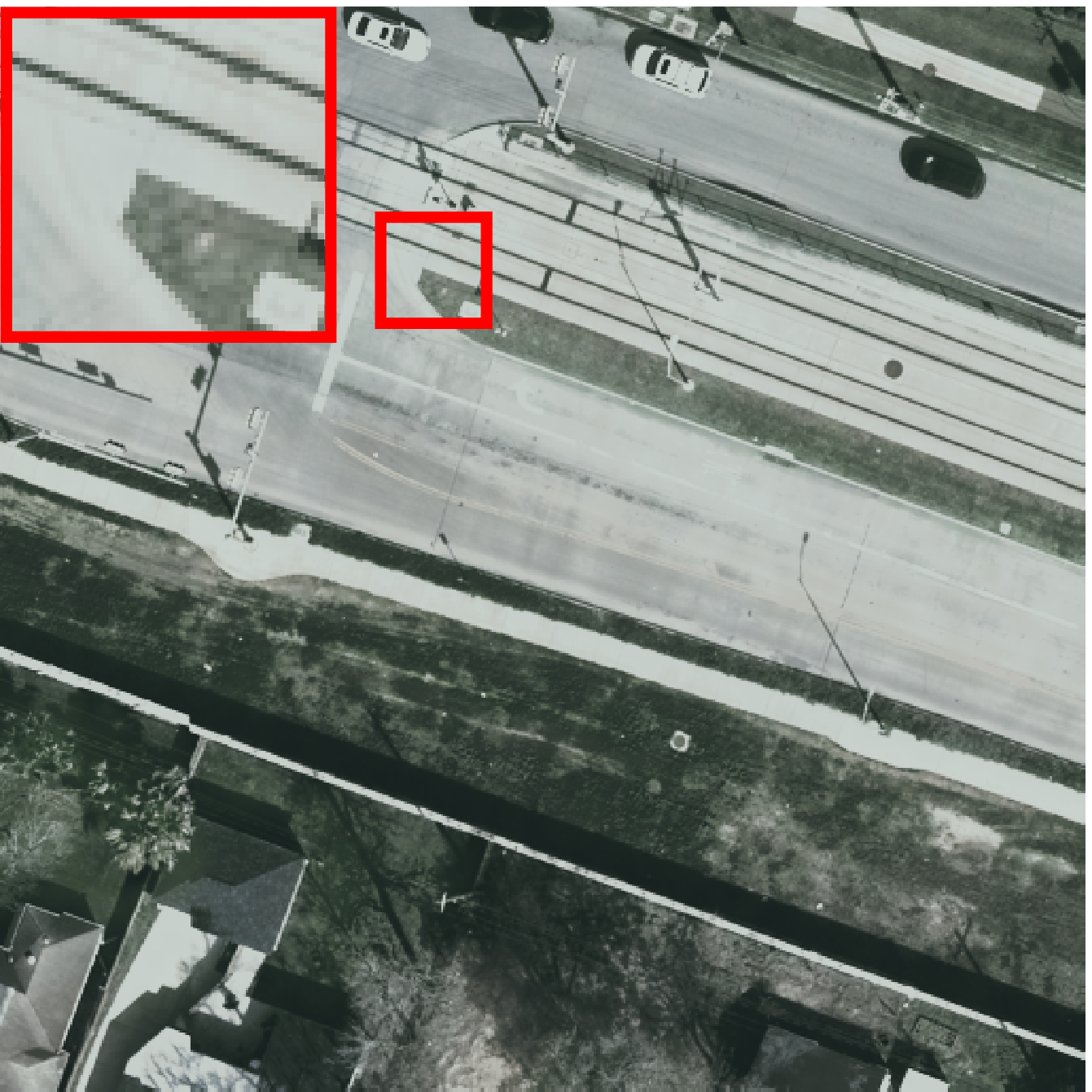}}
\subfigure[] {\includegraphics[width=\myimgsizer in]{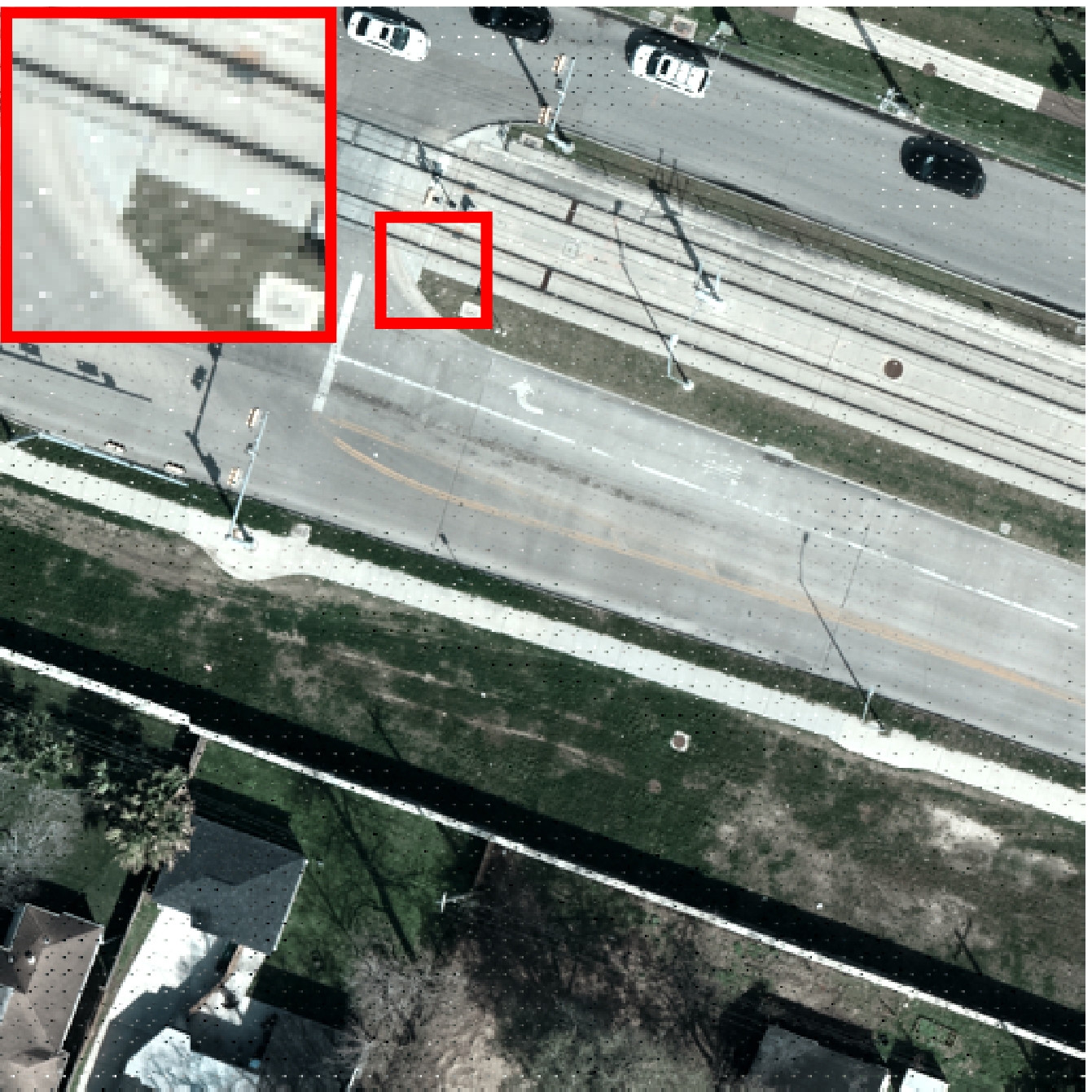}} \\
\subfigure[] {\includegraphics[width=\myimgsizer in]{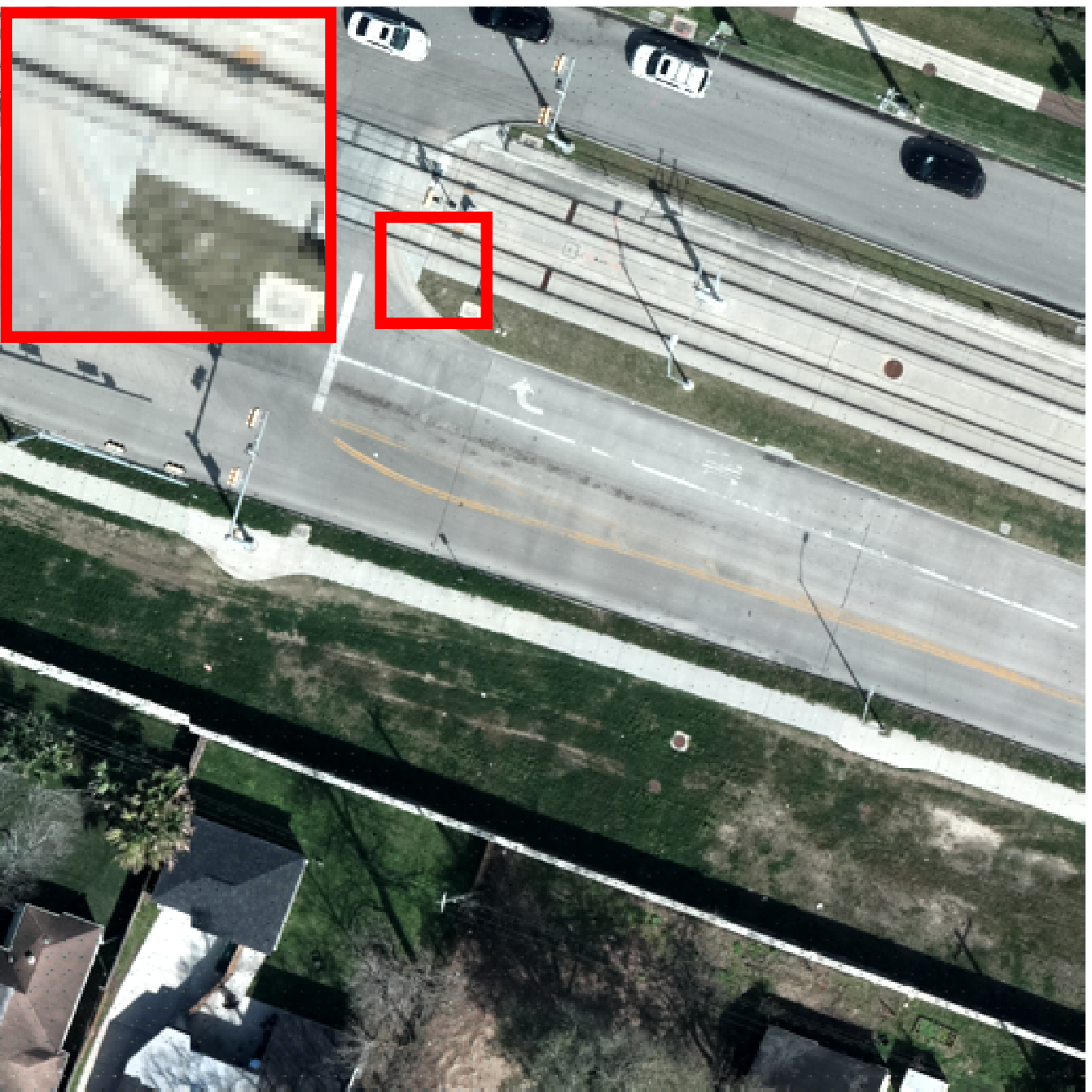}}
\subfigure[] {\includegraphics[width=\myimgsizer in]{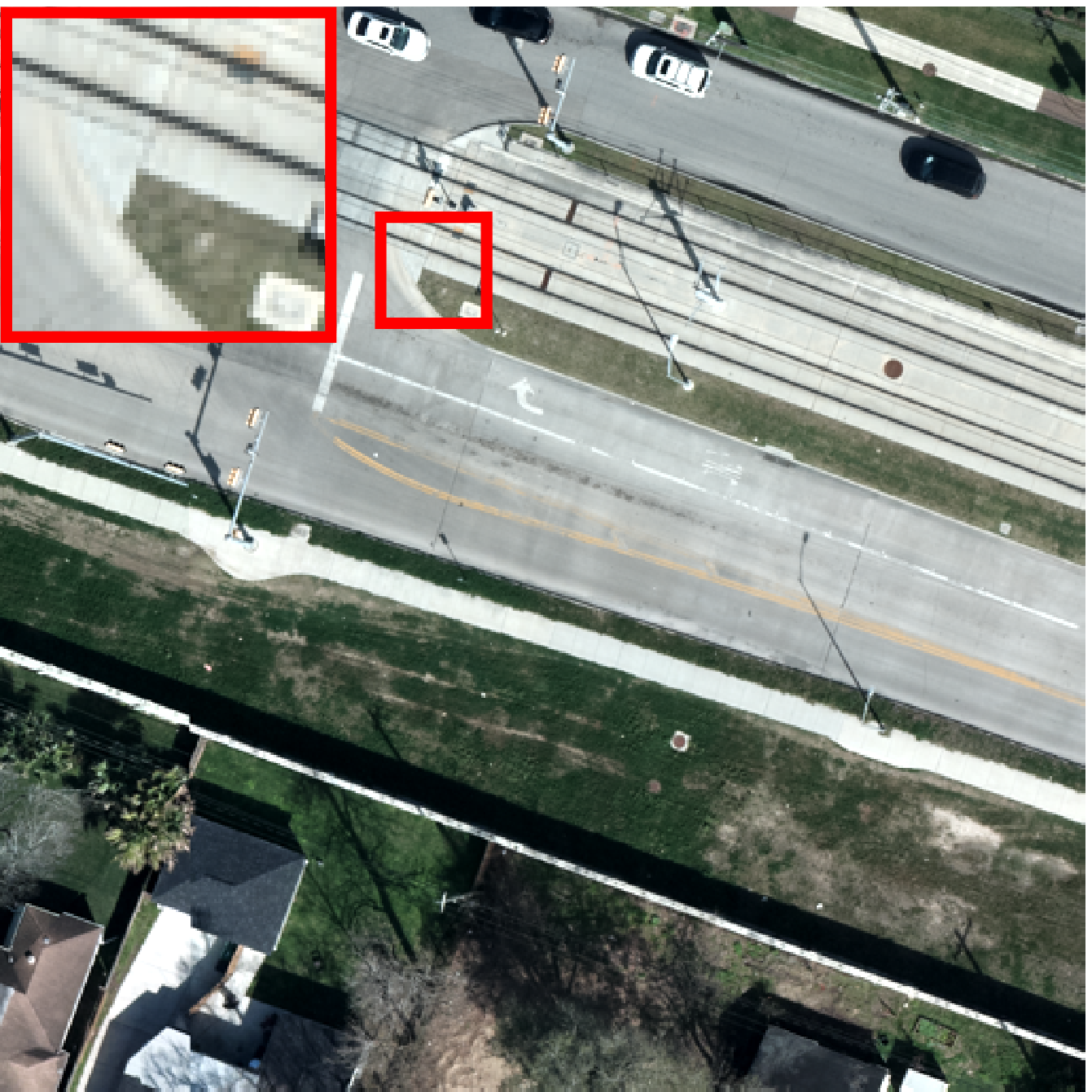}}
\subfigure[] {\includegraphics[width=\myimgsizer in]{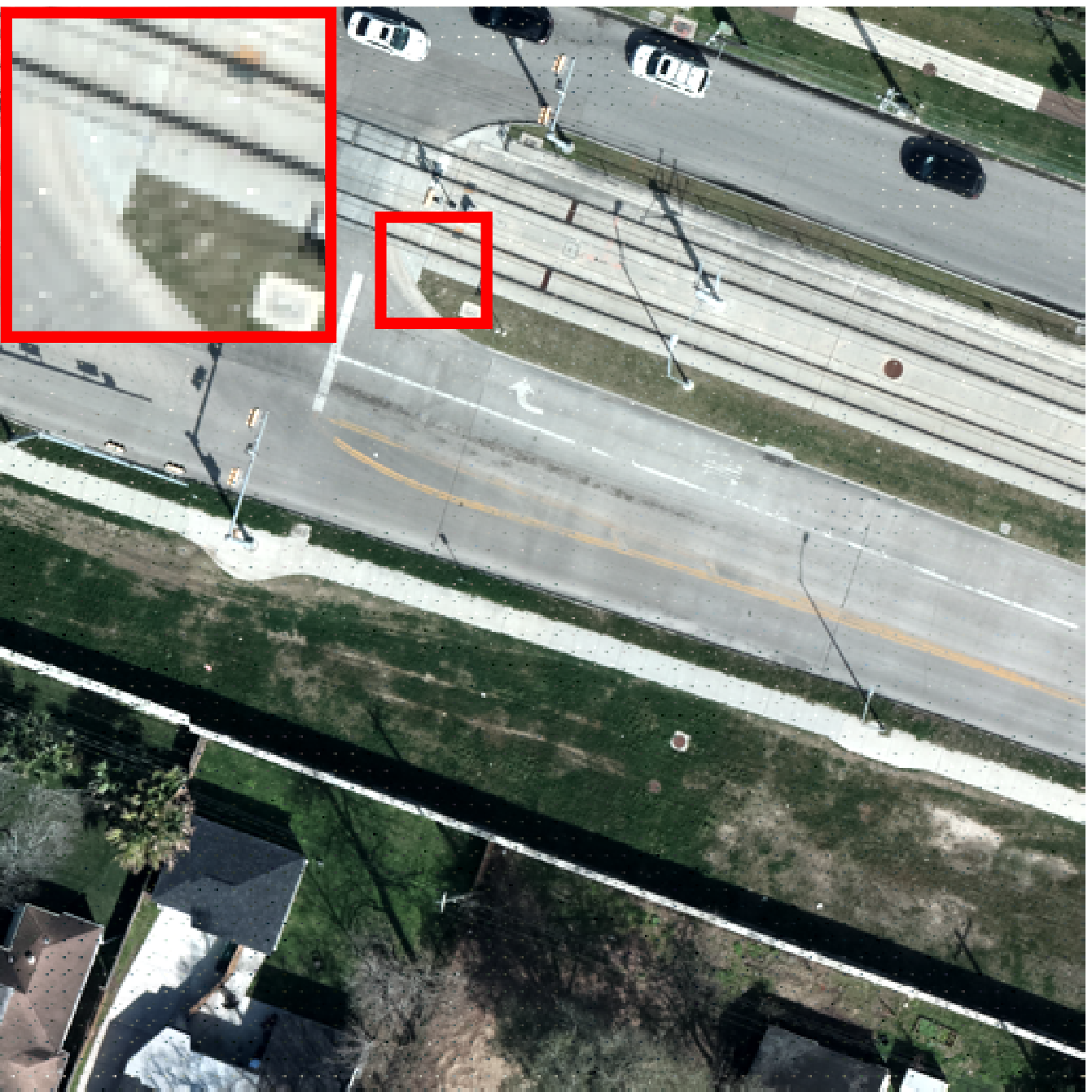}}
\subfigure[] {\includegraphics[width=\myimgsizer in]{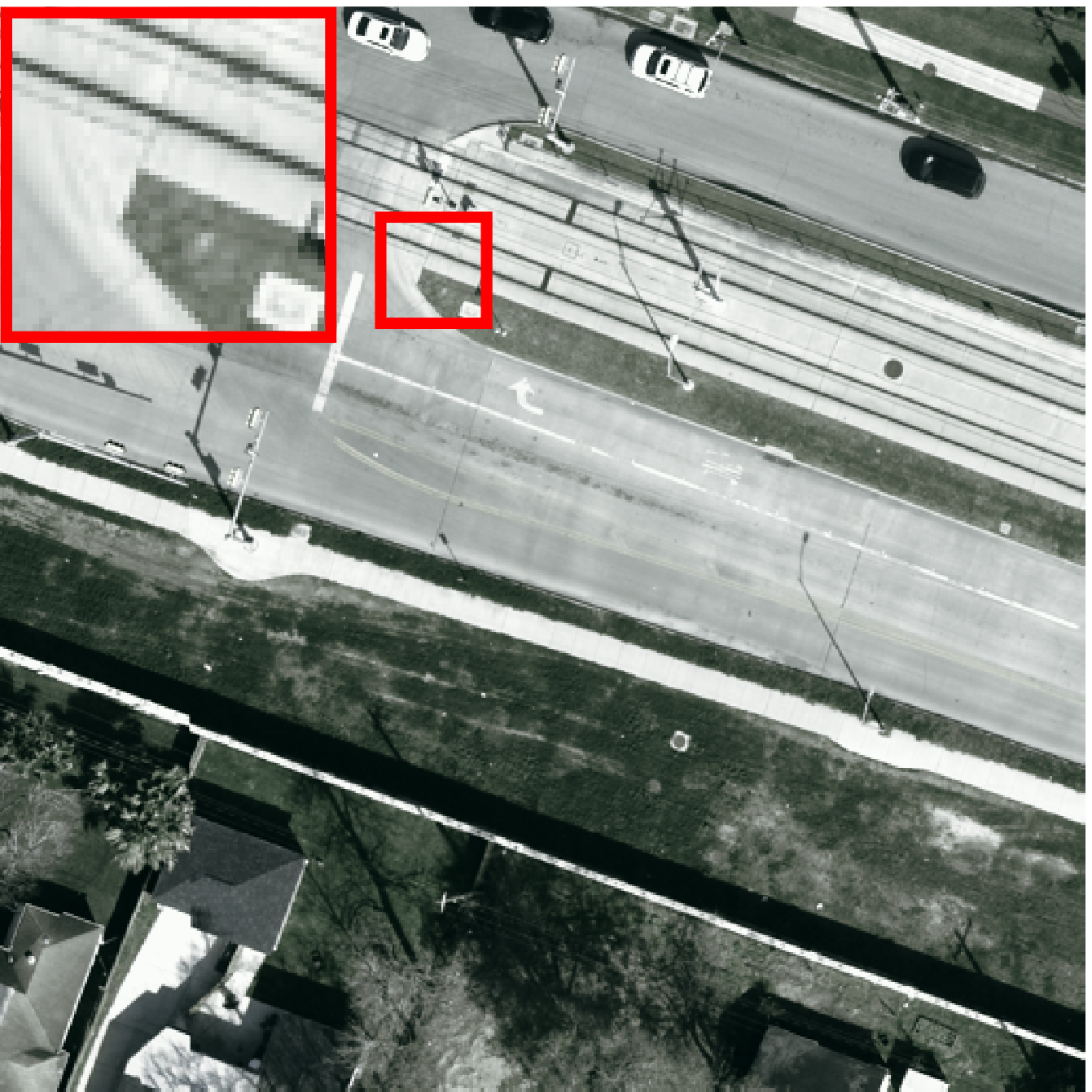}}
\subfigure[] {\includegraphics[width=\myimgsizer in]{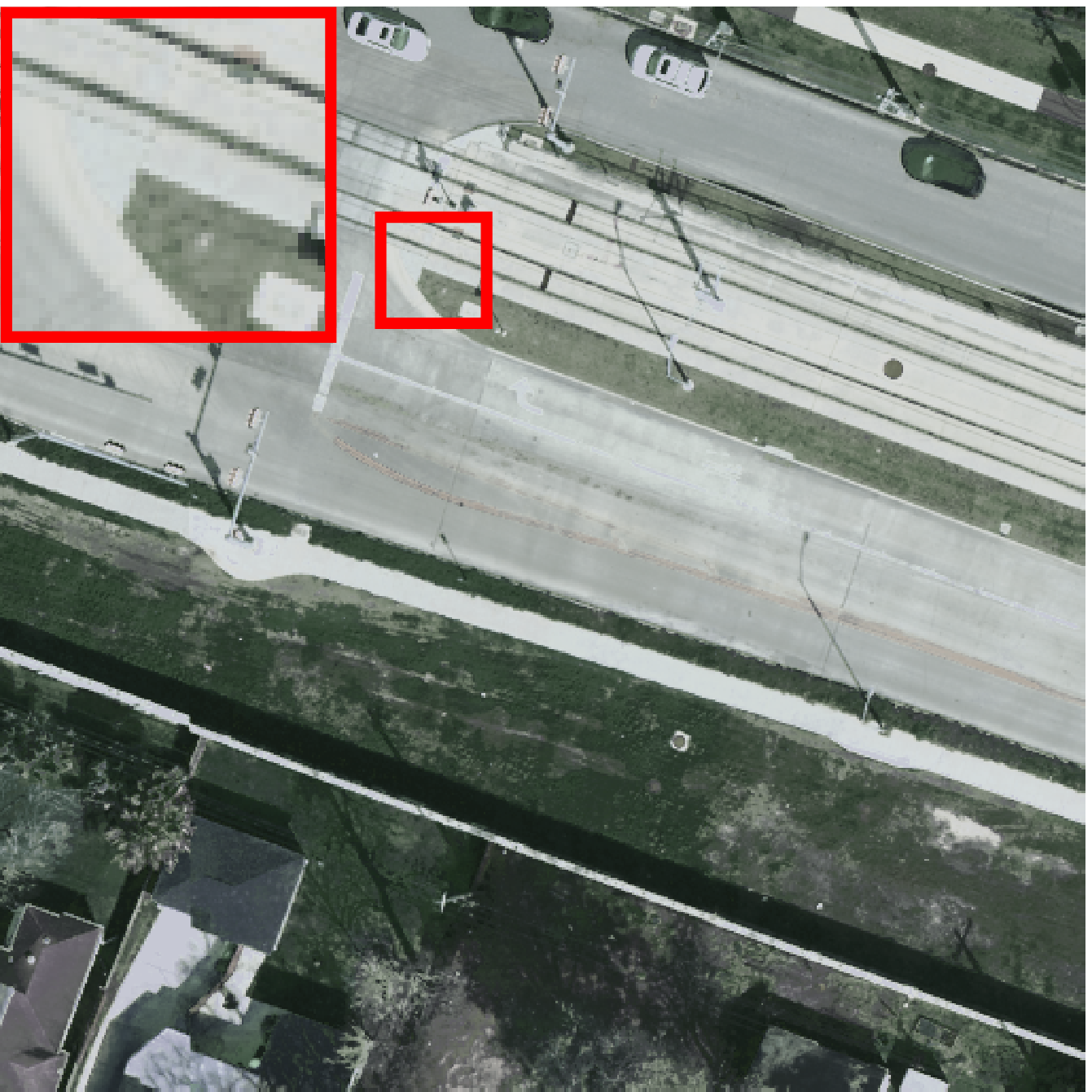}}
\subfigure[] {\includegraphics[width=\myimgsizer in]{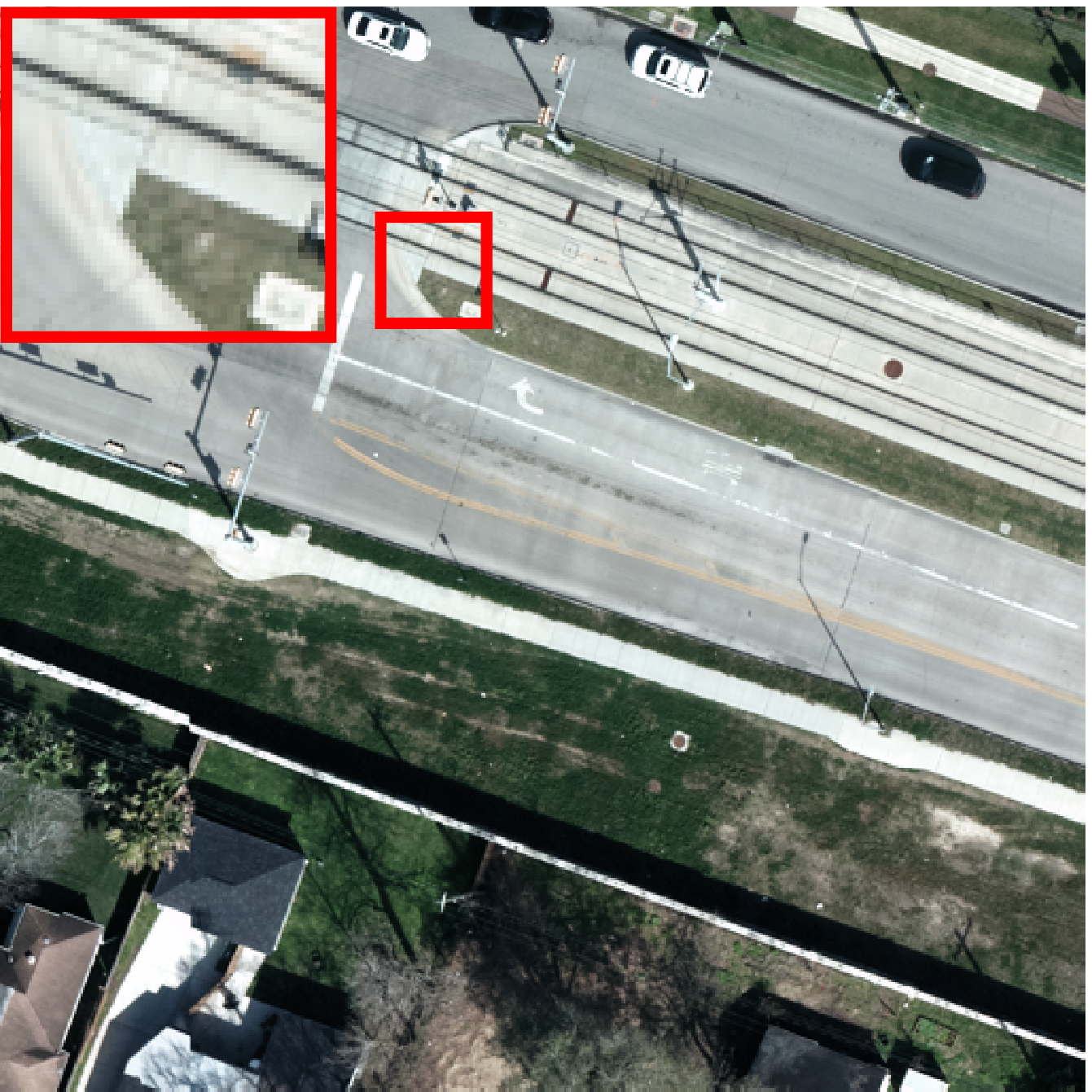}}
\caption{RGB images (with a meaningful region marked and zoomed in 3 times for easy observation) of HSI super-resolution results when applied to real dataset.
(a) HR-RGB image. (b) LR-HSI. (c) SLYV.
(d) CNMF. (e) CSU. (f) NSSR.
(g) HySure. (h) NPTSR. (i) CNNFUS. (j) uSDN. (k) HyCoNet. (l) MIAE. }
\label{fig_real}
\end{figure*}

The University of Houston (UH) dataset released by the 2018 IEEE GRSS Data Fusion Contest \cite{xu2019advanced} is used to evaluate MIAE in practical application. The original data is acquired by the National Center for Airborne Laser Mapping (NCALM), covering the UH campus and its surrounding urban areas. This experiment selects a LR-HSI and a high-resolution RGB (HR-RGB) image from this multi-modal optical remote sensing datasets. The LR-HSI collected by ITRES CASI-1500 sensor contains $4172 \times 1202$
pixels with a spatial resolution of 1 m and 48 spectral bands with a spectral range of 0.38 to 1.05 $\mu m$. The HR-RGB image collected by DiMAC ULTRALIGHT+ sensor contains $83440 \times 24040$ pixels. Take the LR-HSI as a reference, we select an
area of $64 \times 64 \times 48$ as our observation data, and downsample the corresponding area of the HR-RGB to be a $512 \times 512 \times 3$-size image. That is, the resolution ratio is $r=8$.
RGB images of the real dataset and the fusion results of the compared methods mentioned in Section \ref{sec_art} are given in Fig. \ref{fig_real}.
Visually, it can be seen that MIAE, NPTSR and HySure give the good color and brightness results, and the result of the proposed MIAE is much closer to the HR-RGB image.

\section{Conclusion}
\label{sec_con}
This paper has proposed an unsupervised MIAE network for HSI super-resolution. The proposed MIAE involves an implicit autoencoder network and the structures are concise. Firstly, inspired by that performing NMF on the target HR-HSI can facilitate the inference process of super-resolution, the implicit autoencoder network is built on the target HR-HSI by integrating its NMF model, where the two NMF parts, spectral and spatial matrices, are treated as decoder parameters and hidden outputs respectively. The autoencoder network treats each hyperspectral pixel of the target HR-HSI as an individual sample, that is, the network is trained pixel by pixel. Secondly, the `implicit' indicates the input pixel of the autoencoder network is unknown, and thus a pixel-wise fusion model taken the two observed images as inputs is presented to estimate the hidden layer vector directly. The pixel-wise fusion model is simple and effective. Specifically, the LR-HSI is resized to the same size of the target HR-HSI using bilinear interpolation, in order to feed the network pixel by pixel. To break the fixed format of model and provide more flexibility, the gradient descent algorithm is used to solve the pixel-wise fusion model, and the algorithm is reformulated and unfolded to form the encoder network. Finally, the loss function is built on the relationship between the target HR-HSI and the two observed images. With the specific pixel-wise architecture, MIAE can be treated as a kind of manifold prior-based model and can be trained patch by patch to accelerate the training process. Moreover, a blind estimation network is proposed to estimate the PSF and SRF in an unsupervised manner. MIAE has been experimentally tested using three synthetic datasets and one real dataset, and the experimental results demonstrate its effectiveness. Although the results obtained by MIAE are very encouraging, further improvements such as the application of convolutional autoencoder should be pursued in future.


\section*{Acknowledgment}
The authors would like to thank the authors of \cite{wei2015fast,yokoya2011coupled,lanaras2015hyperspectral,dong2016hyperspectral,simoes2015convex,Xu19,dian2021regularizing,uSDN,HyCoNet} for providing their codes.
They would like to thank NCALM and the Hyperspectral Image Analysis Laboratory at UH for providing the UH datasets, and the Image Analysis and Data Fusion Technical Committee of the IEEE GRSS for supporting the annual Data Fusion Contest.

\ifCLASSOPTIONcaptionsoff
  \newpage
\fi

\bibliographystyle{IEEEtran}
\bibliography{IEEEabrv,IAEF}

\begin{thebibliography}{10}
\providecommand{\url}[1]{#1}
\csname url@samestyle\endcsname
\providecommand{\newblock}{\relax}
\providecommand{\bibinfo}[2]{#2}
\providecommand{\BIBentrySTDinterwordspacing}{\spaceskip=0pt\relax}
\providecommand{\BIBentryALTinterwordstretchfactor}{4}
\providecommand{\BIBentryALTinterwordspacing}{\spaceskip=\fontdimen2\font plus
\BIBentryALTinterwordstretchfactor\fontdimen3\font minus
  \fontdimen4\font\relax}
\providecommand{\BIBforeignlanguage}[2]{{%
\expandafter\ifx\csname l@#1\endcsname\relax
\typeout{** WARNING: IEEEtran.bst: No hyphenation pattern has been}%
\typeout{** loaded for the language `#1'. Using the pattern for}%
\typeout{** the default language instead.}%
\else
\language=\csname l@#1\endcsname
\fi
#2}}
\providecommand{\BIBdecl}{\relax}
\BIBdecl

\bibitem{alparone2007comparison}
L.~Alparone, L.~Wald, J.~Chanussot, C.~Thomas, P.~Gamba, and L.~M. Bruce,
  ``Comparison of pansharpening algorithms: Outcome of the 2006 grss
  data-fusion contest,'' \emph{IEEE Transactions on Geoscience and Remote
  Sensing}, vol.~45, no.~10, pp. 3012--3021, 2007.

\bibitem{loncan2015hyperspectral}
L.~Loncan, L.~B. De~Almeida, J.~M. Bioucas-Dias, X.~Briottet, J.~Chanussot,
  N.~Dobigeon, S.~Fabre, W.~Liao, G.~A. Licciardi, M.~Simoes \emph{et~al.},
  ``Hyperspectral pansharpening: A review,'' \emph{IEEE Geoscience and remote
  sensing magazine}, vol.~3, no.~3, pp. 27--46, 2015.

\bibitem{yokoya2017hyperspectral}
N.~Yokoya, C.~Grohnfeldt, and J.~Chanussot, ``Hyperspectral and multispectral
  data fusion: A comparative review of the recent literature,'' \emph{IEEE
  Geoscience and Remote Sensing Magazine}, vol.~5, no.~2, pp. 29--56, 2017.

\bibitem{meng2019review}
X.~Meng, H.~Shen, H.~Li, L.~Zhang, and R.~Fu, ``Review of the pansharpening
  methods for remote sensing images based on the idea of meta-analysis:
  Practical discussion and challenges,'' \emph{Information Fusion}, vol.~46,
  pp. 102--113, 2019.

\bibitem{Dian2021RecentAA}
R.~Dian, S.~Li, B.~Sun, and A.~Guo, ``Recent advances and new guidelines on
  hyperspectral and multispectral image fusion,'' \emph{Information Fusion},
  vol.~69, pp. 40--51, 2021.

\bibitem{Vivone2021A}
G.~Vivone, M.~D. Mura, A.~Garzelli, R.~Restaino, G.~Scarpa, M.~Ulfarsson,
  L.~Alparone, and J.~Chanussot, ``A new benchmark based on recent advances in
  multispectral pansharpening: Revisiting pansharpening with classical and
  emerging pansharpening methods,'' \emph{IEEE Geoscience and Remote Sensing
  Magazine}, vol.~9, pp. 53--81, 2021.

\bibitem{tu2001new}
T.-M. Tu, S.-C. Su, H.-C. Shyu, and P.~S. Huang, ``A new look at ihs-like image
  fusion methods,'' \emph{Information fusion}, vol.~2, no.~3, pp. 177--186,
  2001.

\bibitem{nencini2007remote}
F.~Nencini, A.~Garzelli, S.~Baronti, and L.~Alparone, ``Remote sensing image
  fusion using the curvelet transform,'' \emph{Information fusion}, vol.~8,
  no.~2, pp. 143--156, 2007.

\bibitem{Zhang2017Multispectral}
K.~Zhang, M.~Wang, and S.~Yang, ``Multispectral and hyperspectral image fusion
  based on group spectral embedding and low-rank factorization,'' \emph{IEEE
  Transactions on Geoscience and Remote Sensing}, vol.~55, no.~3, pp.
  1363--1371, 2017.

\bibitem{Zhang2018Exploiting}
L.~Zhang, W.~Wei, C.~Bai, Y.~Gao, and Y.~Zhang, ``Exploiting clustering
  manifold structure for hyperspectral imagery super-resolution,'' \emph{IEEE
  Transactions on Image Processing}, vol.~27, no.~12, pp. 5969--5982, 2018.

\bibitem{Xu19}
Y.~Xu, Z.~Wu, J.~Chanussot, and Z.~Wei, ``Nonlocal patch tensor sparse
  representation for hyperspectral image super-resolution,'' \emph{IEEE
  Transactions on Image Processing}, vol.~28, no.~6, pp. 3034--3047, 2019.

\bibitem{Xue2021Spatial}
J.~Xue, Y.~Zhao, Y.~Bu, W.~Liao, J.~Chan, and W.~Philips, ``Spatial-spectral
  structured sparse low-rank representation for hyperspectral image
  super-resolution,'' \emph{IEEE Transactions on Image Processing}, vol.~30,
  pp. 3084--3097, 2021.

\bibitem{simoes2015convex}
M.~Sim{\~o}es, J.~Bioucas-Dias, L.~B. Almeida, and J.~Chanussot, ``A convex
  formulation for hyperspectral image superresolution via subspace-based
  regularization,'' \emph{IEEE Transactions on Geoscience and Remote Sensing},
  vol.~53, no.~6, pp. 3373--3388, 2015.

\bibitem{wei2015hyperspectral}
Q.~Wei, J.~Bioucas-Dias, N.~Dobigeon, and J.-Y. Tourneret, ``Hyperspectral and
  multispectral image fusion based on a sparse representation,'' \emph{IEEE
  Transactions on Geoscience and Remote Sensing}, vol.~53, no.~7, pp.
  3658--3668, 2015.

\bibitem{dian2019hyperspectral}
R.~Dian and S.~Li, ``Hyperspectral image super-resolution via subspace-based
  low tensor multi-rank regularization,'' \emph{IEEE Transactions on Image
  Processing}, vol.~28, no.~10, pp. 5135--5146, 2019.

\bibitem{liu2020truncated}
J.~Liu, Z.~Wu, L.~Xiao, J.~Sun, and H.~Yan, ``A truncated matrix decomposition
  for hyperspectral image super-resolution,'' \emph{IEEE Transactions on Image
  Processing}, vol.~29, pp. 8028--8042, 2020.

\bibitem{yokoya2011coupled}
N.~Yokoya, T.~Yairi, and A.~Iwasaki, ``Coupled nonnegative matrix factorization
  unmixing for hyperspectral and multispectral data fusion,'' \emph{IEEE
  Transactions on Geoscience and Remote Sensing}, vol.~50, no.~2, pp. 528--537,
  2012.

\bibitem{lanaras2015hyperspectral}
C.~Lanaras, E.~Baltsavias, and K.~Schindler, ``Hyperspectral super-resolution
  by coupled spectral unmixing,'' in \emph{IEEE International Conference on
  Computer Vision}, 2015, pp. 3586--3594.

\bibitem{lin2018convex}
C.-H. Lin, F.~Ma, C.-Y. Chi, and C.-H. Hsieh, ``A convex optimization-based
  coupled nonnegative matrix factorization algorithm for hyperspectral and
  multispectral data fusion,'' \emph{IEEE Transactions on Geoscience and Remote
  Sensing}, vol.~56, no.~3, pp. 1652--1667, 2018.

\bibitem{Wu2020Hyperspectral}
R.~Wu, W.-K. Ma, X.~Fu, and Q.~Li, ``Hyperspectral super-resolution via
  global–local low-rank matrix estimation,'' \emph{IEEE Transactions on
  Geoscience and Remote Sensing}, vol.~58, no.~10, pp. 7125--7140, 2020.

\bibitem{dong2016hyperspectral}
W.~Dong, F.~Fu, G.~Shi, X.~Cao, J.~Wu, G.~Li, and X.~Li, ``Hyperspectral image
  super-resolution via non-negative structured sparse representation,''
  \emph{IEEE Transactions on Image Processing}, vol.~25, no.~5, pp. 2337--2352,
  2016.

\bibitem{veganzones2016hyperspectral}
M.~A. Veganzones, M.~Simoes, G.~Licciardi, N.~Yokoya, J.~M. Bioucas-Dias, and
  J.~Chanussot, ``Hyperspectral super-resolution of locally low rank images
  from complementary multisource data,'' \emph{IEEE Transactions on Image
  Processing}, vol.~25, no.~1, pp. 274--288, 2016.

\bibitem{han2018self}
X.-H. Han, B.~Shi, and Y.~Zheng, ``Self-similarity constrained sparse
  representation for hyperspectral image super-resolution,'' \emph{IEEE
  Transactions on Image Processing}, vol.~27, no.~11, pp. 5625--5637, 2018.

\bibitem{yi2018hyperspectral}
C.~Yi, Y.-Q. Zhao, and J.~C.-W. Chan, ``Hyperspectral image super-resolution
  based on spatial and spectral correlation fusion,'' \emph{IEEE Transactions
  on Geoscience and Remote Sensing}, vol.~56, no.~7, pp. 4165--4177, 2018.

\bibitem{han2020hyperspectral}
X.~Han, J.~Yu, J.-H. Xue, and W.~Sun, ``Hyperspectral and multispectral image
  fusion using optimized twin dictionaries,'' \emph{IEEE Transactions on Image
  Processing}, vol.~29, pp. 4709--4720, 2020.

\bibitem{li2018fusing}
S.~Li, R.~Dian, L.~Fang, and J.~M. Bioucas-Dias, ``Fusing hyperspectral and
  multispectral images via coupled sparse tensor factorization,'' \emph{IEEE
  Transactions on Image Processing}, vol.~27, no.~8, pp. 4118--4130, 2018.

\bibitem{kanatsoulis2018hyperspectral}
C.~I. Kanatsoulis, X.~Fu, N.~D. Sidiropoulos, and W.-K. Ma, ``Hyperspectral
  super-resolution: A coupled tensor factorization approach,'' \emph{IEEE
  Transactions on Signal Processing}, vol.~66, no.~24, pp. 6503--6517, 2018.

\bibitem{Xu2020Hyperspectral}
Y.~Xu, Z.~Wu, J.~Chanussot, and Z.~Wei, ``Hyperspectral images super-resolution
  via learning high-order coupled tensor ring representation,'' \emph{IEEE
  transactions on neural networks and learning systems}, vol.~31, no.~11, pp.
  4747--4760, 2020.

\bibitem{Chen2021Hyperspectral}
Y.~Chen, J.~Zeng, W.~He, X.-L. Zhao, and T.-Z. Huang, ``Hyperspectral and
  multispectral image fusion using factor smoothed tensor ring decomposition,''
  \emph{IEEE Transactions on Geoscience and Remote Sensing}, pp. 1--17, 2021.

\bibitem{Bungert2018Blind}
L.~Bungert, D.~A. Coomes, M.~J. Ehrhardt, J.~Rasch, R.~Reisenhofer, and C.-B.
  Schönlieb, ``Blind image fusion for hyperspectral imaging with the
  directional total variation,'' \emph{Inverse Problems}, vol.~34, no.~4, p.
  044003, 2018.

\bibitem{palsson2017multispectral}
F.~Palsson, J.~R. Sveinsson, and M.~O. Ulfarsson, ``Multispectral and
  hyperspectral image fusion using a 3-d-convolutional neural network,''
  \emph{IEEE Geoscience and Remote Sensing Letters}, vol.~14, no.~5, pp.
  639--643, 2017.

\bibitem{scarpa2018target}
G.~Scarpa, S.~Vitale, and D.~Cozzolino, ``Target-adaptive cnn-based
  pansharpening,'' \emph{IEEE Transactions on Geoscience and Remote Sensing},
  vol.~56, no.~9, pp. 5443--5457, 2018.

\bibitem{yuan2018a}
Q.~Yuan, Y.~Wei, X.~Meng, H.~Shen, and L.~Zhang, ``A multiscale and multidepth
  convolutional neural network for remote sensing imagery pan-sharpening,''
  \emph{IEEE Journal of Selected Topics in Applied Earth Observations and
  Remote Sensing}, vol.~11, no.~3, pp. 978--989, 2018.

\bibitem{zhang2019pan}
Y.~Zhang, C.~Liu, M.~Sun, and Y.~Ou, ``Pan-sharpening using an efficient
  bidirectional pyramid network,'' \emph{IEEE Transactions on Geoscience and
  Remote Sensing}, vol.~57, no.~8, pp. 5549--5563, 2019.

\bibitem{zheng2020hyperspectral}
Y.~Zheng, J.~Li, Y.~Li, J.~Guo, X.~Wu, and J.~Chanussot, ``Hyperspectral
  pansharpening using deep prior and dual attention residual network,''
  \emph{IEEE Transactions on Geoscience and Remote Sensing}, vol.~58, no.~11,
  pp. 8059--8076, 2020.

\bibitem{Jiang2020LearningSP}
J.~Jiang, H.~Sun, X.~Liu, and J.~Ma, ``Learning spatial-spectral prior for
  super-resolution of hyperspectral imagery,'' \emph{IEEE Transactions on
  Computational Imaging}, vol.~6, pp. 1082--1096, 2020.

\bibitem{Zhang2021SSR}
X.~Zhang, W.~Huang, Q.~Wang, and X.~Li, ``Ssr-net: Spatial-spectral
  reconstruction network for hyperspectral and multispectral image fusion,''
  \emph{IEEE Transactions on Geoscience and Remote Sensing}, vol.~59, pp.
  5953--5965, 2021.

\bibitem{Dong2021Remote}
X.~Dong, X.~Sun, X.~Jia, Z.~Xi, L.~Gao, and B.~Zhang, ``Remote sensing image
  super-resolution using novel dense-sampling networks,'' \emph{IEEE
  Transactions on Geoscience and Remote Sensing}, vol.~59, no.~2, pp.
  1618--1633, 2021.

\bibitem{Liu2021A}
D.~Liu, J.~Li, and Q.~Yuan, ``A spectral grouping and attention-driven residual
  dense network for hyperspectral image super-resolution,'' \emph{IEEE
  Transactions on Geoscience and Remote Sensing}, vol.~59, no.~9, pp.
  7711--7725, 2021.

\bibitem{Li2020Hyperspectral}
J.~Li, R.~Cui, B.~Li, R.~Song, Y.~Li, Y.~Dai, and Q.~Du, ``Hyperspectral image
  super-resolution by band attention through adversarial learning,'' \emph{IEEE
  Transactions on Geoscience and Remote Sensing}, vol.~58, no.~6, pp.
  4304--4318, 2020.

\bibitem{Dong2021Generative}
W.~Dong, S.~Hou, S.~Xiao, J.~Qu, Q.~Du, and Y.~Li, ``Generative
  dual-adversarial network with spectral fidelity and spatial enhancement for
  hyperspectral pansharpening,'' \emph{IEEE Transactions on Neural Networks and
  Learning Systems}, pp. 1--15, 2021.

\bibitem{yang2017pannet}
J.~Yang, X.~Fu, Y.~Hu, Y.~Huang, X.~Ding, and J.~Paisley, ``Pannet: A deep
  network architecture for pan-sharpening,'' in \emph{IEEE International
  Conference on Computer Vision}, 2017, pp. 5449--5457.

\bibitem{xie2019hyperspectral}
W.~Xie, J.~Lei, Y.~Cui, Y.~Li, and Q.~Du, ``Hyperspectral pansharpening with
  deep priors,'' \emph{IEEE Transactions on Neural Networks and Learning
  Systems}, vol.~31, no.~5, pp. 1529--1543, 2019.

\bibitem{fu2021deep}
X.~Fu, W.~Wang, Y.~Huang, X.~Ding, and J.~Paisley, ``Deep multiscale detail
  networks for multiband spectral image sharpening,'' \emph{IEEE Transactions
  on Neural Networks and Learning Systems}, vol.~32, no.~5, pp. 2090--2104,
  2021.

\bibitem{Deng2021Detail}
L.-J. Deng, G.~Vivone, C.~Jin, and J.~Chanussot, ``Detail injection-based deep
  convolutional neural networks for pansharpening,'' \emph{IEEE Transactions on
  Geoscience and Remote Sensing}, vol.~59, no.~8, pp. 6995--7010, 2021.

\bibitem{dian2018deep}
R.~Dian, S.~Li, A.~Guo, and L.~Fang, ``Deep hyperspectral image sharpening,''
  \emph{IEEE transactions on neural networks and learning systems}, vol.~29,
  no.~11, pp. 5345--5355, 2018.

\bibitem{shen2019spatial}
H.~Shen, M.~Jiang, J.~Li, Q.~Yuan, Y.~Wei, and L.~Zhang, ``Spatial--spectral
  fusion by combining deep learning and variational model,'' \emph{IEEE
  Transactions on Geoscience and Remote Sensing}, vol.~57, no.~8, pp.
  6169--6181, 2019.

\bibitem{shen2020a}
D.~Shen, J.~Liu, Z.~Xiao, J.~Yang, and L.~Xiao, ``A twice optimizing net with
  matrix decomposition for hyperspectral and multispectral image fusion,''
  \emph{IEEE Journal of Selected Topics in Applied Earth Observations and
  Remote Sensing}, vol.~13, pp. 4095--4110, 2020.

\bibitem{xie2019multispectral}
Q.~Xie, M.~Zhou, Q.~Zhao, D.~Meng, W.~Zuo, and Z.~Xu, ``Multispectral and
  hyperspectral image fusion by ms/hs fusion net,'' in \emph{Proceedings of the
  IEEE Conference on Computer Vision and Pattern Recognition}, 2019, pp.
  1585--1594.

\bibitem{wei2020deep}
W.~Wei, J.~Nie, Y.~Li, L.~Zhang, and Y.~Zhang, ``Deep recursive network for
  hyperspectral image super-resolution,'' \emph{IEEE Transactions on
  Computational Imaging}, vol.~6, pp. 1233--1244, 2020.

\bibitem{Dong2021Model}
W.~Dong, C.~Zhou, F.~Wu, J.~Wu, G.~Shi, and X.~Li, ``Model-guided deep
  hyperspectral image super-resolution,'' \emph{IEEE Trans Image Process},
  vol.~30, pp. 5754--5768, 2021.

\bibitem{dian2021regularizing}
R.~Dian, S.~Li, and X.~Kang, ``Regularizing hyperspectral and multispectral
  image fusion by cnn denoiser,'' \emph{IEEE Transactions on Neural Networks
  and Learning Systems}, vol.~32, pp. 1124--1135, 2021.

\bibitem{zhang2021deep}
L.~Zhang, J.~Nie, W.~Wei, Y.~Li, and Y.~Zhang, ``Deep blind hyperspectral image
  super-resolution,'' \emph{IEEE Transactions on Neural Networks and Learning
  Systems}, vol.~32, pp. 2388--2400, 2021.

\bibitem{uSDN}
Y.~Qu, H.~Qi, and C.~Kwan, ``Unsupervised sparse dirichlet-net for
  hyperspectral image super-resolution,'' in \emph{2018 IEEE/CVF Conference on
  Computer Vision and Pattern Recognition}, 2018, pp. 2511--2520.

\bibitem{FusionNet}
Z.~Wang, B.~Chen, R.~Lu, H.~Zhang, and P.~K. Varshney, ``Fusionnet: An
  unsupervised convolutional variational network for hyperspectral and
  multispectral image fusion,'' \emph{IEEE Transactions on Image Processing},
  vol.~29, pp. 7565--7577, 2020.

\bibitem{Yao2020Cross}
J.~Yao, D.~Hong, J.~Chanussot, D.~Meng, X.~Zhu, and Z.~Xu, ``Cross-attention in
  coupled unmixing nets for unsupervised hyperspectral super-resolution,'' in
  \emph{European Conference on Computer Vision}, 2020, pp. 208--224.

\bibitem{GDD}
T.~Uezato, D.~Hong, N.~Yokoya, and W.~He, ``Guided deep decoder: Unsupervised
  image pair fusion,'' in \emph{European Conference on Computer Vision}, 2020,
  pp. 87--102.

\bibitem{HyCoNet}
K.~Zheng, L.~Gao, W.~Liao, D.~Hong, B.~Zhang, X.~Cui, and J.~Chanussot,
  ``Coupled convolutional neural network with adaptive response function
  learning for unsupervised hyperspectral super resolution,'' \emph{IEEE
  Transactions on Geoscience and Remote Sensing}, vol.~59, no.~3, pp.
  2487--2502, 2021.

\bibitem{Lee1999LearningTP}
D.~D. Lee and H.~Seung, ``Learning the parts of objects by non-negative matrix
  factorization,'' \emph{Nature}, vol. 401, pp. 788--791, 1999.

\bibitem{Qu2019uDASAU}
Y.~Qu and H.~Qi, ``udas: An untied denoising autoencoder with sparsity for
  spectral unmixing,'' \emph{IEEE Transactions on Geoscience and Remote
  Sensing}, vol.~57, pp. 1698--1712, 2019.

\bibitem{Qian2020Spectral}
Y.~Qian, F.~Xiong, Q.~Qian, and J.~Zhou, ``Spectral mixture model inspired
  network architectures for hyperspectral unmixing,'' \emph{IEEE Transactions
  on Geoscience and Remote Sensing}, vol.~58, no.~10, pp. 7418--7434, 2020.

\bibitem{Palsson2021Convolutional}
B.~Palsson, M.~O. Ulfarsson, and J.~R. Sveinsson, ``Convolutional autoencoder
  for spectral–spatial hyperspectral unmixing,'' \emph{IEEE Transactions on
  Geoscience and Remote Sensing}, vol.~59, no.~1, pp. 535--549, 2021.

\bibitem{ranchin2000fusion}
T.~Ranchin and L.~Wald, ``Fusion of high spatial and spectral resolution
  images: the arsis concept and its implementation,'' \emph{Photogrammetric
  engineering and remote sensing}, vol.~66, no.~1, pp. 49--61, 2000.

\bibitem{wei2015fast}
Q.~Wei, N.~Dobigeon, and J.-Y. Tourneret, ``Fast fusion of multi-band images
  based on solving a sylvester equation,'' \emph{IEEE Transactions on Image
  Processing}, vol.~24, no.~11, pp. 4109--4121, 2015.

\bibitem{xu2019advanced}
Y.~Xu, B.~Du, L.~Zhang, D.~Cerra, M.~Pato, E.~Carmona, S.~Prasad, N.~Yokoya,
  R.~Hansch, and B.~Le~Saux, ``Advanced multi-sensor optical remote sensing for
  urban land use and land cover classification: Outcome of the 2018 ieee grss
  data fusion contest,'' \emph{IEEE Journal of Selected Topics in Applied Earth
  Observations and Remote Sensing}, vol.~12, no.~6, pp. 1709--1724, 2019.

\end{thebibliography}

\end{document}